\documentclass[%
 reprint,
 superscriptaddress,
 amsmath,amssymb,
 aps,
prb,
floatfix,
longbibliography 
]{revtex4-2}

\usepackage[utf8]{inputenc}
\usepackage{amsmath}
\usepackage{amssymb}
\usepackage{xcolor}
\usepackage{float}
\usepackage{graphicx}
\usepackage{url}
\usepackage{hyperref}
\usepackage{physics}
\definecolor{C0}{HTML}{1F77B4}
\definecolor{C1}{HTML}{FF7F0E}
\definecolor{C2}{HTML}{2CA02C}
\definecolor{C3}{HTML}{D62728}
\definecolor{C4}{HTML}{9467BD}
\definecolor{C5}{HTML}{8C564B}
\definecolor{C6}{HTML}{E377C2}
\definecolor{C7}{HTML}{7F7F7f}
\definecolor{C8}{HTML}{BCBD22}
\definecolor{C9}{HTML}{17BECF}

\newcommand{\tocless}{}

\def\DHLhksqrt#1#2{%
\setbox0=\hbox{$#1\sqrt{#2\,}$}\dimen0=\ht0
\advance\dimen0-0.2\ht0
\setbox2=\hbox{\vrule height\ht0 depth -\dimen0}%
{\box0\lower0.4pt\box2}}

\newcommand{\mymatrix}[4]{%
  \begin{matrix}
    #1 & #2 \\
    \rule{0pt}{4pt}#3 & #4
  \end{matrix}
}

\begin{document}
\title{Local interference algorithms for observables in large random circuits}

\author{Philippe Suchsland}
\affiliation{Google Quantum AI}
\email{suchsland@google.com}

\author{Brayden Ware}
\affiliation{Google Quantum AI}
\email{braydenware@google.com}

\author{Vladislav D. Kurilovich}
\affiliation{Google Quantum AI}

\author{Vadim Smelyanskiy}
\affiliation{Google Quantum AI}

\author{Kostyantyn Kechedzhi}
\affiliation{Google Quantum AI}

\date{\today}

\begin{abstract}
In quantum dynamics, it is generally believed that interference effects lead to a sign problem for classical Monte Carlo sampling methods, rendering sampling ineffective. 
In this paper we show that this is generically not the case for the task of simulating the evolution of observables in random instances of circuit dynamics.
Specifically, we consider two examples of observables which give large signals despite the randomness of the dynamics: first order out-of-time-ordered correlators (OTOCs) in unstructured random circuits without conservation laws and time-ordered charge-charge correlators (TOCs) in  particle number-preserving random circuits.
We show that these observables can be approximated by classical stochastic processes and that these approximations can be systematically improved with a series expansion that incorporates coherent corrections associated with clusters of gates.
We design and implement algorithms that combine Monte Carlo sampling and tensor network methods to approximate the series.
We observe numerically that the computational complexity of approximating TOCs and OTOCs scales polynomially in the circuit size and in the target precision.
We study the accuracy of these algorithms by directly mapping their ensemble-averaged errors to partition functions of replica statistical models, allowing for perturbative analysis and numerics in $1$d brickwork geometries.
Our theoretical analysis explains the empirical observation of polynomial complexity of these algorithms and is consistent with polynomial complexity extending to arbitrarily small precision and to the thermodynamic limit.

\end{abstract}
\maketitle

\section{Introduction}

Classically simulating the dynamics of \emph{arbitrary} quantum circuits requires resources that grow exponentially in qubit number or circuit depth\footnote{Assuming the standard complexity conjecture that $\mathsf{BQP} \nsubseteq \mathsf{BPP}$} --- but does this hardness extend also to \emph{typical} instances of quantum dynamics? 
While exponential quantum advantage has been generally expected to be a robust feature of quantum dynamics, theoretical evidence for this is strongest for sampling tasks. The boundaries of classical computation for the simulation of physical observables remain poorly defined.

The expectation that typical quantum dynamics is exponentially hard is based on two well-established ideas: that generic evolution quickly spreads states across Hilbert space, and that most states in Hilbert space are incompressible. 
These expectations can be made rigorous in the context of random quantum circuits, particularly when gates are chosen uniformly with respect to the Haar measure~\cite{Potter2022, Fisher_2023}. 
Theoretical understanding of random circuit dynamics has led to exact models of the growth of entanglement entropy~\cite{nahum_2017, zhou_2019}, operator entanglement entropy~\cite{jonay_2018}, temporal entanglement~\cite{Foligno_2023}, and magic~\cite{niroula_2024, turkeshi_2025}, each supporting claims of exponential cost for representing the resulting evolved states or operators as tensor networks~\cite{Verstraete_2004, Zwolak_2004, Prosen_2007, schuch_2008} or as superpositions of stabilizer states~\cite{Bravyi_2016, bravyi_2019, leone_2022}. 

However, hardness results for classically simulating the evolution of the entire wavefunction do not generically translate into hardness results for simulating observables. 
A variety of algorithms that target the evolution of local observables have been proposed, based on the expectation that thermalization dampens sensitivity of local observables to global details of the quantum states~\cite{altman_ehud_computing_2018, 2021-Lerose-Abanin-InfluenceMatrixFloquetDynamics, White_2018, Rakovsky_2020, Kvorning_2022, Parker_2019, rudolph2023classical, beguvsic2023converged,schuster2026probingclassicalcomplexityquantum}. 
These \emph{local information} algorithms implement evolution of operators or density matrices, represented for example using tensor networks~\cite{White_2018, Rakovsky_2020, Kvorning_2022} or as sparse superpositions of Pauli operators~\cite{Parker_2019, White_2023, rudolph2023classical, beguvsic2023converged}, while truncating operator components with large spatial support to manage the growth of complexity.
This strategy can be studied rigorously in the context of random quantum circuits, where it has been shown to lead to efficient algorithms for estimating local observables in unstructured Haar random circuit ensembles~\cite{angrisani2024classicallyestimatingobservablesnoiseless} and in charge-conserving circuit ensembles~\cite{von_Keyserlingk_2022}.
The existence of these algorithms challenges the search for dynamics with robust, repeatable, and exponential-scaling quantum advantage.

\begin{figure*}
    \centering
    \includegraphics[scale=0.33]{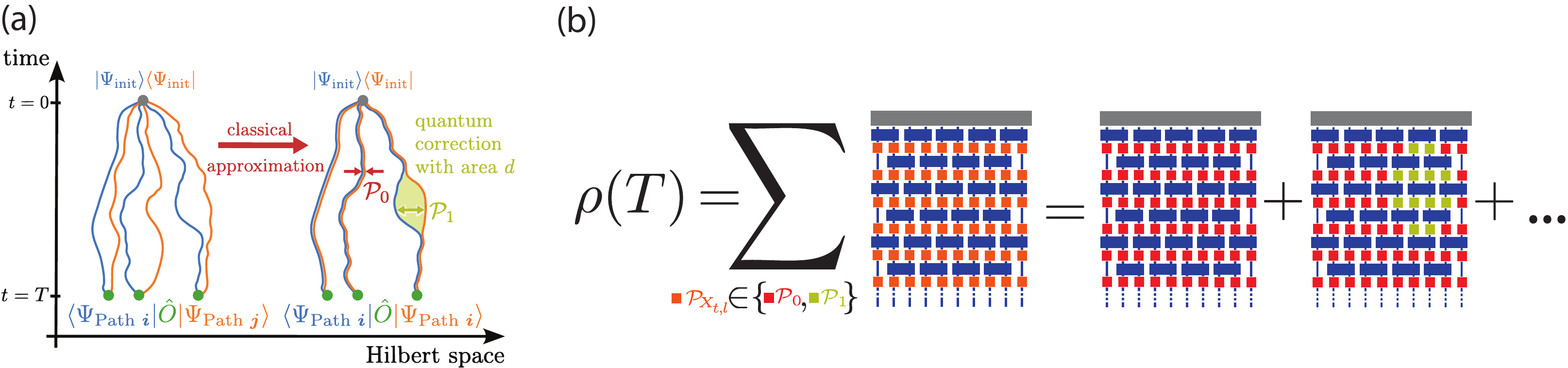}
    \caption{(a) The TOCs and OTOCs considered in this paper can be expressed as path integrals, with contributions in the sum associated with a forward-evolved trajectory (blue) and backward-evolved trajectory (orange) which coincide at the final time. In circuits with randomly drawn gates, these path integrals are dominated by terms with completely paired trajectories, the collection of which has a classical stochastic description. In this work, we consider terms involving local deviations from paired trajectories. We show that such contributions are both efficient to evaluate and sufficient for accurate computation in large random circuit instances. (b) Decomposition of the density matrix, $|\Psi_\mathrm{init}(T)\rangle\langle \Psi_\mathrm{init}(T)|$ for TOCs and $B(T)\otimes B(T)$ for OTOCs, into parts. The density matrix $\rho(T)$ is decomposed into a sum over projection configurations. We refer to $X_{t,l}=0$ at all $t,l$, i.e., $\rho$ is projected onto its diagonal, as the classical signal. The insertion of the projector $\mathcal{P}_0$ is equivalent to a reduction to the sum over diagonal elements, $|a\rangle \langle a|$ for TOCs, $\boldsymbol\sigma_a\otimes\boldsymbol\sigma_a$ for OTOCs, with prefactors $p_a$ so that the sum can be efficiently computed via sampling.
    }
    \label{fig:overview_results}
\end{figure*}

Given these challenges, one natural direction for the search for quantum advantage is the measurement of non-local observables. 
The out-of-time-ordered correlator (OTOC) has been proposed as a promising candidate due to both experimental accessibility and indications that it is challenging for conventional tensor network algorithms~\cite{Mi_2021, 1802.00801, 1901.05793}.
OTOCs probe the ballistic spreading of local perturbations into scrambled global information~\cite{vonKeyserlingk_2018};
for example, the first-order infinite temperature OTOC $\mathrm{Tr}(B(t) M B(t) M)$ measures the growth of the evolved local operator $B(t)$ by probing whether the operator commutes with a spatially separated probe operator $M$. 
The truncations of the operator evolution $B(t)$ used in local information algorithms deliberately introduce bias towards small Pauli weight components, leading to qualitatively incorrect operator spreading.

To fix this bias, Ref.~\onlinecite{abanin2025constructiveinterferenceedgequantum} introduced Monte Carlo operator evolution methods which instead manage the growth of complexity by randomly sampling Pauli string components to keep at each step. 
In these methods, qualitatively correct ballistic spreading of the operator is reconstructed in the average over many trajectory samples for any amount of truncation --- even when only one Pauli component is kept at each time step.
However, operator components from distinct random trajectories cannot interfere at later times; this discarding of interference introduces approximation error whenever sampling is used, even in the limit of infinite samples.
The trade-off between computational complexity and accuracy of several Monte Carlo operator evolution methods was studied numerically in Ref.~\cite{abanin2025constructiveinterferenceedgequantum}, but the question of whether these methods could both be efficient and have controlled approximation error remained open.

In this work, we resolve this question by introducing a new class of algorithms, which we dub \emph{local interference} algorithms.  %
These methods employ Monte Carlo trajectory evolution, but also add a scheme to correct for the removal of interference caused by sampling.
In the first part of this paper, we explain the unifying principle behind these algorithms: interference effects can be sorted by a degree of locality, measured by a set of locations in space-time that participate in the interference, with more local interference effects having larger influence on the value of observables.
We show that incorporating interference corrections up to a fixed cut-off volume $d$ and discarding less local corrections leads to algorithms that are both efficient in run-time and increasingly accurate as $d$ is increased.
This strategy applies equally well to time-ordered correlators (TOCs) as to OTOCs, by replacing operator evolution with Pauli basis sampling by sampling of states in a computational basis; we consider both applications throughout the paper.

In the second part of this paper, we study the locality of interference corrections for OTOCs in $1$d Haar random circuits and for TOCs in $U(1)$-conserving random circuits. Our key result in both cases is a prediction that the size of the interference corrections to the observable scales as a stretched-exponential $\exp(-\sqrt{d})$ in the space-time volume $d$ of the sites that participate in the interference. This result is argued for through a detailed understanding of the statistical models that govern the ensemble-averaged moments of Haar random circuit dynamics, and confirmed with explicit numerical computations  of the ensemble-averaged moments.
The rapid decay of interference corrections in their volume implies that algorithms that compute only local corrections can achieve accurate approximations.

In the final part of the paper, we design and analyze specific local interference algorithms for the computation of $1$D TOCs and OTOCs. 
The first algorithm uses exact tensor network contraction on large circuit blocks, with sampling used at the boundary between distinct circuit blocks, along with an inclusion-exclusion subtraction scheme for correcting missed interference effects.
The second method uses no sampling at all, but instead uses truncated MPS evolution of the master equation associated with the interference-corrected trajectory ensemble.
In both cases, the empirical performance of the algorithms suggests complexity that is polynomial in circuit size and in approximation accuracy when applied to random circuit OTOCs or TOCs.
These results highlight the breadth of the design space available for local interference algorithms and point the way towards applications to more general quantum dynamics applications.

\begin{figure*}
    \centering
    \includegraphics[width=\linewidth]{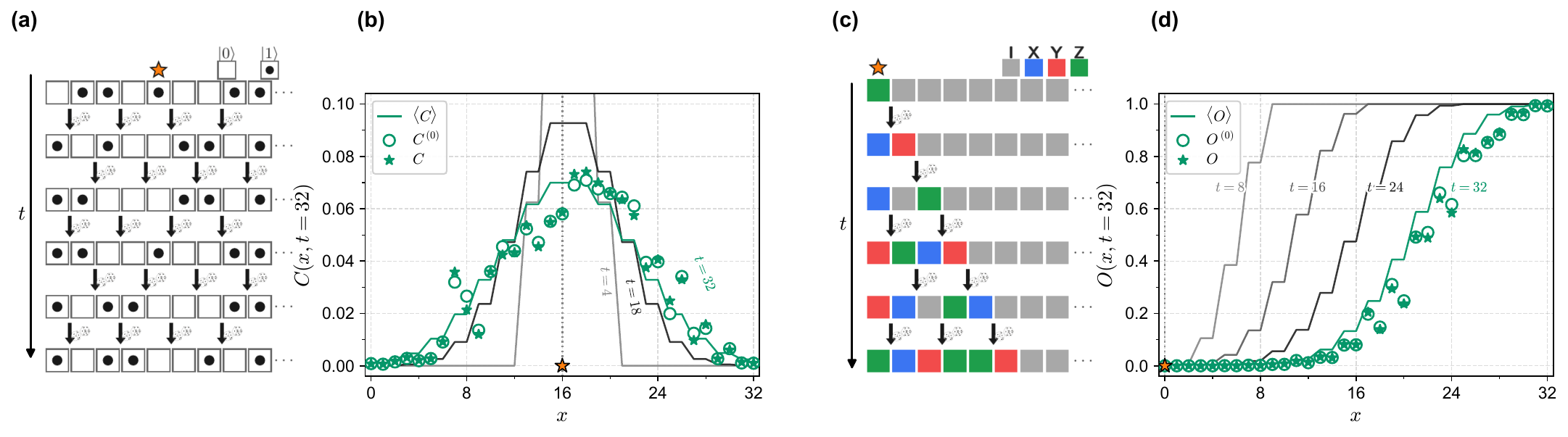}
    \caption{Diagonal approximation for TOCs and OTOCs. (a) For TOCs of $U(1)$-conserved circuits, the diagonal approximation $C^{(0)}$ describes the classical random hopping of charged particles, known as the simple symmetric exclusion process. The hopping rates are specific to each circuit instance. (b) For a single circuit instance of depth $T=32$ and size $L=33$, the values of the correlator $C(x, t)$ (stars) and compared to the corresponding diagonal approximation $C^{(0)}(x, t)$ (circles). (c) For OTOCs, the diagonal approximation $O^{(0)}(x, t)$ instead corresponds to a stochastic description of operator growth in the Pauli basis. (d) Comparison of the diagonal approximation $O^{(0)}(x, t)$ and the exact OTOC $O(x, t)$ in a single circuit instance of depth $T=32$ and size $L=33$.
    }
    \label{fig:montecarlo}
\end{figure*}

\section{Stochastic trajectory sampling and interference corrections}
\label{sec:motivation}

\subsection{Diagonal approximation}\label{sec:montecarlo}

The core of our approach to simulating TOCs and first-order OTOCs is Monte Carlo sampling of operator components. We consider first a maximal version of this sampling algorithm, keeping only a single Pauli string component at each time step. This results in a classical random walk for the evolution of the operator $B(t)$ over Pauli strings. For TOCs, we have an analogous evolution by sampling the evolved wavefunction in the computational basis, yielding a classical random walk among computational basis configurations.
To illuminate the relation of these classical stochastic evolutions to the original quantum evolution, we use a path integral expansion of the TOC and OTOC, showing that in each case they can be expressed through a complex-weighted sum over all pairs of trajectories in the classical evolution, as pictured in Fig.~\ref{fig:overview_results} (a). The stochastic classical evolution reduces in this picture to a sum over only diagonal trajectory pairs that match throughout the evolution. We show below that these diagonal trajectory pairs in the path integral naturally have positive amplitudes. This motivates an interpretation of these diagonal trajectory pairs as saddle-points in a path integral representation, and suggests that local fluctuations around these saddle points will dominate the entire sum.

To be concrete, we write these path integrals explicitly for OTOCs and TOCs. Consider circuits $U(T) = u_T u_{T-1} \ldots u_1$ with $T$ layers and acting on $L$ total qubits, with each layer composed of two-qubit gates. %
For the TOC $C(x, T)$ at time $T$ and position $x$, 
\begin{gather}
C(x, T) = \mathrm{Tr}[U^{\dagger}(T) \sigma^z_{x} U(T)\sigma^z_{0}]/2^L,
\end{gather}
we insert resolutions of the identity in the computational basis between each time step, yielding a sum over pairs of trajectories $\mathbf{y}, \mathbf{y'}$:
\begin{equation}
\begin{gathered}
C(x, T) =2^{-L}\sum_{\mathbf{y}, \mathbf{y'}} (-1)^{y_{0,0} + y_{T, x}} \delta_{\mathbf{y}_T, \mathbf{y'}_T} \delta_{\mathbf{y}_0, \mathbf{y'}_0} \left[ \vphantom{\prod_t} \right.\\\prod_{t}   \left. \bra{\mathbf{y'}_{t-1}} u^{\dagger}_{t} \ket{\mathbf{y'}_{t}}\bra{\mathbf{y}_{t}} u_{t} \ket{\mathbf{y}_{t-1}} \vphantom{\prod_t} \right].
\end{gathered}
\label{eq:TOCpathintegral}
\end{equation}
Here, $\mathbf{y}_t$ and $\mathbf{y'}_t$ denote the computational basis states inserted between the steps of the forward and backwards evolutions $U$ and $U^{\dagger}$, respectively. The diagonal trajectory pairs are defined precisely as those where $\mathbf{y}_t=\mathbf{y'}_t$ for all times $t$.
By restricting to everywhere matched trajectory pairs, we obtain the following diagonal approximation for $C(x, T)$:
\begin{equation}
\begin{gathered}%
    C^{(0)}(x, t) = \sum_{\{\mathbf{y}\}} s(\mathbf{y}) p(\mathbf{y}_t|\mathbf{y}_{t-1})\cdots p(\mathbf{y}_1|\mathbf{y}_0),
\end{gathered}
\label{eq:TOCdiagonal}
\end{equation}
where $s(\mathbf{y})=(-1)^{y_{t, x} + y_{0,0}}$.
Crucially, this diagonal part encodes a classical stochastic Markov process with transition probabilities $p(\mathbf{y}_{t}|\mathbf{y}_{t-1})=|\langle \mathbf{y}_{t}|u_{t}|\mathbf{y}_{t-1}\rangle|^2$; by unitarity of $u_t$, these probabilities are all positive and sum to one. 

If we restrict to charge-conserving circuit dynamics built from two-qubit gates, the specific stochastic process that is encoded describes particles randomly hopping among sites, in a manner known as the \emph{simple symmetric exclusion process} or SSEP. For a given random circuit instance, the hopping probabilities are disordered in space and time, as they are determined by the randomly chosen gates.
An illustration of this stochastic process is shown in Fig.~\ref{fig:montecarlo}(a). By further introducing an average over the circuits taken from a suitable Haar ensemble of charge-conserving circuits, the average correlation function $\langle C(x, t) \rangle$ is also described by SSEP, but with uniform hopping coefficients~\cite{Khemani_2018}. Notably, when averaged over the ensemble the diagonal approximation is exact. In either the clean SSEP for the ensemble average or the disordered SSEP derived from a given circuit instance, the dynamics of the correlation function shows diffusive spreading, as illustrated in Fig.~\ref{fig:montecarlo}(b).

The diagonal approximation for the OTOC is formed in an analogous manner. For simplicity, we always consider the case where the butterfly operator $B$ and measurement operator $M$ are Pauli-$Z$ operators on single sites, creating the OTOC $O(x, T)$  at time $T$ and position $x$:
\begin{gather}
O(x, T) = \frac{1}{2^L}\mathrm{Tr}[U^{\dagger}(T) \sigma^z_{x} U(T) \sigma^z_{0} U^{\dagger}(T) \sigma^z_{x} U(T)\sigma^z_{0}].
\end{gather}
By rearranging and expanding $B(t) = U(t) \sigma^z_0 U^{\dagger}(t)$ into complete Pauli operator basis $\boldsymbol\sigma_{t}$ at each intermediate time $t$, we obtain
\begin{equation}
\begin{gathered}
O(x, T) =\sum_{\boldsymbol{\sigma}, \boldsymbol{\sigma'}} \delta_{\boldsymbol\sigma_T, \boldsymbol\sigma'_T} s(\boldsymbol\sigma) \delta_{\boldsymbol\sigma_0, \sigma^z_0} \delta_{\boldsymbol\sigma'_0, \sigma^z_0}  \left[ \vphantom{\prod_t} \right.\\\prod_{t}   \left. \mathrm{Tr} (\boldsymbol\sigma'_{t} u_{t} \boldsymbol\sigma'_{t-1} u^{\dagger}_t)\mathrm{Tr} (\boldsymbol\sigma_{t} u_{t} \boldsymbol\sigma_{t-1} u^{\dagger}_t)/4^L\vphantom{\prod_t} \right],
\end{gathered}
\label{eq:OTOCpathintegral}
\end{equation}
where the sign $s(\boldsymbol\sigma)$ takes the value $1$ if $\boldsymbol\sigma_T$ commutes with $\sigma^z_x$ and $-1$ otherwise.
Once again, the diagonal approximation is defined through the restricted sum where the trajectories $\boldsymbol\sigma, \boldsymbol\sigma'$ are everywhere identical:
\begin{equation}%
    O^{(0)}(x, T) = \sum_{\{\boldsymbol\sigma\}} s(\boldsymbol\sigma) p(\boldsymbol\sigma_T|\boldsymbol\sigma_{T-1})\cdots p(\boldsymbol\sigma_1|\boldsymbol\sigma_0),
\label{eq:OTOCdiagonal}
\end{equation}
where the probabilities $p(\boldsymbol\sigma_t|\boldsymbol\sigma_{t-1}) = \mathrm{Tr} (\boldsymbol\sigma_{t} u_{t} \boldsymbol\sigma_{t-1} u^{\dagger}_t)^2/4^L$ are once again positive and sum to $1$ by unitarity of $u_{t}$. The boundary condition restricts this sum so that each trajectory starts with the initial butterfly operator $\boldsymbol\sigma_0=\sigma^z_0$.

Like for the case of TOC, the diagonal approximation encodes a classical stochastic Markov process. This process mimics the properties of the full quantum operator growth: the trajectory $\boldsymbol\sigma_{t}$ randomly walks among Pauli strings whose endpoints move ballistically outward from the initial site, as pictured in Fig.~\ref{fig:montecarlo}(c). When averaged over Haar random circuits, the classical stochastic process governing $O^{(0)}(x, t)$ simplifies even further: the weights of all non-identity Pauli strings become identical in the average, with the average OTOC $\langle O(x, t) \rangle$ being completely described by the dynamics of the right-most non-identity site in the Pauli string~\cite{Rakovszky_2018}. However, for individual circuit instances the stochastic hopping rates differ among sites and among the distinct Pauli strings, allowing the approximation $O^{(0)}(x, t)$ to closely mimic the exact values of $O(x, t)$ in each circuit instance, as shown in Fig.~\ref{fig:montecarlo}(d).

\subsection{Formalism for interference corrections}\label{sec:formalism}

Remarkably, the diagonal approximation is quite accurate for typical random circuit instances, despite the fact that it completely replaces the quantum evolution with a classical evolution. This motivates a simple strategy for improving the approximation: instead of restricting the sums in the path integrals Eq.~\eqref{eq:TOCpathintegral} and Eq.~\eqref{eq:OTOCpathintegral} to exactly identical trajectory pairs, we include trajectory pairs which deviate from each other a small amount. We notate these restricted sums as follows:
\begin{equation}
O^{(d)}(x, t) =\sum_{\text{valid }|\boldsymbol{\sigma}-\boldsymbol{\sigma'}|\leq d} s(\boldsymbol\sigma) A(\boldsymbol\sigma, \boldsymbol\sigma')\vphantom{\prod_t},
\end{equation}
where $A(\sigma, \sigma')$ is the amplitude in Eq.~\ref{eq:OTOCpathintegral} for the trajectory pair, and where $|\sigma - \sigma'|$ is the number of space-time locations in which the trajectories differ from each other. A completely analogous notation is used for TOCs.

In fact, we find it helpful to use an even more fine-grained grouping of the interference terms, which group trajectories by the exact set of places where they differ. We show now that this grouping can be given a physical interpretation related to the sensitivity of the terms to noise events that occur at heralded locations in space-time.
We start with U(1)-conserving circuits and extend the formalism to OTOCs next. To simplify the notion of pairing paths, we use a density matrix description $\rho = \sum_{a,b}\langle a|\rho|b\rangle |a\rangle \langle b|$ with bitstrings $a,b$ in the computational basis so that paired paths $a=b$ correspond to the diagonal matrix elements of $\rho$. More precisely, we define the Hamming distance of the matrix element $|a\rangle\langle b|$ as $||a-b||_1$, where $||a-b||_1$ denotes the Hamming distance between the bitstrings $a$ and $b$. In this basis, we can obtain single-qubit projectors on the Hamming distance $\alpha=0,1$ subspace:
\begin{align}
    \mathcal{P}_\alpha[\rho] &= \rho/2 +(-1)^\alpha \sigma^z \rho \sigma^z/2.
\end{align}
Using the projectors $\mathcal{P}_\alpha[\rho]$, the phase flip channel with rate ${\gamma}$.%
can be written as
\begin{align}\label{eq:noise_channel_proj}
    \mathcal{E}[\rho] &= (1-\gamma) \rho + \gamma \sigma^z \rho \sigma^z \nonumber\\
                      &= \mathcal{P}_0[\rho] +(1-2\gamma)\mathcal{P}_1[\rho].
\end{align}
For multiple qubits, we can analyze the dephasing sensitivity of a multi-qubit-state by creating multiqubit projectors $\mathcal{P}^{\otimes{L}}_{\vec{\alpha}} = \mathcal{P}_{\alpha_1} \otimes \mathcal{P}_{\alpha_2}\ldots \otimes \mathcal{P}_{\alpha_L}$ and using that the projectors are a resolution of the identity $\mathcal{P}_0+\mathcal{P}_1=\mathcal{I}$, which implies $\sum_{\vec{\alpha}} \mathcal{P}^{\otimes{L}}_{\vec{\alpha}} = \mathcal{I}^{\otimes L}$. 

This enables a decomposition of density matrices $\rho$ using these projectors, Fig.~\ref{fig:overview_results} (b). For instance, at the single-qubit level, we decompose the single qubit density matrix $\rho= I/2 + \sum_\nu \rho_\nu \sigma^\nu$ with Pauli matrices $\sigma^\nu$ into the diagonal and off-diagonal part $\rho \equiv \mathcal{P}_0[\rho] +\mathcal{P}_1[\rho]$. The diagonal part $\rho_0=\mathcal{P}_0[\rho]=I/2 + \rho_z \sigma^z$ has Hamming distance $0$, is invariant under the phase flip channel $\mathcal{E}[\rho_0] = \rho_0$, and has no coherence in that sense. Analogously, the off-diagonal part $\mathcal{P}_1[\rho]=\rho_x \sigma^x + \rho_y \sigma^y$ has Hamming distance $1$, is maximally sensitive in terms of its trace norm~\footnote{For $\gamma \leq 1/2$, given any $O=o_I I + o_x\sigma^x + o_y \sigma^y + o_z\sigma^z$ with $|O|^2=1$, then $\Gamma = |\mathcal{E}[O]|^2 = o_I^2+o_z^2 + (1-2\gamma)[o_x^2+o_y^2] \geq (1-2\gamma) = |\mathcal{E}[\mathcal{P}_1[O]]|^2 / |\mathcal{P}_1[O]|^2= \Gamma_{\mathrm{off-diagonal}}$}, and thus can be regarded as maximally coherent. This establishes a direct relation between the Hamming distance and sensitivity to decoherence induced by phase flip noise. Similarly, we can decompose multiqubit density matrices using $\mathcal{P}^{\otimes L}_{\vec{\alpha}}$. 

The diagonal approximation of the path integral form of the TOC $C(x, t)$ is equivalent to the phase-incoherent dynamics where $\mathcal{P}_0$ is inserted after every gate.
Specifically, for a circuit $U$ of the form
\begin{align}
    U = u_{T}\ldots u_2 u_1,
\end{align}
where each layer $u_t$ is composed of two-site gates $V_{j,j+1}$ as
\begin{align}
    u_{t} = \begin{cases}  \prod_{j} V_{2j,2j+1} & t \, \mathrm{even} \\ \prod_{j} V_{2j+1,2j+2} & t \, \mathrm{odd}
    \end{cases},
\end{align}
we can write the evolution of the density matrix with the layer $u_t$ by defining the superoperator $\mathcal{U}_t$
\begin{align}
    \rho(t) = \mathcal{U}_t[\rho(t-1)] = u_t \rho(t-1) u_t^\dagger.
\end{align}
Then the whole time evolution of the initial density matrix $\rho_\mathrm{init}$ can be written as
\begin{align}
    \rho(T) = (\mathcal{U}_{T} \circ \cdots \circ \mathcal{U}_1)[\rho_\mathrm{init}].
\end{align}

For OTOCs, the same formalism can be applied by identifying the Hamming distance between two Paulistrings $\boldsymbol{\sigma}$, $\boldsymbol{\sigma}'$ as that of the matrix element $\boldsymbol{\sigma} \otimes \boldsymbol{\sigma}'$. Hence, the single site projector onto the Hamming distance $0$ subspace is for OTOCs
\begin{align}
    \mathcal{P}_0[\sigma^a \otimes \sigma^b] = 4^{-1}\!\!\!\!\!\!\sum_{c\in \{0,x,y,z\}} \!\!\!\!\!\!\sigma^c \sigma^a\sigma^c  \otimes \sigma^c \sigma^b\sigma^c = \delta_{a,b}  \sigma^a \otimes \sigma^a.
\end{align}
Generalizing the time evolution operator for OTOCs $\mathcal{U}_t[\boldsymbol{\sigma} \otimes \boldsymbol{\sigma}'] = u_t^\dagger  \boldsymbol{\sigma}u_t \otimes u_t^\dagger  \boldsymbol{\sigma}'u_t$ and identifying $\rho(t) \equiv B(t)\otimes B(t)$, allows to use the same formalism for TOCs and OTOCs in the following.

We can now decompose $\rho(T)$ into \textit{parts} $
\rho_X$, labeled by the matrix $X\in\{0,1\}^{T\times L}$,
using the resolution of the identity
$\mathcal{I}=\mathcal{P}_{0}+\mathcal{P}_{1}$ as follows
\begin{align}
    \rho(T) &= ( \cdots \circ (\mathcal{P}_{0}+\mathcal{P}_{1})^{\otimes L}\circ \mathcal{U}_2\circ(\mathcal{P}_{0}+\mathcal{P}_{1})^{\otimes L}\circ\mathcal{U}_1)[\rho_\mathrm{init}], \nonumber\\
    &= \sum_{X_{t,l} \in \{0,1 \}}\rho_{X}(T), \label{eq:path_resolved_density_matrix}
\end{align}
where we use the definition
\begin{align}
    \rho_{X}(T) &=(\mathcal{P}_{X_T}^L\circ \mathcal{U}_T\circ \cdots \circ \mathcal{P}_{X_2}^L\circ \mathcal{U}_2\circ\mathcal{P}_{X_1}^L\circ\mathcal{U}_1)[\rho_\mathrm{init}] \nonumber
\end{align}
and the $L$-site projectors $\mathcal{P}^L_{X_t}=\mathcal{P}_{{X}_{t,1}} \otimes \ldots \otimes \mathcal{P}_{{X}_{t,L}}$. The decomposition is shown in Fig.~\ref{fig:overview_results} (b). The resulting projected density matrix or part $\rho_X$ is either diagonal $X_{t,l}=0$ or off-diagonal $X_{t,l}=1$ at every site $l$ at every time step $t$.
Note that decomposing the density matrix in terms of Pauli strings, $\mathcal{P}_0$ on a site projects onto all Pauli strings with either Pauli matrices $I$ or $\sigma^z$ at that site, while $\mathcal{P}_1$ projects onto $\sigma^y$ or $\sigma^x$.

The remainder of the terms can be characterized by the summed Hamming distance $d$, which is the total number of off-diagonal projectors $\mathcal{P}_1$ in the part $\rho_X$
\begin{align}
d(X)=\sum_{t,l} {X}_{t,l} \label{eq:vlad}.
\end{align}
When applied to the computation of the TOC $C(x, t)$, the term containing $\rho_X$ consists exactly of the terms in the path integral in which the trajectory pairs $y, y'$ deviate at the space-time locations $X$.
Thus, the Hamming distance $d$ directly measures the sensitivity of such terms to dephasing noise. The accuracy of the diagonal approximation in $C(x, t)$ is in other words an indication that for typical random circuits, charge-charge correlators are insensitive to dephasing noise.

The same discussion can be applied in a nearly identical way to OTOCs. However, in this case the noise channel $\mathcal{P}_0$ is identified as the projector of two-copy density matrices onto the diagonal subspace $I \otimes I, \sigma^x \otimes \sigma^x,  \sigma^y \otimes \sigma^y,  \sigma^z \otimes \sigma^z$. We can identify this noise channel as copy-correlated depolarizing noise -- that is, noise which is generated by inserting randomly chosen Pauli errors into the circuit dynamics $U(t)$, so that they appear identically among the two copies of the evolution in $B(t) \otimes B(t)$. In this case, the accuracy of the diagonal approximation shown in Fig.~\ref{fig:montecarlo} shows that OTOCs are insensitive to the insertion of copy-correlated depolarizing noise.

\subsection{Series expansion in Hamming distance}\label{ssec:stoch_indep}

This motivates grouping the contributions to the TOCs and OTOCs by the total Hamming distance $d$.

We define a decomposition of the density matrix according to Hamming distance $d$
\begin{align} \label{eq:coherence_decomp}
    \rho(T) &= \sum_{d} \rho_d(T),\quad \rho_d(T)=\sum_{X \text{ with } d(X) = d} \rho_{X}(T). %
\end{align}
In order to evaluate the decoherence (i.e., phase flip) sensitivity for a measurement

$\langle O \rangle_{\rho(T)}=\mathrm{Tr}[\rho(T) O]/2^L$ with observable $O$, we truncate the sum in Eq.~\eqref{eq:coherence_decomp} to $d\leq D$ yielding an approximation to the density matrix
\begin{align}\label{eq:density_matrix_truncation}
    \rho(T)\approx\rho_{D}(T) = \sum_{d=0}^D \rho_d(T).
\end{align}
We quantify the quality of the approximation by computing the difference between $\langle O \rangle_{\rho_D(T)}$ and the exact result $\langle O \rangle_{\rho(T)}$. Squaring and averaging this difference by integrating the unitaries $V$ over the random ensemble they are selected from (denoted by an overline) yields the mean squared error (MSE), which quantifies the overall approximation quality. In App.~\ref{app:stoch_indep_path}, we show that if the Kraus operators of the projectors $\mathcal{P}_\alpha$ form a subgroup of the group over which the two-qubit gates are averaged, then the parts $\rho_X$ are stochastically independent
\begin{align}
\overline{\rho_{X_1}\otimes\rho_{X_2}}=\delta_{X_1,X_2}\overline{\rho_{X_1}\otimes \rho_{X_1}}.
\end{align}
Consequently, the MSE reduces to a sum over the second moments with Hamming distance $d$ 
\begin{align} \label{eq:mse_coherence_resolved}
    \mathrm{MSE}(T) &= \overline{ \left|\langle O \rangle_{\rho(T)} -\langle O \rangle_{\rho_D(T)}\right|^2} = \sum_{d=D+1} \overline{\langle O \rangle_{\rho_{d}(T)}^2}.
\end{align}
As every part $\rho_X$ of the density matrix provides an independent correction to the classical part $\rho_0$, including any additional part just reduces the MSE monotonically by the respective second moment of that part $\overline{\langle O \rangle_{\rho_X(T)}^2}$. Intuitively, this monotonic behavior can be understood by noting that every insertion of a projector $\mathcal{P}_0$ yields an additional average: it does not change the circuit mean, but reduces the variance. The stochastic independence of the different parts motivates the analysis of the density matrix in terms of its individual parts, as in Eq.~\eqref{eq:path_resolved_density_matrix}. %

Similarly, the Hamming distance resolved density matrices $\rho_d(T)$, Eq.~\eqref{eq:coherence_decomp}, are stochastically independent, allowing us to analyze the decoherence sensitivity of $\rho(T)$ by decomposing it into $\rho_d(T)$. Most importantly, this makes the approximation $\rho(T)\approx\rho_D(T)$ controlled in the sense that the MSE decreases monotonically in $D$, motivating including corrections to $\rho_0(T)$ based on their Hamming-distance $d$, as will be discussed in Sec.~\ref{ssec:addition_algo}.

We proceed with a systematic study of the decoherence sensitivity by resolving the variance
\begin{align}
    \mathrm{Var}\langle O \rangle_{\rho_d}=\overline{\langle O \rangle_{\rho_{d}}^2}-\delta_{d,0}\overline{\langle O \rangle_{\rho_{0}}}^2 \label{eq:resolved_var}
\end{align}
in terms of the Hamming distance. We study the variances rather than the second moments as all variances $\mathrm{Var}\langle O \rangle_{\rho_d}$ follow the same scaling with time, namely $T^{-5/2}$ for TOCs and coincide with the second moments except for $d=0$, where the second moment decays more slowly: $\overline{\langle O \rangle_{\rho_{0}}^2}\sim T^{-1}$.

\section{Random circuit analysis of interference corrections}\label{sec:randomcircuit}

The targets for our algorithms are the values of OTOCs and TOCs in individual random circuit instances. The average value of these observables over a Haar random circuit ensemble is easily computed with a simple classical model in each case---both OTOCs and correlators experience strong circuit-to-circuit fluctuations.
To characterize circuit-to-circuit fluctuations, we perform the statistical mechanics mapping in the problem of computing the second moment of the observables.  
Both for OTOCs and correlation functions, the problem acquires a form of the dynamics of Markovian random walkers.
These dynamics are tractable both numerically and analytically.
The mapping allows us to show that, in both cases, the signal variance decays as a power-law of the circuit depth~$T$:
\begin{equation}
    \mathrm{Var}\,C(T) \propto \frac{1}{T^{5/2}},\quad \mathrm{Var}\,{\rm OTOC}(T) \propto \frac{1}{T^{1/2}},
\end{equation}
where OTOC is assumed to be evaluated at the wavefront.
Interference contributions only enhance the prefactor in these proportionalities.
We note that the scaling of the fluctuations of OTOCs was previously predicted in Ref.~\cite{nahum2018} and further analyzed in Ref.~\cite{abanin2025constructiveinterferenceedgequantum}. We expand on the latter analysis further in this paper, showing a direct agreement between analytical calculations and numerics.

We proceed with introducing a quantitative measure for the role of the interference. As a starting point, we note that the diagonal approximation is, in fact, identical to considering a modified problem where a maximal dephasing noise is present on all qubits. Dephasing noise with finite strength $\delta$ suppresses off-diagonal components of the system's density matrix exponentially.
Formally, we express this by decomposing the fluctuations of the signal---measured by the variance ${\rm Var}\,C(T,\delta)$---according to its dephasing sensitivity:
\begin{align}\label{eq:hamming_dist_dephasing_sens}
     \frac{\mathrm{Var}\,C(T,\delta)}{\mathrm{Var}\,C(T,\delta=0)} = \sum_{d=0}^\infty r_d(T) e^{-\delta d},
\end{align}
so that the fraction of the variance $r_d$ decays exponentially in $d$ and $\delta$. The variable $d$ may be thought of as a measure for the space-time volume sensitive to noise; it corresponds to the total distance between the two nearly paired paths. 
A similar equation also applies to OTOCs, with an appropriately generalized notion of dephasing sensitivity.
In each case, we show that $r_d$ decays rapidly with $d$. This specifies in which way the interference corrections are small.
 
According to Eq.~\eqref{eq:hamming_dist_dephasing_sens}, the weakness of interference contributions in these problems can also be viewed as an insensitivity to dephasing noise. 
Similar approaches to exploiting  insensitivity to noise for simplifying  classical simulations of correlators appear in several prior works, in the realm of matrix-product states~\cite{Rakovsky_2020,von_Keyserlingk_2022,White_2018}, Pauli string expansion~\cite{schuster_2024_paulistrings_noise,angrisani2024classicallyestimatingobservablesnoiseless}, and entanglement~\cite{Noh_2020}. However, the resulting algorithms in these papers take very different forms from those discussed here, which heavily exploit sampling of classical Markov processes. Additionally, none of these papers apply the concept to OTOCs, which were previously thought to be challenging for classical algorithms.

The first key result is that for all cases considered of algebraically decaying $\mathrm{Var}\,C(T)$ in random circuits in the long-time-limit, $r_d$ becomes independent of time.
This means that the interference corrections do not build up over time; in other words, the value $C$ does not become more noise-sensitive with increasing circuit depth. The noise-insensitive term $r_0$ dominates, and $r_d$ decays rapidly with $d$. In particular, for charge-charge correlators in $U(1)$ Haar-random circuits we find $r_d \sim e^{-\sqrt{d}}$. 
A parallel result appears for the signal of OTOCs in Haar-random circuits. In this case, we reformulate OTOCs as correlation functions in a system of two independent copies of the Hilbert space, undergoing the joint evolution $U(t)\otimes U(t)$. The OTOC is then shown to be insensitive to a depolarizing noise channel that acts identically on both copies. 
We find that in the vicinity of the operator front, $x \sim v_B T \pm \mathcal{O}(\sqrt{T})$, $r_d$ again decays as a stretched exponential $r_d \sim e^{-\sqrt{d}}$ and is asymptotically independent of time. In contrast, far behind the front $x \ll v_B T$, the signal is highly sensitive to global interference effects; however, the signal is exponentially small in this domain. 

\subsection{TOCs in charge-conserving circuits}\label{sec:result_charge_cons_circuits}

Here, we study correlation functions $C=\mathrm{Tr}[\sigma^z_{0}(T)\sigma^z_{0}]/2^L$ with $O=\rho_\mathrm{init} = \sigma^z_{0}$. Results for expectation values can be found in App.~\ref{app:expectation_values_u1_circuits}. While the initial operator $\sigma^z_0$ is not formally a density matrix, this does not affect any previous or following statements. We use that $\sigma^z_0$ can be written as a sum of density matrices $\sigma^z_0 = \sum_a s^a |a\rangle \langle a|$ in terms of computational basis states $|a\rangle$ and a sign $s^a=\pm 1$.

As discussed before, the charge conservation provides a natural basis to study decoherence in, so that phase flip noise commutes with the charge operator $Q^z$. In the first part of this section, we will show numerically that quantum corrections in this basis decay as $e^{-1.8 \sqrt{d}}$, where $d$ is the total Hamming distance between paths or the area enclosed by them, see Fig.~\ref{fig:overview_results} (a). Based on this insight we propose a classical algorithm including quantum corrections with increasing $d$ systematically. Note that both the initial state $\rho_\mathrm{init} = \sigma^z_0$ and the final $\sigma^z_0$-probe are diagonal and therefore invariant under phase flip noise. Hence, parts $\rho_X(T)$ with off-diagonal projectors $\mathcal{P}_1$ outside the lightcone vanish $\rho_X(T)=0$.

In the second part of this Sec.~\ref{ssec:analytical_dynamics_u_1}, we present an analytical analysis. Details are provided in App.~\ref{ssec:analytcal_u1_appendix}. It shows how the charge conservation causes correlations to decay polynomially in time: The circuit averaged correlation function $\overline{\mathrm{Tr}[ \sigma^z_0(T)\sigma^z_0]}$ decays as $T^{-1/2}$ and the circuit-to-circuit variance decays as $T^{-5/2}$ for large time. Furthermore, we will argue that parts with a single quantum correction, as shown in Fig.~\ref{fig:overview_results} (a), also generate fluctuations scaling as $T^{-5/2}$. Parts with multiple separate interactions scale as $T^{-4}$ at best, and hence, are subleading. Hence, charge-correlator fluctuations in random circuits are generated by interactions localized in space and time and not spread out. This motivates the development of the classical algorithm exploiting the fact that interactions are rare events in space and time.

\subsubsection{Numerical Results}
We analyze numerically the behavior of the variance $\mathrm{Var}\langle O \rangle_{\rho_d}=\overline{\langle O \rangle_{\rho_{d}}^2}-\delta_{d,0}\overline{\langle O \rangle_{\rho_{0}}}^2$ generated by the parts of the density matrix with Hamming distance $d$, Eq.~\eqref{eq:coherence_decomp}.

\begin{figure}[H]
    \centering
    \includegraphics{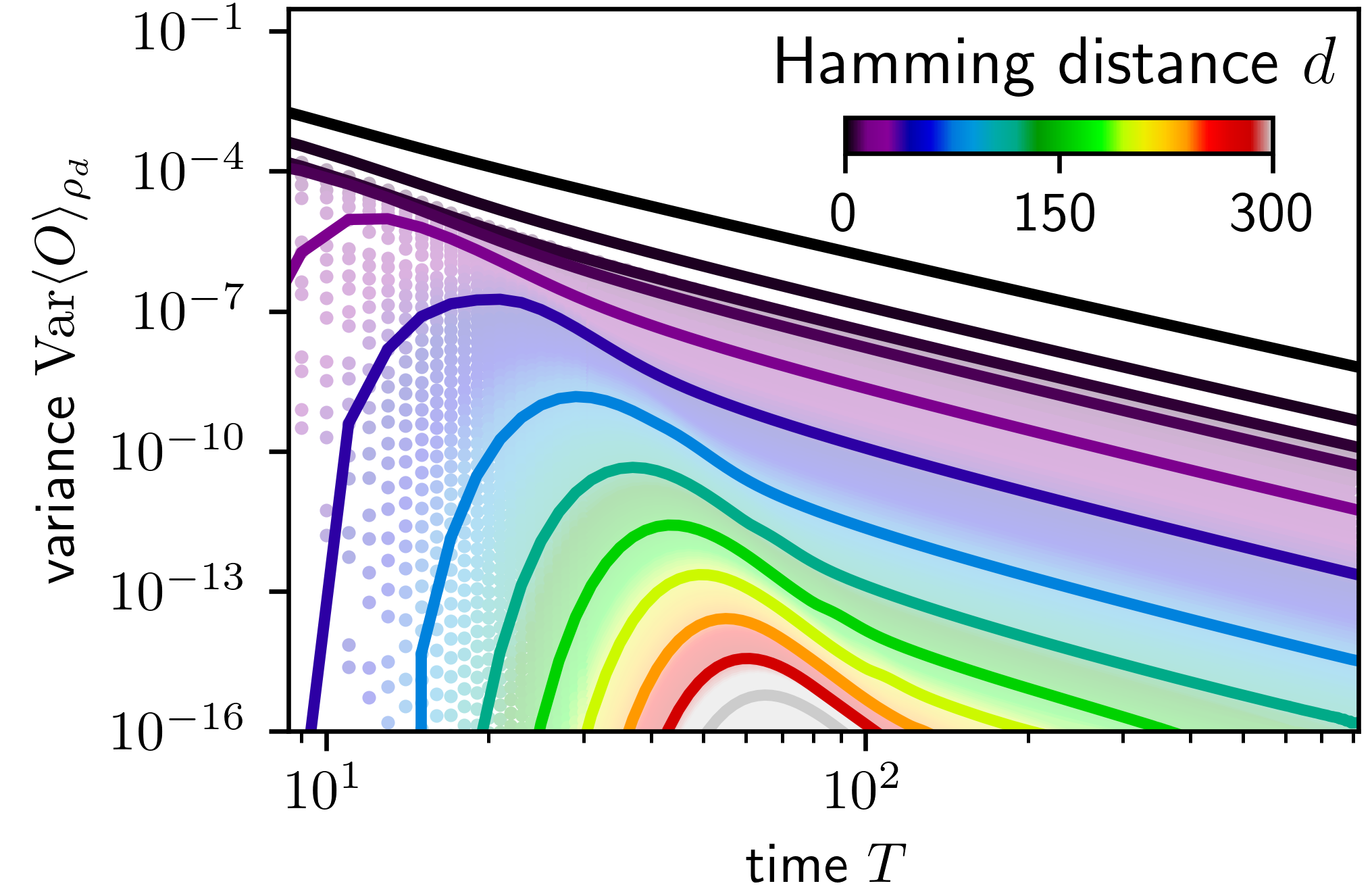}
    \includegraphics{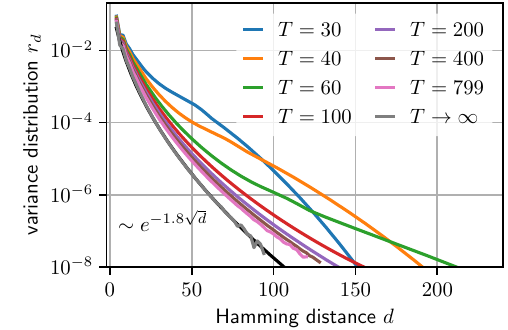}
    \caption{Hamming distance resolved variance $\mathrm{Var} \langle O \rangle_{\rho_d(T)}$, Eq.~\eqref{eq:resolved_var}, with $O=\rho_\mathrm{init}=\sigma^z_0$ for Haar-random circuits with charge conservation acting on qubits shown in two different ways. Top: Variance $\mathrm{Var}\langle O \rangle_{\rho_{d}}$ as a function of time for different $d$. Bottom: Variance distribution $r_d=\mathrm{Var}\langle O \rangle_{\rho_{d}}/\mathrm{Var}\langle O \rangle_{\rho}$, Eq.~\eqref{eq:hamming_dist_dephasing_sens}, as a function of total coherence at different times and the extrapolation to the long-time-limit. The results were obtained simulating the action of the averaged gates $\mathcal{U}_t$, inserting $\mathcal{P}_0 + e^{i\omega}\mathcal{P}_1$ everywhere and Fourier-transformation in $\omega$, see also App.~\ref{app:expectation_values_u1_circuits} and \ref{ssec:analytcal_u1_appendix}. We use itensor~\cite{itensor} and TEBD with a cutoff of $10^{-20}$ and system size $L=200$.}
    \label{fig:u1_coherence_analysis}
\end{figure}

The results are shown in Fig.~\ref{fig:u1_coherence_analysis} (top). As a function of time, we find $\mathrm{Var}\langle O \rangle_{\rho_d}$ to first increase and approach its maximum at around $T_\mathrm{max}\sim d^{0.6}$, App.~\ref{ssec:variance_behaviour_u1}. After assuming its maximum value, the variance decreases initially faster but converges to a decay of $T^{-2.5}$ for large times.

This behavior suggests the following intuitive picture, which is supported further by analytical considerations in Sec.~\ref{ssec:analytical_dynamics_u_1} and numerical studies in Sec.~\ref{sec:numerical_results_algorithm}. 

For a given $d$, at small times, there are not enough space-time points to insert $d$ projectors $\mathcal{P}_1$, the butterfly cone is too small to accommodate that Hamming distance $d$ so that the variance $\mathrm{Var}\langle O \rangle_{\rho_d}$ vanishes. 

For intermediate times, there appear to be two competing mechanisms. With increasing time, the size of the butterfly cone increases, increasing the number of parts which can have Hamming distance $d$, which in turn increases the variance $\mathrm{Var}\langle O \rangle_{\rho_d}$. However, there appears to be an optimal density of off-diagonal projectors $\mathcal{P}_1$: when increasing the time further, the variance decays faster than in the long-time-limit. Hence, when increasing the time and also the space between projectors $\mathcal{P}_1$, it suppresses the variance stronger than the increasing number of parts increases it. We hypothesize this to be a ``quantum'' effect - the enhancement of the variance at short times above the long-time decay $cT^{-2.5}$ with some constant $c$. The large time variance decay of $T^{-2.5}$ is the same as for the classical variance $\overline{\langle O \rangle_{\rho_{0}}^2}-\overline{\langle O \rangle_{\rho_{0}}}^2$. This suggests that it is mostly classical effects governing the decay of the variance at late times: the Hamming distance $d$ occurs only in one space-time region of duration $\mathcal{O}(1)$, while for most of the evolution of the order $\mathcal{O}(T)$ classical dynamics dominate yielding the long-time scaling as $T^{-2.5}$.

The same conclusions can be made when taking a slightly different perspective on the same results. We analyze how the total variance is distributed across different Hamming distances $d$ for different times by calculating the ratios $r_d = \mathrm{Var}\langle O \rangle_{\rho_d}/\mathrm{Var}\langle O \rangle_\rho$. Note that $\sum_d r_d=1$. The results are shown in Fig.~\ref{fig:u1_coherence_analysis} (bottom).

The variance $\mathrm{Var}\langle O \rangle_{\rho_d}$ decays sub-exponentially in $d$ up to a transition point, which increases with time, after which the contributions decay over-exponentially. This over-exponential decay again reflects that the Hamming distance is upper bounded by the lightcone volume and unlikely to exceed the smaller butterfly cone, strongly suppressing parts $\rho_X$ with Hamming distance $d(X)$ larger than the size of the butterfly cone. For large times, we find $\mathrm{Var}\langle O \rangle_{\rho_d}$ to converge to a stretched exponential distribution over $d$ with
\begin{align}\label{eq:long_time_ex}
    r_d = \frac{\mathrm{Var}\langle O \rangle_{\rho_d}}{\mathrm{Var}\langle O \rangle_\rho} \stackrel{T\rightarrow \infty}{\rightarrow} \mathrm{const} \, e^{-1.8 \sqrt{d}}  + \mathcal{O}(T^{-0.5})
\end{align}
via extrapolation. This is striking: The Hamming distance distribution is independent of time and the size of the butterfly-cone. For a fixed rate of phase flip errors, the relative error $\epsilon_\mathrm{r}$ approaches a constant
\begin{align}
    \epsilon_\mathrm{r}^2 = \mathrm{MSE_\gamma(T)}/\mathrm{Var}\langle O\rangle_{\rho(T)}\stackrel{T\rightarrow \infty}{\rightarrow} \mathrm{const},
\end{align}
where $\mathrm{MSE}_\gamma$ refers to the mean squared error obtained for noisy correlation values with phase flip channels at every site in every time step, Eq.~\eqref{eq:noise_channel_proj}. Stated differently, despite more and more phase flip errors occurring with time, the relative error does not increase. This observation may again be explained with the Hamming distance occurring within a single region in the lightcone in every part rather being uniformly distributed. It encourages the development of a classical algorithm truncating the Hamming distance and in that sense, coherence and is presented in Sec.~\ref{sec:algorithm}. 

Here, we can already evaluate the approximation error of the truncation and evaluate the significance of the large Hamming-distance parts with strong decoherence sensitivity. We calculate the relative approximation error for $\rho\approx \rho_D$, Eq.~\eqref{eq:density_matrix_truncation}, using the long-time extrapolation results Eq.~\eqref{eq:long_time_ex} 
\begin{align}
    \epsilon_\mathrm{r}^2&=\frac{\mathrm{MSE}_D(T)}{\mathrm{Var}\langle O \rangle_{\rho(T)}} %
    \stackrel{T\rightarrow \infty} \rightarrow (1.1\sqrt{D}+0.6)e^{-1.8\sqrt{D}}, \label{eq:truncation_error}
\end{align}
and, hence, expect a scaling of $D \sim \log(1/\epsilon_\mathrm{r})^2$. 

Already the classical part $\rho_0$ yields about $90$\% of the full variance in the long-time-limit, i.e., the approximation $\rho(T)\approx \rho_0(T)$ yields a high relative accuracy, see App.~\ref{ssec:convergence_classical_part}. 

We conclude that on large scales, the hydrodynamics governing $\langle O \rangle_{\rho(T)}$ in a typical Haar-random quantum circuit with a U(1) conservation law is dominated by the classical dynamics.

\subsubsection{Analytical Dynamics and Calculation}\label{ssec:analytical_dynamics_u_1}

We continue by analyzing the structure of the variance in random U(1) circuits analytically. It yields an argument for the variance being created within small space-time regions. Here, we sketch the results; the details of the calculations can be found in App.~\ref{ssec:analytcal_u1_appendix}.

In order to calculate the MSE, Eq.~\eqref{eq:mse_coherence_resolved}, we calculate the circuit-to-circuit average and variance of $\mathrm{Tr}[U \sigma^z_0 U^\dagger \sigma^z_0]$ in the continuous space and time limit for infinitely large systems. The average is found to be
\begin{align}
    \mathrm{Tr}[\overline{U(T) \sigma^z_i U(T)^\dagger}\sigma^z_j]/2^L = G(T,i-j)
\end{align}
where the propagator
\begin{align}
    G(T,i-j) = \frac{e^{\frac{-(i-j)^2}{2T}}}{\sqrt{2\pi T}}
\end{align}
follows from the $\sigma^z_i$ performing a random walk under the circuit average as $\overline{V_{i,i+1}\sigma^z_i V_{i,i+1}^\dagger} = \sigma^z_i/2 + \sigma^z_{i+1}/2$. Hence, the circuit averaged correlation function decays as $\mathrm{Tr}[\overline{U(T) \sigma^z_0 U(T)^\dagger}\sigma^z_0]/2^L = (2\pi T)^{-1/2}$. Next, we calculate the second moment of the correlation function over the circuit ensemble as
\begin{align}
    \overline{ \mathrm{Tr} \left[U \sigma^z_0U^\dagger\sigma^z_0 \right] \mathrm{Tr}\left[ U \sigma^z_0U^\dagger \sigma^z_0\right]},
\end{align}
which can be rewritten in terms of a single trace over a doubled Hilbert space
\begin{align}\label{eq:two_copies}
    \mathrm{Tr}\left[ \overline{(U \sigma^z_0U^\dagger \otimes U\sigma^z_0U^\dagger )} \sigma^z_0 \otimes \sigma^z_0\right].
\end{align}
In this doubled Hilbert space, the two $\sigma^z$ perform random walks independently and ``interact'' when they are in the same location, yielding variance contributions. These contributions may be classical or quantum, where the latter are projected out by applying $\mathcal{P}_0$. As an approximation, we can calculate the variance from a single classical interaction summed over the interaction location $x$ and time $t$
\begin{align}
    \mathrm{Var}{\langle O\rangle_{\rho_0}}&\approx \frac{1}{12}\sum_t \sum_x [\partial_x G(t,x)]^2[\partial_x G(T-t,x)]^2 \nonumber\\
    &\propto \frac{1}{T^{5/2}}, \label{eq:class_lowest_order}
\end{align}
the scaling then follows from dimensional analysis, App.~\ref{ssec:analytcal_u1_appendix}.

Now, we consider the random walkers interacting classically a second time which we separate into two cases. First, if the second interaction process does not occur close to the first one (i.e. $\mathcal{O}(T)$ away), these processes yield a contribution to the variance scaling as $1/T^4$. Second, if the additional interaction process occurs close (i.e. $\mathcal{O}(1)$) to the existing one, the structure of Eq.~\eqref{eq:class_lowest_order} still applies. Hence, the two interactions within a small region yield a variance contribution $\propto T^{-5/2}$, which is however smaller than the single interaction contribution due to a small interaction strength of $1/12$. Adding more processes within a small region of order $\mathcal{O}(1)$ yields a correction exponentially small in the number of interactions, see App.~\ref{ssec:analytcal_u1_appendix}. In summary, only processes where the interactions happen in vicinity of each other yield a contribution $\propto T^{-5/2}$ to the classical variance with $d=0$.

The same applies to processes with a finite $d>0$. These now involve terms, where the two walkers $\sigma^z_{j} \otimes \sigma^z_{j+1}$ interact, resulting in terms like $\sigma^+_{j}\sigma^-_{j+1}\otimes \sigma^-_{j}\sigma^+_{j+1}$, with $\sigma^\pm_j = (\sigma^x_j \pm i \sigma^y_j)/2$. The $\sigma^+_{j}\sigma^-_{j+1}\otimes \sigma^-_{j}\sigma^+_{j+1}$ then yield a finite Hamming distance $d$ and are projected out by $\mathcal{P}_0$, giving rise to their interpretation as being quantum. Similar to the classical processes, these interactions are subleading as they have a small prefactor, require the walkers to meet, and effectively couple to the gradients of the distributions $\partial_x G$, like in Eq.~\eqref{eq:class_lowest_order}. The interaction structure being analogous to the classical case yields the $T^{-5/2}$ decay also for terms with $d>0$ in Fig.~\ref{fig:u1_coherence_analysis}: The processes where all of the quantum interactions occur in the vicinity to each other are dominant, processes with quantum interactions distributed over the whole lightcone are subleading in time. The parts in which the quantum interactions happen within vicinity to each other then yield variance contributions decaying as $T^{-2.5}$ similar to a single classical interaction. We speculate that there may be further cancellation effects, so that the variance contribution for a given $d$ decay initially faster than $T^{-2.5}$ after having approached its maximum.

This perspective also suggests a smaller variance contribution from terms with large Hamming distance $d$, also discussed as operator backflow from the perspective of Pauli string expansion~\cite{von_Keyserlingk_2022}. Large Hamming distances $d$ correspond to long lived or many quantum walkers $\sigma^{q}_j = \sigma^+_j \otimes \sigma^-_j$. As  $\sigma^{q}_j$ performs both, a random walk as well as three-body interactions by creating or annihilating further $\sigma^z$-random walkers, the larger $d$ the more likely is the creation of further particles. To contribute to the variance, all of them have to be annihilated again and reduced to $\sigma^z_0\otimes \sigma^z_0$ in order to not vanish with the final trace probe in Eq.~\eqref{eq:two_copies}. This may be the competing process yielding a decrease of the variance contribution with time despite the number of possible processes given $d$ growing.

In Sec.~\ref{sec:algorithm}, we devise an algorithm exploiting this insight: the quantum corrections are dominated by those where the quantum processes occur in the vicinity of each other.

For free fermionic random circuits we find the opposite behavior. Rather than being short lived, quantum walkers $\sigma^{q}_j$ are long lived and provide the dominant contribution to the variance decaying as $1/T^{1.5}$ compared to the faster decaying contribution $1/T^{2.5}$ discussed above for classical interactions, see App.~\ref{ssec:free_ferm} for more details.

\subsection{OTOCs in Haar random circuits}\label{sec:otocs}

The diagonal approximation approach is not only useful for computing correlation functions --- it can also be generalized to out-of-time ordered correlators (OTOCs). 
These objects are used to study global properties of quantum dynamics such as scrambling and chaos. Many classical approaches for computing local correlation functions do not work for OTOCs~\cite{abanin2025constructiveinterferenceedgequantum,Mi_2021}. In this section, we show that first-order OTOCs in random circuits are reasonably well approximated by a `classical' diagonal approximation that ignores interference terms and that can be efficiently sampled.
We note that, while effective for the \textit{first-order} OTOC, the diagonal approximation does not capture the \textit{higher-order} OTOCs faithfully~\cite{abanin2025constructiveinterferenceedgequantum}.

For simplicity, we consider only 1D brickwall circuit geometries with each two-qubit gate being drawn randomly from the Haar distribution. The OTOC is defined by
\begin{align}\label{eq:OTOC_definition}
    \mathrm{OTOC}(x, t) = \mathrm{Tr}[ \sigma^z_{x_0} U^{\dagger} \sigma^z_{x} U \sigma^z_{x_0} U^{\dagger} \sigma^z_x U]/2^L,
\end{align}
where $U=U(t)$ is a brickwall circuit of depth $t$, $x_0$ is the left-most site of the 1D geometry and $L$ is the system size. 
In this form, the OTOC signal decays from $1$ to $0$ as the location $x$ is moved across the operator front of the evolved butterfly operator $U \sigma^z_{x_0} U^{\dagger}$. This operator front travels ballistically ($x \sim x_0 + v_B T$, where the velocity $v_B$ is known as the butterfly velocity) and spreads diffusively~\cite{Khemani_2018,nahum2018}.

While the circuit-averaged behavior of the operator front is simply described, individual circuit instances have large fluctuations in their OTOC fronts that depend on the specific details of that circuit. Near the front of the evolving operator $\sigma^z_{x_0}(t)$, the size of these fluctuations decays with $T$ as the very slow power-law, ${\rm Var}\,\mathrm{OTOC}(x \sim v_B T, T) \propto T^{-\frac12}$, as argued in Ref.~\cite{nahum2018} by considering the effect of circuit-to-circuit fluctuations of the butterfly velocity $v_B$. The scaling of these fluctuations was confirmed using numerical calculations in Ref.~\cite{abanin2025constructiveinterferenceedgequantum}. In App.~\ref{sec:appendix_otoc}, we go beyond this and detail a precise perturbative calculation, controlled by the inverse of the local Hilbert space dimension $q$, that predicts the asymptotic behavior of the OTOC fluctuations.

As in the case of the $\mathrm{U}(1)$ conserving circuits considered previously, we find that the OTOC values for individual circuits can be approximately predicted using a diagonal approximation.
In the case of the OTOC, the diagonal approximation is based on sampling in the operator (rather than state) space.
Specifically, the approximation consists of pairing the trajectories of the two Heisenberg-evolved operators $U^\dagger \sigma_x^z U$. Further, we show that space-time local quantum corrections can be added to the diagonal approximation to capture the remainder of the circuit-to-circuit OTOC fluctuations. These quantum corrections, as in the previous section, are assigned a Hamming distance counting the total number of deviations of the operator trajectories from each other. 
We numerically compute the size of these quantum corrections and empirically find their decay to be a stretched-exponential in Hamming distance.

\subsubsection{Hamming distance in the operator space} \label{subsubsec:hamming_distance}
Let us first introduce the notion of the Hamming distance in the operator space.
To this end, we reinterpret operators $O$ as states $|O)$ in an operator-Hilbert space with scalar product $(O_1|O_2) = \mathrm{Tr}[O_1^\dagger O_2]/2^L$. We introduce the super operators $A \tilde{\otimes} B$ acting as
\begin{align}
    A \tilde{\otimes} B |O) = |A O B).
\end{align}
Hence, the two-site unitaries $V$ evolving an operator become $V_{j,j+1}=W^\dagger\tilde{\otimes} W$ with $W$ being a Haar-random gate.
Then we can write
\begin{align} 
    \mathrm{OTOC}(T) = (\sigma^z_{x_0}(T)| \sigma^z_x \tilde{\otimes} \sigma^z_x| \sigma^z_{x_0}(T)),
\end{align}
where the time evolution is now generated by folded gates.
The coherence perspective then follows by introducing operator density matrices as $|O_1)(O_2|$.  

Since the initial operator of the OTOC is a Pauli operator and the final operator as well, a natural choice of basis to consider coherence in are Pauli strings
\begin{align}
    P = \prod_{j=1}^L \sigma^{\alpha_j}_j.
\end{align}
With this choice of basis, the single site projector onto the diagonal part of the operator-density matrix is
\begin{align}
    \mathcal{P}_0\left[|\sigma^{\alpha})(\sigma^\beta|\right] &= \frac{1}{4}\sum_{\gamma \in \{0,x,y,z\} } |\sigma^\gamma \sigma^\alpha \sigma^\gamma )(  \sigma^\gamma \sigma^\beta \sigma^\gamma| \nonumber \\
    &= \delta_{\alpha,\beta}|\sigma^{\alpha})(\sigma^\alpha|.
\end{align}
Correspondingly $\mathcal{P}_1[\rho] = \rho-\mathcal{P}_0[\rho]$ projects onto the part of the operator-density matrix with local Hamming distance 1. 
The summed Hamming distance between two operator trajectories is a sum over space and time points of the local Hamming distances, in full analogy with Eq.~\eqref{eq:vlad}. 
With this recasting of operators as states in an operator Hilbert space and introducing superoperators, all previous statements apply. In particular, the time-evolved operator density matrix can be written as a sum over parts, each of them yielding a stochastically independent contribution to the circuit-to-circuit OTOC variance. 

\subsubsection{Numerical Results}\label{sec:numerical_results_for_OTOCs}
The analogous analysis for the variance of OTOCs resolved according to Hamming distance uses a statistical model constructed from averaging four copies of the Haar random gates. This results in a configuration space with $4! = 24$ states per site, according to the Weingarten calculus and as detailed in Ref.~\cite{abanin2025constructiveinterferenceedgequantum}. 
Similar to the previous analysis, we use a matrix product state numerical implementation of the statistical model. Relevant details of the statistical model are detailed in App.~\ref{sec:appendix_otoc}.

\begin{figure*}
    \centering
    \includegraphics[scale=1]{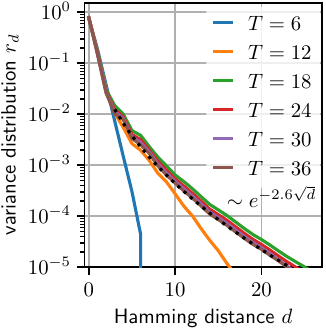}
    ~
    \includegraphics[scale=1]{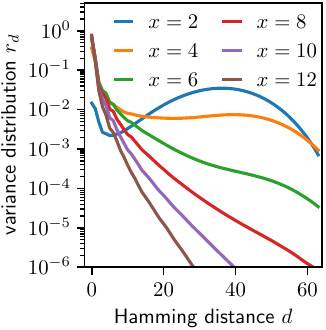}
    ~
    \includegraphics[scale=1]{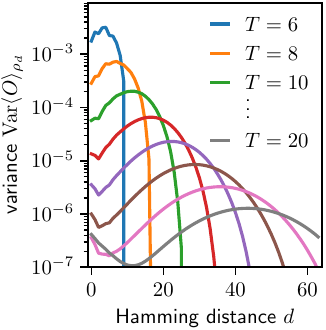}
    \caption{Resolution of the OTOC signal into components with total Hamming distance $d$, reconstructed via Fourier transform with period $64$. Left: Contributions to the variance decay as a stretched exponential in Hamming distance for OTOCs located at the peak of the operator front $x=x_p$, with a decay rate that saturates in time, indicating that circuit-to-circuit fluctuations of OTOCs can be approximated in large circuits without scaling $d$. The dotted line indicates the fitting result to the $T=36$ data. Center: The contributions to the variance from different Hamming distances at fixed depth $T=16$ with position $x$ varied. The exponential decay exhibited in the left panel at the peak of the operator front also occurs ahead of and across the front, with the decay rate gradually decreasing from the front to the back of the front. Far behind the front, the signal shifts to large Hamming distance. Right: The Hamming distance distribution for OTOCs far behind the operator front, here with $x=x_0$. The peak Hamming distance scales with the space-time volume of the OTOC circuit, indicating that the signal is poorly captured by the classical diagonal approximation with small $d$ quantum corrections. However, the total signal is exponentially small in this regime. Due to the Fourier method used to resolve the Hamming distance, the signal wraps modulo the period $64$, which is visible here for $T \geq 20$. }
    \label{fig:otoc_coherence_resolved}
\end{figure*}

The results for the Hamming distance resolved contributions to OTOC variance are shown in Fig.~\ref{fig:otoc_coherence_resolved}.
Near the peak of the operator front, where circuit-to-circuit fluctuations of the OTOC signal are largest, the Hamming distance resolved OTOC signal is well described by stretched-exponential decay with Hamming distance $d$. The shape of this decay quickly converges to an asymptotic form as depth is increased.

Surprisingly, most of the signal is captured by the lowest Hamming distances, even for very deep circuits.
In App.~\ref{sec:appendix_otoc}, we complement this numerical observation by an analytical calculation that is controlled by a small parameter $1 / q$, where $q$ is a local Hilbert space dimension.
We find $r_{d=0} = 1-c / q^2$ with coefficient $c \sim 1$.
This shows that for $q \gg 1$ the majority of the signal is captured by the diagonal approximation; the interference effects are suppressed by the operator scrambling within a large local Hilbert space.

Additionally, we investigate what happens as one moves the measurement qubit in the OTOC across the front. The size of the total OTOC signal decays as a Gaussian as one moves away from the front peak; nonetheless, we can investigate the relative sizes of distinct Hamming distance contributions. Far ahead of the operator front, the variance contribution appears to decay exponentially as a function of the coherence. 
Far behind the front, deep within the physical lightcone, the biggest contribution to the coherence appears at a large Hamming distance that grows with circuit volume. This result is shown in Fig.~\ref{fig:otoc_coherence_resolved} (right).

Together, these results show that the dominant contributions to the OTOC values of individual circuits are described classically, with small and space-time local quantum corrections. Far behind the front, where the operators are extremely well scrambled, quantum corrections once again dominate. At the same time, the OTOC signal in this case becomes exponentially small. Interpolating between these two limits, we see a tradeoff between signal size and complexity. 

\subsubsection{Short correlations in the recursive description of OTOC}

Fig.~\ref{fig:otoc_coherence_resolved} suggests that the circuit specific OTOC may be described classically using reduced memory. This can be understood with the help of the theory of the circuit averaged OTOC and its variance as we discuss in the following. 

In the notations of Sec.~\ref{subsubsec:hamming_distance}, the expansion of the butterfly operator in the Pauli basis reads $|B(t))=\sum_P b_P(t)|P)$, and the corresponding operator density matrix reads $\rho_t=\sum_{P, P^\prime} b_P(t) b_{P^\prime}(t) |P)(P^\prime|$, with the evolution by layers of the random circuit described by $\rho_t = \mathcal{V}_t\rho_{t-1}= V_t \rho_{t-1}V^\dagger_t$. Here $V_t$ is the $t$-th layer of two qubit gates $V_{j,j+1}$. A useful reference is the average of the density matrix over the ensemble of circuits. At long times it approaches a uniform mixture of non-identity Pauli strings $\overline{\rho_\infty} \rightarrow  \frac{1}{4^L-1}\sum_{P\neq I} |P)(P|$, which is stationary under random circuit evolution, while
the identity Pauli string $|I)(I|$ is a distinct trivial stationary state.
Application of a random brickwork circuit to a long chain initialized in the state  $\rho_0 =|\sigma_{x_0}^z)(\sigma_{x_0}^z|$ spreads the initial operator $\sigma_{x_0}^z$ to encompass the whole chain. 

A description of this operator spreading process is given by the probability distribution of the location of the right edge (or front) that separates the two circuit average stationary states, represented by a shell distribution vector $\mathbf{p}_t$, with elements,
\begin{gather}
p_{t}(j) = \mathrm{Tr}_{\mathrm{op}} \left[\Pi_{t,j}\rho_t\right] = \sum_{P: \;\textrm{R}(P)=j}|b_P(t)|^2, \nonumber
\end{gather}
where $R(P)$ denotes rightmost site with a non-identity Pauli factor in the string $P$, $\Pi_{t,j}$ is the projector onto Pauli strings with $R(P)=j$. The trace $\mathrm{Tr}_{\mathrm{op}}$ is taken in the Pauli string basis.

We introduce a reset projector onto the uniform state (written $\tilde{\mathcal{P}}_t$ to distinguish it from Pauli projection),
\begin{gather}
\tilde{\mathcal{P}}_t [\rho_t] = \sum_j \mathrm{Tr}_{\mathrm{op}} \left[\Pi_{t,j}\rho_t\right] \frac{\Pi_{t,j}}{\mathrm{Tr}_{\mathrm{op}} \left[\Pi_{t,j}\right]}. \label{eq:Pi_tj}
\end{gather}
The evolution projected by $\tilde{\mathcal{P}}_t$ at every time step $t$ yields a deterministic Markov process updating $\mathbf{p}_t$ that describes the circuit-averaged OTOC and can be efficiently simulated numerically. This description is not sufficient to describe the circuit specific evolution, which requires tracking evolution in the orthogonal complement (residual) subspace $\tilde{\mathcal{Q}}_t\rho = (I - \tilde{\mathcal{P}}_t)\rho$. The latter decomposes into Pauli-diagonal population differences and Pauli off-diagonal coherences. Introducing the resolution of the identity $I=\tilde{\mathcal{P}}_t+\tilde{\mathcal{Q}}_t$ at every time step separates the evolution in these two subspaces. In particular, $t-s$-step propagation in the orthogonal residual space is described by,
\begin{gather*}
\mathcal{L}_{t,s}=(\tilde{\mathcal{Q}}_t\mathcal{V}_t)...(\tilde{\mathcal{Q}}_{s+1}\mathcal{V}_{s+1}), \; \mathcal{L}_{t,t}=I.
\end{gather*}
In such terms the evolution of the density matrix reads,
\begin{gather}
\rho_t = \mathcal{L}_{t,0}\rho_0+\sum_{s=1}^t\sum_i p_s(i)\frac{\mathcal{L}_{t,s}\Pi_{s,i}}{\mathrm{Tr}_{\mathrm{op}}\left[\Pi_{s,i}\right]}, \label{eq:rho_t_vs_p_s(i)}\\
p_t(j) = \mathrm{Tr}_{\mathrm{op}}\left[\Pi_{t,j}\mathcal{V}_t\mathcal{L}_{t-1,0}\rho_0\right]+\sum_{s=1}^{t-1}\sum_i K_{t,s}(j,i)p_s(i),\nonumber%
\end{gather}
where we defined above $\rho_0 =|B)(B|$, and the OTOC is computed via $O=\mathrm{Tr}_{\mathrm{op}}\left[\sigma^z_{x}\tilde{\otimes}\sigma^z_{x} \rho_T\right]$. The signed kernel,
\begin{gather*}
K_{t,s}(j,i) = \mathrm{Tr}_{\mathrm{op}}\left[\Pi_{t,j}\mathcal{V}_t\mathcal{L}_{t-1,s}\Pi_{s,i}\right]/\mathrm{Tr}_{\mathrm{op}} \left[\Pi_{s,i}\right],
\end{gather*}
describes an excursion of depth $h=t-s-1\geq 0$ that propagates for $t-s-1$ consecutive steps in the orthogonal residual space before returning to the projected space at layer $t$. At $h=0$, $K_{s+1,s}(j,i)$ is a gate independent transition matrix generating circuit averaged propagation of the operator front. At $h=1$, $K_{s+2,s}(j,i)$ captures the classical Pauli-diagonal circuit-to-circuit fluctuations, whereas the kernels with $h\geq2$ account for multi-step memory and include Pauli off-diagonal interference terms.  

The effect of the random circuit is to suppress correlations 
mediated by long excursions in the orthogonal residual space. Specifically, the mean-square column norm of an excursion of depth $h\geq 1$  is expected to be bounded by an exponential,
\begin{gather}
 \overline{\bigl(\sum_j |K_{s+h+1,s}(j,i)|\bigr)^2}
     \le \textrm{poly}(h) e^{- \lambda_K h},\label{eq:K_bound}
\end{gather}
with $h$-, $L$-, and $T$- independent constant $\lambda_K>0$. %

Such exponential suppression of depth $h$ excursions in the orthogonal space implies that an accurate  approximation, $O^{[k]}$, for the circuit-specific OTOC can be obtained using the recursion relations truncated to excursions of depth $h\leq k$. We conjecture that within the diffusive operator front window $|x-x_0-v_BT|\leq a \sqrt{2DT}$ (where $\mathrm{Var}(O) \sim T^{-1/2}$) the truncated recursion approximates the exact OTOC, $O^{[k]}$, with a relative mean squared error exponentially small in $k$,
\begin{gather}
 \frac{\overline{|O^{[k]}-O|^2}}{ \mathrm{Var}(O)
       }
       \le \Lambda_Oe^{- \lambda_O k}, \label{eq:finite_memory_otoc_estimate_bound}
\end{gather}
with $k$-, $L$-, and $T$- independent constants $\Lambda_O, \lambda_O>0$. A depth independent cutoff $k$ therefore is sufficient for fixed relative accuracy.

\subsubsection{Short range correlation in the limit of $q \gg 1$}

The bound Eq.~(\ref{eq:finite_memory_otoc_estimate_bound}) can be motivated by analyzing the variance of the OTOC, a circuit average of two copies (replicas) of OTOC, that is described by the four-copy $S_4$ statistical mechanics model introduced in Appendix~\ref{sec:appendix_otoc}.

In this four-copy model we therefore track two replicas of the right-moving edge (operator front) of the same random circuit. The residual excursions are described by the paired projection $\tilde{\mathcal{Q}}_t \otimes \tilde{\mathcal{Q}}_t$. In the circuit average model this projection annihilates the configurations that factorize into a product of two circuit average OTOC configurations. Far to the left behind the operator front the four-copy state approaches the stationary state $\tau=(12)(34)$ that corresponds to the product of two copies of OTOCs each far behind the front. Far to the right ahead of the front the state is identity. 
In the large-$q$ limit, configurations of the intermediate region are spatial domains of permutations $\sigma \in S_4$ separated by domain walls~\cite{zhou_2019}. Under application of gate layers interiors of uniform domains remain stationary and only the domain walls evolve as weakly correlated biased random walks propagating to the right together with the front. At each time step the statistical weight of the domain walls scales as $q^{-D}$, where $D=\sum_x d(\sigma_x,\sigma_{x+1})$ and $d(\sigma_x,\sigma_{x+1})$ is the transposition distance across the domain wall between qubits $x$ and $x+1$. The stationary states far behind and ahead of the front constrain $D\geq d(\tau,I)=2$. The bias for the front to move right suppresses backward steps in comparison to forward steps by $\mathcal{O}(q^{-2})$ which constrains the entropy of such biased random walk configurations to a small correction $\mathcal{O}(q^{-2})$ to its free energy.  

 When the two replica fronts are separated, their evolution proceeds entirely in the subspace of $S_2\otimes S_2 = \{I, (12), (34), \tau\}$ and factorizes into a product of two independent circuit-average evolutions. Residual excursions are generated only when the two fronts collide exciting permutations outside of $S_2\otimes S_2$ that survive the projection $\tilde{\mathcal{Q}}_t \otimes \tilde{\mathcal{Q}}_t$. The simplest excursion event corresponds to a single bubble domain of state $\sigma$ bounded by two domain walls that starts at a collision and ends at a collision $h$ layers later. The statistical weight of the leading order off-diagonal domain %
 can be approximately computed from a pair of biased random walks describing the domain wall configurations. The result scales with excursion depth $h$ as,
\begin{equation}
\sim e^{-\kappa(q) h},
\label{eq:vc_domain_wall_cost}
\end{equation}
with the exponent
$\kappa \approx 2\ln q +\mathcal{O}(q^{-2})$ approximately equal to the cost of two independent domain walls and the effect of entropy is a small correction.

The $1/q$ corrections to the single domain configurations remain small also for the physical qubits $q=2$, where $\kappa(q)$ remains positive. Statistical weight of such excursions determine the value of the Kernel $K_{t,s}(j,i)$ suggesting that it is also bounded by an exponent in $h$. Its estimate includes more complicated excursions with multiple domains with possible inhomogeneous interior. Such configurations are suppressed by the weight of the extra domain walls $q^{-D}$. We therefore expect that for sufficiently large $q$ summing over such configurations does not change the exponential scaling in $h$, which leads us to conjecture Eq.~(\ref{eq:K_bound}).

\subsubsection{OTOC variance at fixed Hamming distance}

This domain-wall picture also implies the stretched exponential scaling of $r_d$, the variance of OTOC conditioned on the Hamming distance $d$. Non-zero Pauli Hamming weight is generated exclusively within the residual excursions. Within such an excursion Pauli twirling projects onto Pauli-diagonal component with weight $p_{\mathrm{diag}}\simeq q^{-2}$ and Pauli off-diagonal with weight $1-p_{\mathrm{diag}}$. A fixed Hamming weight $d$ therefore corresponds to a bubble domain of space time area $A_d \sim d/(1-p_{\mathrm{diag}})$. For the two biased random-walk domain walls to sweep an are $A_d$ the bubble must have depth $h_d\sim \sqrt{A_d}$. The statistical weight of such a bubble domain determines the scaling of a single domain contribution to $r_d$
\begin{gather}
r_d \propto e^{-\lambda_d \sqrt{d}},
\end{gather}
with a positive constant $\lambda_d>0$. In Fig.~\ref{fig:otoc_coherence_resolved}~(left)  $d$ corresponds the total Hamming distance over the course of evolution $d=\sum_i^md_i$ that could be a combination of several excursions of depth $h_i\sim\sqrt{d_i}$. Since the total excursion depth scales as $\sum_i h_i \sim \sum_i\sqrt{d_i}\geq \sqrt{d}$ a single domain dominates and the stretched-exponential decay observed at the operator front in Fig.~\ref{fig:otoc_coherence_resolved}~(left), where $\lambda_d \approx 2.6$ for qubits ($q = 2$) is consistent with large excursions with Hamming distance close to $d$ dominating the value of $r_d$. Finally,  because a dominant excuriosn of depth $k$ carries hamming weight $D\sim k^2$ the tail scaling with $e^{-\lambda_d \sqrt{D}}\sim e^{-\lambda_O k}$ is consistent with the conjectured bound in Eq.~(\ref{eq:finite_memory_otoc_estimate_bound}).

\section{Algorithms for interference corrections}\label{sec:algorithm}

Based on the insights in the previous sections, we explore possible classical algorithms in three steps. First, we discuss a direct sampling algorithm for computing $O^{(d)}$ or $C^{(d)}$ -- the approximation of the OTOC or TOC which includes all interference corrections up to a fixed volume $d$.
This algorithm works by sampling according to the diagonal approximation trajectories and then reweighting the observable by evaluating the interference corrections for each sampled trajectory.
This gives a generic algorithm which applies independently of the circuit geometry, and which gives a polynomial runtime in circuit size for any fixed $d$. By combining the exponential increase of cost with $d$ with the stretched-exponential decay with $d$ of the contributions to TOCs or OTOCs in $1$d, this gives an overall run-time bound that is quasi-polynomial in accuracy $\epsilon$.

Next, we analyze a more efficient sampling algorithm based on the observation that the quantum interactions typically occur within a single region, i.e., the most relevant parts $\rho_X$ are those where all $\mathcal{P}_1$ occur within one region. This algorithm is based on partitioning the circuit into components; exact tensor network contraction is used within components, and sampling is used at the boundary between components. To correct for the error at the boundaries, we combine the results from $d+1$ disjoint partitions of the circuit using an inclusion-exclusion principle, which we show reintroduces all interference corrections of size $d$ or smaller. Empirical analysis of this algorithm shows polynomial scaling in both circuit size and accuracy.

Finally, we show that sampling can be removed entirely from the algorithm by using a truncated matrix product state (MPS) to directly compute $O^{(d)}$ summed over all trajectories. Using numerics in $1$d, the empirical cost of the MPS algorithm is shown to grow only as a stretched-exponential in $d$; as a consequence, the run-time that is predicted is polynomial in accuracy.

\subsection{Monte Carlo with interference corrections}\label{ssec:icmc}

\begin{figure*}
    \centering
    \includegraphics[width=\linewidth]{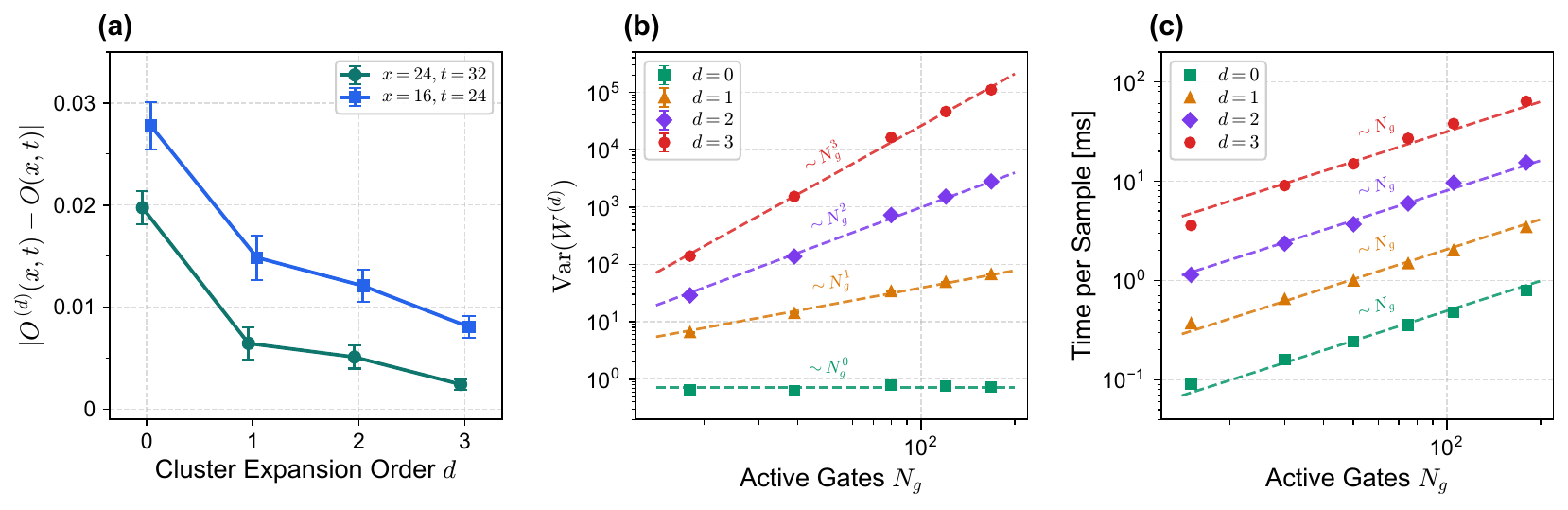}
    \caption{Empirical performance of the interference corrected Monte Carlo algorithm of Sec.~\ref{ssec:icmc} for the computation of OTOCs in $1$d Haar-random brickwork circuits. (a) The average error in the predicted OTOC value $O(x, t)$ drops as the order of the approximation $d$ is increased. (b) The number of samples needed at each order is $\mathcal{O}(N_g^d/\varepsilon^2)$, where $N_g$ is the number of gates in the circuit and $\varepsilon$ is the statistical error achieved on the Monte Carlo measurement of $O^{(d)}$. (c) The time to compute a sample scales linearly in $N_g$ by exploiting a connected cluster expansion.
    }
    \label{fig:icmc}
\end{figure*}

We can exploit the decay of the TOC and OTOC contributions with the Hamming distance $d$ to efficiently compute the series of approximations $C^{(d)}(x, t)$ for the TOC or $O^{(d)}(x, t)$ for the OTOC. We show here that even a naive sampling algorithm can compute these corrections with at most polynomial scaling $\mathcal{O}(N_g^{d+1})$ in circuits with $N_g$ gates. 

To derive the form of the algorithm, we first group the terms in the sum of the full path integral Eq.~\ref{eq:OTOCpathintegral} by the first trajectory $\boldsymbol\sigma$ in the trajectory pair, and restrict the sum of $\boldsymbol\sigma'$ to nearby trajectories with Hamming distance $|\boldsymbol\sigma - \boldsymbol\sigma'| \leq d$: 
\begin{gather*}
O^{(d)}(x, t) =\sum_{\boldsymbol{\sigma}} s(\boldsymbol\sigma)  p(\boldsymbol\sigma)
\!\!\!\sum_{\text{valid }\boldsymbol{\sigma'} | |\sigma'-\sigma|\leq d} \left[ \vphantom{\prod_t} \frac{A(\boldsymbol\sigma, \boldsymbol\sigma')}{p(\boldsymbol\sigma)} \vphantom{\prod_t} \right],
\end{gather*}
where $A(\boldsymbol\sigma, \boldsymbol\sigma')=\prod_t \mathrm{Tr} (\boldsymbol\sigma'_{t} u^{\dagger}_{t} \boldsymbol\sigma'_{t-1} u_t)\mathrm{Tr} (\boldsymbol\sigma_{t} u^{\dagger}_{t} \boldsymbol\sigma_{t-1} u_t)/4^L$ and where $p(\boldsymbol\sigma)$ is the classical trajectory probability of Eq.~\ref{eq:OTOCdiagonal}. 
The algorithm then proceeds by sampling trajectory $\boldsymbol\sigma$ according to the classical stochastic process with probabilities $p(\boldsymbol\sigma)$, while reweighting the observable $s(\boldsymbol\sigma)$ by the interference-corrected weight $W(\sigma) = \Sigma_{|\sigma'-\sigma|\leq d} A(\boldsymbol\sigma, \boldsymbol\sigma')/p(\boldsymbol\sigma)$.

To compute the weights, we first fix a set of space-time locations $X = \{(x, t)\}$ for which  $\sigma'_{x, t} \neq \sigma_{x,t}$ whenever $(x, t) \in X$, and $\sigma'_{x, t} = \sigma_{x,t}$ whenever $(x, t) \in X^c$, so that $W(\sigma) = \sum_X W_X(\sigma)$. Since the size of the subset is restricted $|X|\leq d$, the total number of such subsets is bounded by polynomial scaling $N_g^d$ in circuits with $N_g$ gates. Second, we note that in circuits made of two-qubit gates, the amplitude and probabilities factors in the numerator $A(\sigma, \sigma')$ and denominator $p(\sigma)$ cancel everywhere except for gates attached to the space-time sites in $X$. Finally, we note that removing the restriction to $\sigma'_{x, t} \neq \sigma_{x,t}$ within $X$ allows for computing 
$$
V_X(\sigma) =\sum_{X' \subseteq X} W_{X'}(\sigma)
$$
with a single tensor network contraction over only the gates that are attached to the region $X$, using the trajectory $\sigma$ as the boundary conditions. Thus, we can use the principle of inclusion-exclusion to reconstruct $W_X$ from $V_X$:
$$
W_X(\sigma) =\sum_{X' \subseteq X} (-1)^{|X| - |X'|}V_{X'}(\sigma).
$$
Finally, we can construct the total weight 
\begin{equation}
\begin{aligned}
W(\sigma) &= \sum_{|X|\leq d}W_X(\sigma)
\end{aligned}
\label{eq:weight}
\end{equation}
Thus, we can compute the weights using $\mathcal{O}(N_g^{d})$ tensor network contractions, where each network has at most $d$ two-qubit gates in it. 

The efficiency of the sampling method also depends on the sample variance of the estimated observable. As our method has no correlations between sampled trajectories, it is enough to bound the variance of $W(\sigma) s(\sigma)$ over the ensemble of trajectories $\sigma$. A naive bound that expands this variance through the sum over clusters $\langle W_X(\sigma) W_Y(\sigma) \rangle$, where $\langle \rangle$ denotes an average over $\sigma$, yields an upper bound on the circuit-size dependence as $\mathcal{O}(N_g^{2d})$. The maximum magnitude of the variance $\langle W_X(\sigma) W_Y(\sigma) \rangle$ is circuit-size independent, as the clusters are finite and do not scale with the circuit size.
Together, these naive bounds yield an overall runtime for approximating $O^{(d)}(x, t)$ to statistical accuracy $\varepsilon$ that scales as $\mathcal{O}(N_g^{3d}/\varepsilon^2)$.

In fact, the bound can be reduced significantly through the lens of a connected cluster expansion. This leads to a significant simplification for computing the weights: $W_X$ for disconnected clusters $X$ splits into an exact product of weights on connected subclusters, $$W_X = \prod_{C_j \text{ conn.}} W_{C_j}.$$  Additionally, the cross terms of the weights in the variance bound $\langle W_X(\sigma) W_Y(\sigma) \rangle$ are only non-zero if each connected component of $X$ overlaps with each connected component of $Y$. This restricts the count of pairs of contributing clusters to the variance to $\mathcal{O}(N_g^d)$. 
To illustrate the convergence and run-time cost of this algorithm, we applied it to the approximation of OTOCs in $1$d random brickwork circuits. In Fig.~\ref{fig:icmc}(b), we see directly that the variance of the sample weights over the trajectories scales as $N_g^d$, as predicted by the connected cluster argument. In Fig.~\ref{fig:icmc}(c), we see the run-time cost per sample of a specific implementation for the computation of the weights, which approximately shows the predicted $\mathcal{O}(N_g)$ scaling by exploiting the cluster connectivity.

Conventional wisdom suggests that Monte Carlo sampling of dynamics trajectories for real-time quantum evolution is subject to an exponentially challenging sign problem, which should manifest as an exponential growth in the sample variance for trajectory amplitudes.
This algorithm does not directly contradict that. Instead, by imposing a cutoff on the size $d$ of the interference corrections, we cap the growth of the sample variance to polynomial in the circuit size. As $d$ is increased to its maximum of the total circuit volume, the sample variance returns to exponential growth.

However, when combined with the rapid decay of the approximation error $|O^{(d)}(x,T) - O(x, T)|$, the algorithm becomes useful. 
With stretched-exponential decay of $|O^{(d)}(x,T) - O(x, T)|$, we get an overall run-time with quasi-polynomial accuracy scaling $\mathcal{O}(N_g^{|\log(1/\epsilon)|^2})$. For first-order OTOCs, this is already a dramatic speedup over prior expectations of exponential scaling; however, we will show in the following sections that even this run-time can be further improved upon.

\subsection{Sliced Monte Carlo Algorithm} \label{ssec:addition_algo}

A simple way to improve the accuracy is to merge several gates into a bigger one and thereby apply the diagonal projection less often. We refer to this as a \textit{partition} of the circuit. In App.~\ref{ssec:sampling_grouped_gates}, we provide a detailed explanation of the sampling algorithm for partitions. Forming partitions allows a trade-off between accuracy and computational complexity: Merging gates increases the computational complexity for sampling from $p(a_t|a_{t-1})$, and it increases the accuracy because it corresponds to including more parts. All these parts $\rho_X$, with $\mathcal{P}_0$ or $\mathcal{P}_1$ at the legs connecting merged gates, are stochastically independent, see App.~\ref{app:stoch_indep_path}. This yields a monotonic accuracy improvement; thus, merging provides a controlled level of approximation.

Next, we combine $R$ partitions to obtain a better approximation reliably. We label a partition by $Y$ with $Y_{t,l}\in\{0,2\}$ and define $\mathcal{P}_2=\mathcal{I}$ in analogy to parts labeled through $X$. As $\mathcal{P}_2=\mathcal{P}_0+\mathcal{P}_1$, $\rho_Y$ is a sum of several $\rho_X$. We have shown in Sec.~\ref{ssec:stoch_indep} that parts $\rho_X$ are stochastically independent; each additional part included decreases the MSE. Therefore, the strategy for combining several partitions $Y_1,Y_2,\cdots,Y_R$ is to decompose them into parts $\rho_X$. Then, we iteratively only add those parts in $Y_r$ that are not yet included in $Y_1,\cdots,Y_{r-1}$, ensuring that the MSE decreases monotonically with every additional partition.

This can be formalized as follows: given two partitions $Y_1$ and $Y_2\neq Y_1$, the density matrices $\rho_{Y_1}$ and $\rho_{Y_2}$ are not stochastically independent. This can be corrected by subtracting $\rho_{Y_1 \odot Y_2}$, with the partition $\left(Y_1 \odot Y_2\right)_{t,l} =(Y_1)_{t,l}(Y_2)_{t,l}/2\in\{0,2\}$. Hence, $Y_1 \odot Y_2$ contains all projectors $\mathcal{P}_0$ of $Y_1$ and $Y_2$. This way, $\rho_{Y_1}$ and $\rho_{Y_2}-\rho_{Y_1 \odot Y_2}$ are stochastically independent and no part $\rho_X$ is included twice. Hence, the MSE of
\begin{align}\label{eq:combining_collections}
    \rho_{Y_1 \cup Y_2} = \rho_{Y_1}+\rho_{Y_2}-\rho_{Y_1 \odot Y_2}
\end{align}
is just the sum over all parts that are not included in either $Y_1$ or $Y_2$. As a result, $\rho_{Y_1 \cup Y_2}$ yields an MSE less than or equal to the one obtained from using $\rho_{Y_1}$ or $\rho_{Y_2}$ alone. Note that evaluating $\rho_{Y_1 \odot Y_2}$ is expected to be no more difficult than evaluating $\rho_{Y_1}$ or $\rho_{Y_2}$ via Monte Carlo sampling since $\rho_{Y_1 \odot Y_2}$ contains additional projectors $\mathcal{P}_0$ compared to $\rho_{Y_1}$ or $\rho_{Y_2}$.

The procedure Eq.~\eqref{eq:combining_collections} can be iterated to obtain $\rho_{Y_1 \cup Y_2 \cup \cdots \cup Y_R}$ at the cost of applying the sampling algorithm for all $2^R-1$ partitions summing up to $\rho_{Y_1 \cup Y_2 \cup \cdots \cup Y_R}$.

\subsubsection{Runtime analysis}
In this section, we discuss one possible partitioning scheme for the addition algorithm. This scheme allows us to make precise statements about the scaling of the computational cost with accuracy based on the findings in Fig.~\ref{fig:u1_coherence_analysis}.

One possibility for partitions capturing all parts with Hamming distance $d(X)$ up to $d(X)=R-1$ is a horizontal slicing pattern like $Y_1$ in Fig.~\ref{fig:horiz_slicing_eff} for one-dimensional systems. An algorithmically simpler, but less efficient diamond shaped pattern is discussed in App.~\ref{ssec:numerical_results_algorithm_diamond}. We generate $R$ partitions $\{Y'_r\}_{r=1}^{R}$, where $Y'_r$ is generated by shifting $Y_1$ in space by $r-1$, e.g., $Y'_3=Y_2$ in Fig.~\ref{fig:horiz_slicing_eff}. Note that this construction can also be generalized to any spatial dimension.

We proceed with showing that we capture all parts with $d(X)< R$. Combining all $R$ partitions, we miss all parts that are not fully contained in any of the partitions, i.e., a part $X$ is missed if there is a $t,l$ so that $X_{t,l}=1$ and $(Y_r')_{t,l}=0$ for all partitions $r$. Since at all $t,l$ there is at most one partition with $(Y_r')_{t,l}=0$, a part is only missed if it has $X_{t,l}=1$ in at least $R$ spacetime locations $(t,l)$. This implies a Hamming distance of $1$ at each $(t,l)$ with $X_{t,l}=1$, see App.~\ref{ssec:analytcal_u1_appendix}. Therefore, any missed part has a Hamming distance of $d(X) \geq R$.

Several factors contribute to the computational cost to achieve a relative accuracy $\epsilon_r$. Calculating the sampling probabilities $p(a_t|a_{t-1})$ requires applying all $TL$ gates to statevectors of $R$ qubits, yielding a scaling of $TL e^{\lambda R}$ with some $\lambda$. The sampling has to be repeated $\epsilon_r^{-2} T^\alpha$ times in order for the stochastic variance be of order $\epsilon_r$ relative to the algebraically decaying signal variance decaying as $T^{-\alpha}$. In total, we have to perform the sampling procedure for $2^{R}-1$ partitions and adding the results via the inclusion/exclusion principle enhances the variance by $2^{R}-1$, so that the total computational cost scales as
\begin{align}
    T \, L e^{\lambda R}\epsilon_r^{-2} T^\alpha 4^R.
\end{align}
Note that we expect the sample variance to also decay with the Hilbert space size and reducing the sampling cost, which we here omit for simplicity.

We have $R\sim D$ and $D\sim \log(\epsilon_\mathrm{r})^2$ from Eq.~\eqref{eq:truncation_error}. Hence, the scaling of the computational cost in the relative accuracy $\epsilon_\mathrm{r}$ is upper bounded by
\begin{align}
     T^{\alpha+1} L e^{(\lambda+\log(4)) \log(\epsilon_\mathrm{r})^2}\epsilon_r^{-2}.\label{eq:computation_cost_scaling}
\end{align}
Note that from the analytical calculation we expect the decoherence sensitivity to be further suppressed in higher dimensions for the charge correlators in $U(1)$ Haar random circuits, so that we expect this algorithm to work better for higher dimensions. The algorithm captures the essence of the quantum corrections: small regions with decoherence sensitivity. In the following we provide analytical arguments and demonstrate numerically that this yields the error of the algorithm to decrease significantly faster than Eq.~\eqref{eq:computation_cost_scaling} in practice.

\subsubsection{Numerical Results}\label{sec:numerical_results_algorithm}
While the scaling presented in the previous section is based on the numerical results presented in Fig.~\ref{fig:u1_coherence_analysis}, in this section we combine the insights from the analytical calculation Sec.~\ref{ssec:analytical_dynamics_u_1} with a numerical study of the algorithms accuracy. This suggests that the computational cost of the slicing algorithm scales polynomial with accuracy.

\begin{figure}
    \centering
    \includegraphics[scale=0.9]{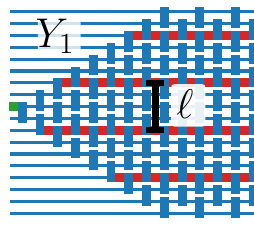}
    \qquad
    \includegraphics[scale=0.9]{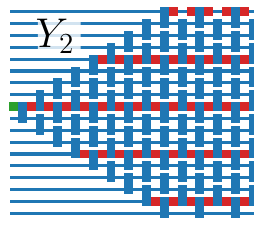}
    \includegraphics[scale=0.49]{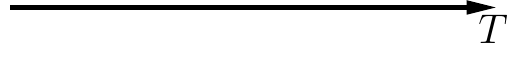}
    \includegraphics[scale=0.49]{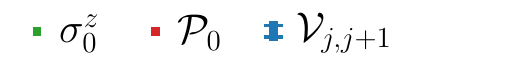}
    \caption{One possibility for circuit partitions in 1+1D depicted similar to Fig.~\ref{fig:overview_results} (b). The initial density matrix $\rho_\mathrm{init}=\sigma^z_0$ (green) is evolved with the superoperators $\mathcal{V}_{j,j+1}$ (blue) consisting of the two-qubit gates $V_{j,j+1}$. We insert projections $\mathcal{P}_0$ (red) along horizontal lines with distance $\ell$ and translate the pattern in space: %
    Here, we generate $R$ partitions by shifting $Y_r$ by $\ell(r-1)/R$, here we show $Y_2$ for $R=2$ (right). These two or more partitions are then combined via iterating Eq.~\eqref{eq:combining_collections}.}
    \label{fig:horiz_slicing_eff}
\end{figure}

\begin{figure}
    \centering
    \includegraphics[scale=0.8]{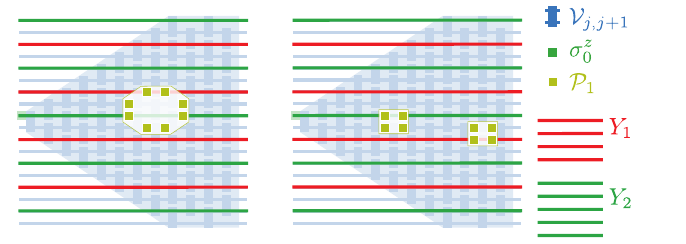}
    \caption{Two examples for parts $\rho_X$, Eq.~\eqref{eq:path_resolved_density_matrix}, which are not captured when combining the partitions shown in Fig.~\ref{fig:horiz_slicing_eff} via Eq.~\eqref{eq:combining_collections}. Left: parts with one large region of off-diagonal projections $\mathcal{P}_1$ having overlap with $\mathcal{P}_0$s in all, $Y_1$ and $Y_2$, so that the part is not captured. Right: parts which have two separate regions of $\mathcal{P}_1$s with overlap with $\mathcal{P}_0$s in $Y_1, Y_2$. When combining up to $6$ partitions, we find the large regions to be the dominant contributions for small $\ell$, but with increasing $\ell$ the low Hamming weight, disconnected parts become more relevant Fig.~\ref{fig:performance_alg_vert} and App.~\ref{app:detailed_perform_heuristic_slicing}.}
    \label{fig:horiz_failure_parts}
\end{figure}

We can be more efficient compared to the approach in the previous section, which required us to apply the Monte Carlo sampling procedure for $2^R-1$ partitions. Here, we focus on just combining two, four or six different partitions, shown in Fig.~\ref{fig:horiz_slicing_eff}, so that sampling has to be applied to a total of maximally $2^6-1=63$ partitions. That means we do not capture low Hamming distance parts where at least two, four or six quantum interactions in the form of a $\mathcal{P}_1$ occur on all partitions we are combining, see Fig.~\ref{fig:horiz_failure_parts}. However, based on the analytical insights, we expect their MSE contribution to be small, the MSE is dominated by contributions with a single region of $\mathcal{P}_1$s, depicted in the left panel of Fig.~\ref{fig:horiz_failure_parts}. This expectation is verified numerically in the following and further substantiated in App.~\ref{ssec:numerical_results_algorithm_diamond}.

\begin{figure}
    \centering
    \includegraphics[scale=1]{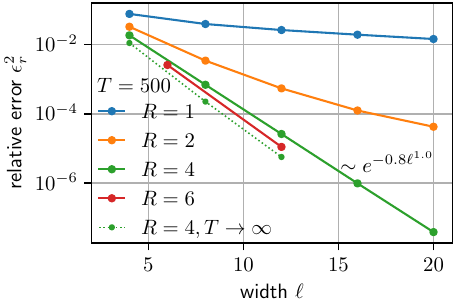}
    \caption{Relative error (squared) $\epsilon_\mathrm{r}^2$, Eq.~\eqref{eq:truncation_error}, for the addition algorithm combining $R$ horizontal partitions for different $\ell$ at $T=500$ (solid) and extrapolating $T\rightarrow \infty$ (dotted). Results for one (blue), combining two (orange), four (green) or six (red) partitions $Y_1,\ldots,Y_6$, Fig.~\ref{fig:horiz_slicing_eff}. A fit to the four partition data (green, solid) yields the scaling $e^{-0.8 \ell^{1.0}}$, finite size effects are discussed in App.~\ref{app:detailed_perform_heuristic_vert_slicing}.}
    \label{fig:performance_alg_vert}
\end{figure}

The performance of the sampling algorithm combining two to six partitions in terms of the relative error $\epsilon_\mathrm{r}$ is shown in Fig.~\ref{fig:performance_alg_vert}.
For a given number of partitions $R$, the error initially decreases fast with the spacing $\ell$, but eventually saturates, requiring to increase the number of partitions $R$ to achieve a further increase in accuracy. This indicates that parts with multiple interaction events become relevant: With increasing $\ell$, the algorithm captures single interaction events better, so that the convergence at fixed $R$ is caused by parts with multiple interaction events. Indeed, increasing $R$ and, that way, capturing separate interaction events better, yields significant increase in accuracy. This approach enables us to simulate the charge correlator in $U(1)$-Haar random circuits for arbitrary large times with a relative RMSE of less than $10^{-3}$ by partitioning the circuit into chunks of $\ell=20$ qubits.

Achieving such a small error at relatively small $\ell$ makes a rigorous scaling analysis numerically demanding. With four partitions, up to $\ell<20$ we find an exponential decrease of the error as $e^{-0.8 \ell^{1.0}}$, yielding in this regime a computational cost of
\begin{align}
    N_\mathrm{samples}  L T (\epsilon_\mathrm{r})^\beta, \label{eq:scaling_simplified_alg_hor}
\end{align}
with some exponent $\beta$.

However, with increasing $\ell$, we expect that more partitions $R$ need to be combined to further improve the accuracy, see also App.~\ref{app:detailed_perform_heuristic_vert_slicing}.

At the same time, the analytical arguments provided before in Sec.~\ref{ssec:analytical_dynamics_u_1} and presented in Ref.~\cite{von_Keyserlingk_2022} suggest the polynomial scaling to hold indefinitely. The contribution of parts is expected to decrease exponentially with the maximal extent $\Delta$ of an interaction process $e^{-\lambda_\Delta \Delta}$ like the one depicted in Fig.~\ref{fig:horiz_failure_parts} (left) as it involves an in $
\Delta$ exponentially large Hilbert space, see also Ref.~\cite{von_Keyserlingk_2022} Sec.~5 D. Furthermore, the contribution of parts is also expected to decrease exponentially with the number $K$ of separate interaction events $e^{-\lambda_K K}$, Sec.~\ref{ssec:analytical_dynamics_u_1}. Using the previous setup and combining $R=\ell$ partitions captures up to $K<R$ interaction processes with width up to $\Delta < R=\ell$. This suggests the accuracy to scale as $\epsilon_r \sim e^{-\mathrm{min}(\lambda_\Delta,\lambda_K)R}$ with a computational cost $e^{(\lambda + \log(2))R}$, which means the sampling algorithm scales polynomial in $\epsilon_r$. Further details are provided in App.~\ref{app:detailed_perform_heuristic_slicing}.

\begin{figure*}
    \centering

    \includegraphics[width=\linewidth]{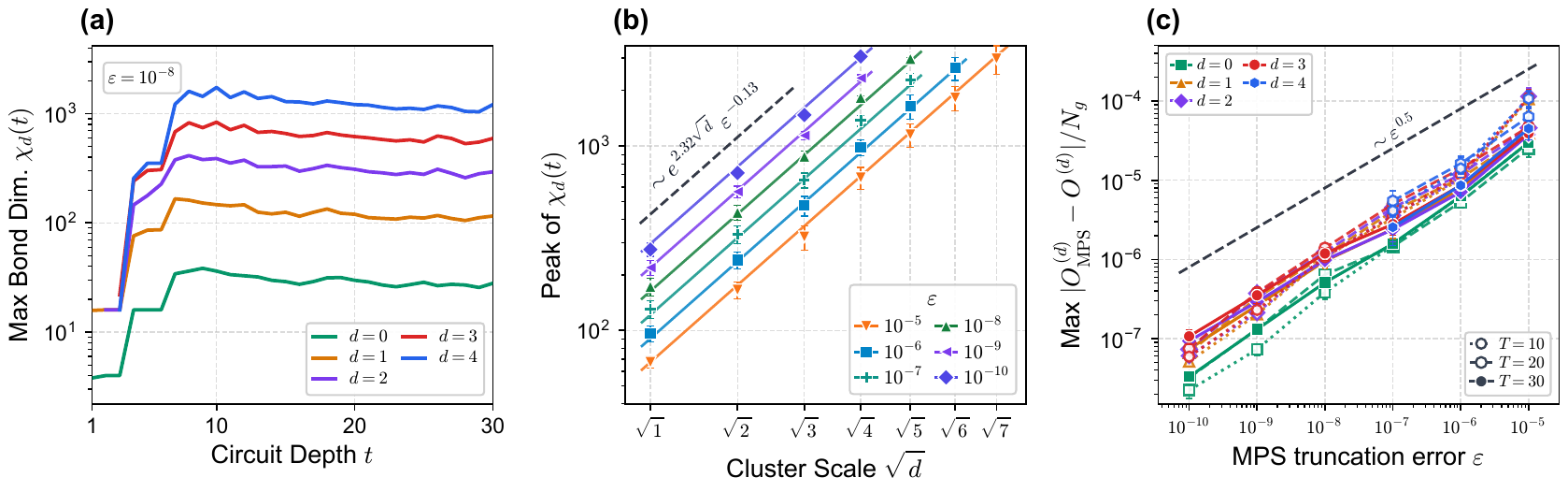}

    \caption{Empirical performance of the MPS algorithm for computing the $d^{\mathrm{th}}$-order approximation to the OTOC, $O^{(d)}(x, t)$. Data shown  is averaged over $10$ random realizations of $1$d Haar random OTOCs on $L=32$ qubits with depth $T=30$. (a) When truncation error is fixed, the bond dimension $\chi_d(t)$ for each MPS $\ket{\psi^{(d)}}$ (maximized over bonds) peaks at a finite-time and then saturates. (b) The size of the peak bond dimension is well described by the fit $\chi_d \sim e^{A \sqrt{d}}/\epsilon^b$, where $A \sim 2.3$ and $b \sim 0.13$. (c) Convergence of the MPS computation with truncation error, measured as an error rate per gate, $|O^{(d)}_{\text{ref}}(x, t) - O^{(d)}_{\text{MPS}}(x, t)|/N_g$, maximized over positions $x$ across the chain and depths $t$ from $0$ to $T$. The reference value uses the same MPS calculation with $\epsilon = 10^{-11}$. The result is consistent with an error that scales as $N_g \varepsilon^{0.5}$.
    }
    \label{fig:mps_alg}
\end{figure*}

\subsection{Matrix product state algorithm}\label{ssec:MPS}

An alternative approach forgoes sampling altogether, using a direct tensor network calculation of $O^{(d)}(x, t)$ for OTOCs (or $C^{(d)}(x, t)$ for TOCs). This is most straightforward for $1$d circuit geometries, where matrix product states (MPS) can be applied. We will see in this section that the implementation of this calculation has empirical run-time scaling that is polynomial in circuit size and in the target accuracy.

The operation of the matrix product state algorithm uses $d+1$ simultaneous MPS states, $\ket{\psi^{(k)}}$ where $k$ runs from $0$ to $d$. Each state uses a local Hilbert space dimension of size $16$, representing a truncated evolution of two simultaneous copies of the evolved butterfly operator $B(t)\otimes B(t)$.
$\ket{\psi^{(0)}}$ is initialized to $B(0) \otimes B(0)$ at time $t=0$, which is a product operator with bond dimension $1$, while the remaining states are initialized at first to $0$.

The evolution proceeds gate by gate, where each gate $V$ is converted into its action on two-copy operators $T = V \otimes V^{\dagger} \otimes V \otimes V^{\dagger}$, which is then split into two components $T_{\text{diag}}$ and $T_{\text{off-diag}}$. $T_{\text{diag}}$ is the projection of $T$ into the span of the diagonal pairs of Pauli matrices $I \otimes I, X \otimes X, Y \otimes Y, Z\otimes Z$. The update is then applied via the rule
\begin{equation}
 \ket{\psi^{(k)}} \leftarrow T_{\text{diag}}  \ket{\psi^{(k)}} + T_{\text{off-diag}}  \ket{\psi^{(k-1)}}, 
\end{equation}
for each $k$ from $k=d$ to $k=0$ in reverse order. By construction, each $\ket{\psi^{(k)}}$ only contains contributions from trajectory pairs that are off-diagonal in the Pauli basis exactly $k$ times. The applications of the partial gates $T_{\text{diag}}  \ket{\psi^{(k)}}$ and $T_{\text{off-diag}}  \ket{\psi^{(k-1)}}$ are done with the standard time-evolution with block decimation (TEBD) algorithm with fixed truncation error, with brings each MPS into canonical form centered at the location of the gate and truncates with singular value decomposition (SVD) to reduce the local bond dimension. The addition of the MPS is done exactly, followed by an SVD sweep to reduce the bond dimension. Each SVD uses a fixed total truncation error $\varepsilon$ to determine the number of states to keep, where $\varepsilon$ is the sum of the squares of discarded singular values.

To illustrate the complexity of this approach to computing OTOCs, we monitor the accuracy of this algorithm and the growth of the MPS bond dimensions throughout the evolution. 
The results are shown in Fig.~\ref{fig:mps_alg}. In this algorithm, we see that each MPS $\ket{\psi^{(k)}}$ shows saturation of the bond dimension to $\chi_k \sim e^{A \sqrt{k}}$. 
The bond dimension of each $\ket{\psi^{(k)}}$ is independent of the total-cutoff $d$; thus, the cost of the MPS algorithm in total at each $d$ is dominated by the MPS with $k=d$, which has the largest bond dimension. 
The saturation of the bond dimension is in stark contrast to calculations that directly apply Heisenberg-evolution of $B(t)$, which require bond dimensions that grow exponentially in time to accurately predict $O(x, t)$. 

In Fig.~\ref{fig:mps_alg}(b), we show that the peak bond dimension is $\chi_d \sim e^{A \sqrt{d}}/\varepsilon^b$, where the empirical fit constants are $A \sim 2.3$ and $b \sim 0.13$. 
Additionally, we show that the predictions for the approximation $O^{(d)}$ to the OTOC converge with decreasing truncation error in Fig.~\ref{fig:mps_alg}(c), with the accuracy scaling as $N_g \varepsilon^{0.5}$ for $1$d circuits with $N_g$ gates. This result is consistent with the expectation from adding the error $|\ket{\psi^{(k)}} - \ket{\psi^{(k)}}_{\text{trunc.}}| \sim \varepsilon^{0.5}$ from each truncation.

The overall approach to approximating OTOCs with accuracy $\epsilon$ via this algorithm requires first picking a value of $d$ such that $|O^{(d)}(x, t) - O(x, t)| < \epsilon/2$, and then picking a truncation error $\varepsilon$ such that $|O^{(d)}(x, t) - O^{(d)}_{\text{MPS}}(x, t)| <\epsilon/2$. Using our empirical measurement in $1$d of the accuracy at order $d$ for Haar random circuits, we see that $d$ should be picked such that $e^{-B \sqrt{d}} \sim \epsilon$. This stretched-exponential precisely matches the stretched-exponential growth of the bond dimension to yield a bond dimension; thus the needed bond dimension will be polynomial in accuracy $\epsilon$.
Specifically, we can derive the overall run-time scaling from the above fits and using the cost scaling $N_g \chi^3$ for the MPS algorithm; this yields an empirical complexity of the form
$$
(N_g)^{1 + 6b}\left(\frac{1}{\varepsilon}\right)^{3A/B + 6b},
$$
where $b \sim 0.13$, $A \sim 2.3$ from the fit in Fig.~\ref{fig:mps_alg}(b) and $B \sim 2.6$ from the fit in Fig.~\ref{fig:otoc_coherence_resolved}.

The success of this approach shows that sampling is not a necessary feature of local interference algorithms; instead, the key feature is the truncation of trajectory pairs which deviate too far from each other. However, sampling may still yield the more successful algorithm in higher dimensional geometries, where even area-law scaling of the state doesn't guarantee an efficient tensor network algorithm.

\section{Conclusions and Outlook}\label{sec:conclusions}

In this work, we have shown that time-ordered correlators and first-order out-of-time-ordered correlators in random circuits are approximately described by a classical stochastic dynamics which is derived by discarding all interference in either the computational basis or Pauli basis, respectively. Additionally, one can recover increasingly accurate approximations to the full quantum dynamics by re-introducing only `local' interference which takes place in contained space-time volumes. This fact leads to a variety of algorithms with reduced classical computational complexity for simulating TOCs and OTOCs. 

The decay of interference effects in $1$-dimensional TOCs and OTOCs can be viewed as a manifestation of quantum chaos via a path-integral formulation of quantum dynamics. In typical circuit instances, operator trajectories that deviate strongly pick up uncorrelated random phases, which in aggregate cancel. This feature of the dynamics is shown to be closely related to insensitivity of these TOCs to dephasing noise and of OTOCs to depolarizing noise that acts in a coordinated way between two independent copies of the evolution. For TOCs, this conceptually overlaps with prior studies of noise-insensitivity of local correlators, which also lead to reduced computational complexity~\cite{Rakovsky_2020}; however, the specific local interference algorithms here are distinct from those implemented in prior works. For efficient classical computation of first-order OTOCs, the local interference paradigm appears essential.

The nature of this decay is studied analytically and numerically through replica statistical models, which map this decay into a competition between domain wall tension and domain wall entropy in an effective ferromagnet. We argue that this competition yields an exponential suppression of interference contributions in their duration, and a stretched-exponential decay in their volume.%

From this basic principle, we derived several algorithms which have different computational properties. First, we showed that a Monte Carlo sampling of the path integral that only incorporates local interference has a sign-problem that is only polynomially large. This generic algorithm is the least practical, but it gives a quasi-polynomial runtime upper bound that is only contingent on having stretched-exponential or faster decay of local interference corrections with volume. This scenario may apply to random circuit TOCs and first-order OTOCs in more than one dimension; we leave this question for future study.

Next, we analyzed variant algorithms which build on this same underlying principle, but also take advantage of more structure in $1$d TOCs and OTOCs to achieve better complexity bounds. A matrix product state variant of the local-interference expansion of the path integral exploits an effective area-law entanglement that applies to the classical stochastic dynamics and to the local interference corrections, achieving an empirically polynomial run-time for $1$d first order OTOCs. Similarly, using an algorithm that combines several sampling partitions of the circuit, we empirically find a runtime for charge correlators that is polynomial in the inverse accuracy.

We note that the results presented in this work were reported in parts in Ref.~\cite{suchsland2025imposing} and built upon in Ref.~\cite{abanin2025constructiveinterferenceedgequantum}, which introduces further adaptations and variations of the Monte Carlo sampling algorithm.

Interesting future directions of research involve extensions to other observables, continuous-time evolution, non-abelian conservation laws, higher dimensional systems, and in particular dynamics that is uniform in space and time. These problems may benefit from the use of these algorithms directly or require the development of new ideas.

\textbf{Related Work:} Upon completion of this work, we became aware of an independent, parallel study~\cite{otoc_poly} on the arXiv. Our conclusions appear to agree where they overlap.

\tocless{\begin{acknowledgments}
We are grateful to Tom O'Brien, Nikita Astrakhantsev, Benjamin Villalonga, Yuri Lensky, Dmitry Abanin, Masud Haque, Sanjay Moudgalya, Curt von Keyserlingk and Tibor Rakovszky for engaging discussions. 
\end{acknowledgments}}

\vspace{0.5em}
\textbf{Statement of AI use.}
Generative AI tools were used during the preparation of this manuscript to assist with copyediting and text polishing. %
We acknowledge conversations with Gemini 3.1-4, Claude Opus 5.1-5.5, and ChatGPT Astra 6 models for the design of the interference corrected Monte Carlo algorithm, and for the analytical OTOC complexity analysis.
All AI-assisted derivations and calculations were independently verified by the authors; all scientific claims, conclusions, and content of this manuscript remain the sole responsibility of the authors.

\newpage
\bibliography{bibliography}

\clearpage %
\onecolumngrid %
\appendix %

\tocless\section{Charge-Conserving Circuit Construction\label{sec:charge_conserv}}
Requiring the two-qubit unitary $V_{j,j+1}$ to preserve charge restricts the matrix elements to be (ordering the states as $|\uparrow\uparrow\rangle,|\uparrow\downarrow\rangle,|\downarrow\uparrow\rangle,|\downarrow\downarrow\rangle$)
\begin{align}
    V_{j,j+1} = e^{i\alpha} \begin{pmatrix} 1 & 0 & 0 & 0 \\ 0 & v_{1,1} & v_{1,2} & 0 \\ 0 & v_{2,1} & v_{2,2} & 0 \\ 0 & 0 & 0 & e^{i\beta} \end{pmatrix}
\end{align}
with
$\begin{pmatrix}
v_{1,1} & v_{1,2} \\ v_{2,1} & v_{2,2}
\end{pmatrix}$
being a Haar-random $2$x$2$ unitary. We omit $e^{i\alpha}$ being a global phase, sample the $2$x$2$ random sub-matrix from the Haar-random ensemble and $\beta$ uniformly from $[0,2\pi)$. This measure is invariant under conjugation with any charge-conserving unitary.

\tocless\section{Stochastic Independence of Parts\label{app:stoch_indep_path}}

In order to evaluate the phase flip sensitivity for an expectation value with observable $O$, we truncate the sum in Eq.~\eqref{eq:coherence_decomp} to $d\leq D$. We compute the difference in the expectation value compared to the exact result. We average the difference by integrating the unitaries $\mathcal{U}_t$ over the group $G$. In the following, the phase flip channel is a subgroup of $G$. We use the previous definition for the summed Hamming distance $d(X)=\sum_{t,l} X_{t,l}$.
\begin{align}
    \mathrm{MSE}(T) &= \int_G \mathrm{d} \mathcal{U}_{T} \cdots \int_G \mathrm{d}\mathcal{U}_1 \mathrm{Tr} \left[ O \left(\sum_d \rho_d(T) - \sum_d^D \rho_d(T)\right)\right]^2\\ &= \int_G \mathrm{d} \mathcal{U}_{T} \cdots \int_G \mathrm{d}\mathcal{U}_1 \left(\sum_{d>D} \mathrm{Tr} \left[ O  \rho_d(T)\right]\right)^2 \\
    &= \sum_{X \text{ with } d(X) > D} \int_G \mathrm{d} \mathcal{U}_{T} \cdots \int_G \mathrm{d}\mathcal{U}_1 \mathrm{Tr} \left(\left[ O  \rho_{X}(T)\right]\right)^2.
\end{align}
Here the last step follows from the stochastic independence of different $\rho_{X}(t)$ which can be seen in two steps. First, we have for the group $G'$ of the unitary $U$ that $\sigma^z G'=G'$ so that
\begin{align}
&\int_G\mathrm{d}\mathcal{U} \left(\mathcal{P}_\alpha \circ \mathcal{U}\right) \otimes \mathcal{U}\\
&=\frac{1}{2}\int_{G'} \mathrm{d}U  \,  ( U \otimes U^\dagger +(-1)^{\alpha} \sigma^z U \otimes U^\dagger \sigma^z ) \otimes  U \otimes U^\dagger \\
&=\int_{G'} \mathrm{d}U  \,  U \otimes U^\dagger  \otimes  U \otimes U^\dagger +(-1)^{\alpha} \int_{G'} \mathrm{d}U  \,  \sigma^z U \otimes U^\dagger \sigma^z \otimes   U \otimes U^\dagger  \\
&=\int_{G'} \mathrm{d}U  \,  U \otimes U^\dagger  \otimes  U \otimes U^\dagger +(-1)^{\alpha} \int_{\sigma^z G'} \mathrm{d}U  \,  U \otimes U^\dagger  \otimes   \sigma^z U \otimes U^\dagger \sigma^z  \\
&=\int_{G'} \mathrm{d}U  \,  U \otimes U^\dagger  \otimes  U \otimes U^\dagger +(-1)^{\alpha} \int_{ G'} \mathrm{d}U  \,  U \otimes U^\dagger  \otimes   \sigma^z U \otimes U^\dagger \sigma^z  \\
&=\int_{G'} \mathrm{d}U  \,  U \otimes U^\dagger  \otimes  \left( U \otimes U^\dagger +(-1)^{\alpha}   \sigma^z U \otimes U^\dagger \sigma^z \right)  \\
&=\int_{G} \mathrm{d}\mathcal{U}  \, \mathcal{U}  \otimes  (\mathcal{P}_\alpha \circ \mathcal{U}).
\end{align}
Using this property, we can follow
\begin{align}
&\int_G\mathrm{d}\mathcal{U}_t \left(\mathcal{P}_\alpha \circ \mathcal{U}_t\right)[\rho(t-1)] \otimes \left(\mathcal{P}_\beta \circ \mathcal{U}_t\right)[\rho(t-1)] \\
&= \int_G\mathrm{d}\mathcal{U}_t \, \mathcal{U}_t[\rho(t-1)] \otimes \left(\mathcal{P}_\beta\circ \mathcal{P}_\alpha \circ \mathcal{U}_t\right)[\rho(t-1)]\\
&= \int_G\mathrm{d}\mathcal{U}_t \,  \mathcal{U}_t[\rho(t-1)] \otimes \left(\mathcal{P}_\alpha \circ \mathcal{U}_t\right)[\rho(t-1)] \delta_{\alpha,\beta}\\
&= \int_G\mathrm{d}\mathcal{U}_t \left(\mathcal{P}_\alpha \circ \mathcal{U}_t\right)[\rho(t-1)] \otimes \left(\mathcal{P}_\alpha \circ \mathcal{U}_t\right)[\rho(t-1)] \delta_{\alpha,\beta},
\end{align}
using $\mathcal{P}_\alpha \circ \mathcal{P}_\beta = \mathcal{P}_\alpha \delta_{\alpha,\beta}$ in the two last steps, respectively.

Applying this relation for every unitary and projector we obtain
\begin{align}
    \overline{\rho_{X_1} \otimes \rho_{X_2}} &=  \int_G \mathrm{d} \mathcal{U}_{T} \cdots \int_G \mathrm{d}\mathcal{U}_1 \rho_{X_1} \otimes \rho_{X_2}\\
    &= \int_G \mathrm{d} \mathcal{U}_{T} \cdots \int_G \mathrm{d}\mathcal{U}_1 \rho_{X_1} \otimes \rho_{X_1} \delta_{X_1,X_2} = \delta_{X_1,X_2}\overline{\rho_{X_1} \otimes \rho_{X_1}}
\end{align}
from which the above stated properties follow.

Note that this property holds for any channel as long as the group we average over is invariant under the application of the channel.

\tocless\section{Haar-random circuits\label{sec:haar_coherence_results}}
In the following, we want to investigate the role of decoherence in the form of the phase flip channel for correlators $\mathrm{Tr}[U\sigma^z_0 U^\dagger \sigma^z_0]/2^L$ in Haar-random circuits without conservation laws. They serve as an example for a typical circuit without any further constraints. 

\tocless\subsection{Weingarten Calculus for Haar-random circuits}
We begin with a short review of Weingarten Calculus adapted to our use case, for a more general explanation, see~\cite{Collins_2022} or its application in operator spreading~\cite{nahum2018}.
As also discussed in the main text, Eq.~\eqref{eq:two_copies}, we can rewrite the calculation of the second moments to an expectation value in a double Hilbert space
\begin{align}
    \overline{\mathrm{Tr}[\sigma^z_0(T)\sigma^z_0]\mathrm{Tr}[\sigma^z_0(T)\sigma^z_0]}= \mathrm{Tr}\left[ \overline{[\sigma^z_0\otimes \sigma^z_0](T)} \sigma^z_0\otimes \sigma^z_0 \right], \label{eq:repeat_two_copies}
\end{align}
where the overline denotes the ensemble average, which here amounts to integrating all gates over the group of Haar-random gates.

Now the whole circuit, and, hence, individual gates $V_{j,j+1}$ appear four times in the expression Eq.~\eqref{eq:repeat_two_copies}, so that the ensemble average involves integrals of the two-site gates $V_{j,j+1}$ over the Haar-random group $G_\mathrm{Haar\, rd}$ of the form
\begin{align}
    P_{j,j+1} = \int_{G_\mathrm{Haar\, rd}} \!\mathrm{d}V_{j,j+1} \, (V_{j,j+1}\otimes V_{j,j+1}^\dagger \otimes V_{j,j+1} \otimes V_{j,j+1}^\dagger). \label{eq:proj}
\end{align}
From the Schur-Weyl duality~\cite{harrow2013churchsymmetricsubspace} we know that we can write
\begin{align}
    P_{j,j+1} = \sum_{\alpha,\beta} W_{\alpha,\beta} \left(|\Pi_\alpha\rangle\langle\Pi_\beta|\right)_{j,j+1},
\end{align}
where
\begin{align}
    |\Pi_\alpha\rangle\in  &\left\{|II\rangle = \frac{1}{q^2}\sum_{a_j,a_{j+1},b_j,b_{j+1}} |a_j a_{j+1}\rangle\otimes |a_j a_{j+1}\rangle\otimes |b_j b_{j+1}\rangle\otimes  |b_j b_{j+1}\rangle,\right.\\
    &\phantom{\{}\left.|SS\rangle = \frac{1}{q^2}\sum_{a_j,a_{j+1},b_j,b_{j+1}}  |a_j a_{j+1}\rangle\otimes |b_j b_{j+1}\rangle\otimes |b_j b_{j+1}\rangle\otimes |a_j a_{j+1}\rangle\right\},\\
    |I\rangle &= \frac{1}{q}\sum_{a,b} |a\rangle\otimes |a\rangle \otimes |b\rangle \otimes |b\rangle,\\
    |S\rangle &= \frac{1}{q}\sum_{a,b} |a\rangle\otimes |b\rangle \otimes |b\rangle \otimes |a\rangle.
\end{align}
with $q$ being the local Hilbert space dimension, which is $q=2$ for qubits. From Eq.~\eqref{eq:proj} and the properties of the Haar-random group it follows that $P$ is a projector, and defining $D_{\alpha,\beta}=\langle \Pi_\alpha|\Pi_\beta\rangle$ we can find the Weingarten matrix $W$ to fulfill
\begin{align} \label{eq:weingarten_matrix}
    P^2 = P \Rightarrow  W D W = W \Rightarrow W = D^{-1}=\begin{pmatrix} 1 & q^{-2} \\ q^{-2} & 1 \end{pmatrix}^{-1} = \frac{q^2}{q^4-1}\begin{pmatrix} q^2 & -1 \\ -1 & q^2 \end{pmatrix}.
\end{align}
Now we can just simulate the dynamics of Eq.~\eqref{eq:repeat_two_copies} in the space spanned by $|I\rangle$ and $|S\rangle$. In this folded space, the initial operator $\sigma^z_0\otimes \sigma^z_0$ becomes on site $0$
\begin{align}
    |\rho_\mathrm{init}\rangle_0 = \sum_{a,b} (-1)^a(-1)^b |a\rangle \otimes |a\rangle \otimes |b\rangle \otimes |b\rangle, \qquad |\rho_\mathrm{init}\rangle_{j\neq 0} = |I\rangle_j.
\end{align}
The application of the first gate projects into the $|S\rangle,|I\rangle$ subspace, in which
\begin{align}
    \langle I|\rho_\mathrm{init}\rangle_0 = 0, \langle S|\rho_\mathrm{init}\rangle_0 = 1 \quad \mathrm{and} \quad \langle I|\rho_\mathrm{init}\rangle_{j\neq 0} = q^2, \langle S|\rho_\mathrm{init}\rangle_{j\neq 0}=1.
\end{align}
The same holds for the final state as $O=\rho_\mathrm{init}$. Hence, the simulation of Eq.~\eqref{eq:repeat_two_copies} corresponds to simulating the evolution of the initial state $|\cdots I_{-1} S_0 I_1 \cdots \rangle$ under the evolution of Eq.~\eqref{eq:weingarten_matrix} and calculating the overlap with the final state $|\cdots I_{-1} S_0 I_1 \cdots \rangle$. For example, we find
\begin{align}
    P_{0,1} | SI \rangle = \frac{q^2}{q^4-1}\left(q^2|SS\rangle\langle SS|- |SS\rangle\langle II|-|II\rangle\langle SS|+q^2|II\rangle\langle II|\right)| SI \rangle = \frac{q}{q^2+1}\left(|SS\rangle + |II\rangle\right).
\end{align}
For a TEBD simulation we use $|S'\rangle = q |S\rangle$ as a basis, making the simulation more stable, so that all necessary evolution rules are
\begin{align}
    P | S'I \rangle = \frac{1}{q^2+1}|S'S'\rangle+ \frac{q^2}{q^2+1}|II\rangle, \quad P | S'S' \rangle = | S'S' \rangle \quad P | II \rangle = | II \rangle.
\end{align}
In order to evaluate the Hamming distance contributions, we insert $\mathcal{P}_0 + \mathcal{P}_1 e^{i \omega}$, Eq.~\eqref{eq:noise_channel_proj}, everywhere and Fourier transforming the result in $\omega$. In this basis, their respective action can be calculated to be
\begin{align}
    \mathcal{P}_0 |I\rangle &= |I\rangle, \,  &\mathcal{P}_0 |S'\rangle &= \frac{q}{q+1}|I\rangle+\frac{1}{q+1}|S'\rangle,\\
    \mathcal{P}_1 |I\rangle &= 0, \,  &\mathcal{P}_1 |S'\rangle &= \frac{-q}{q+1}|I\rangle+\frac{q}{q+1}|S'\rangle.
\end{align}

\tocless\subsection{Numerical Results}
\begin{figure}[H]
    \centering
    \includegraphics[scale=1]{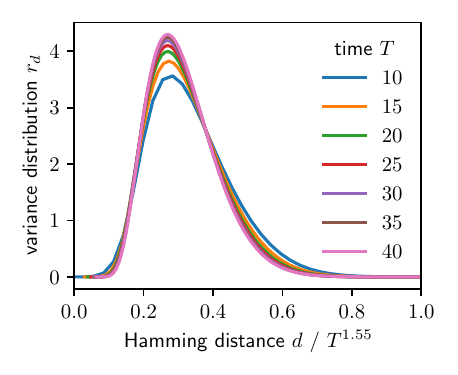}
    \includegraphics{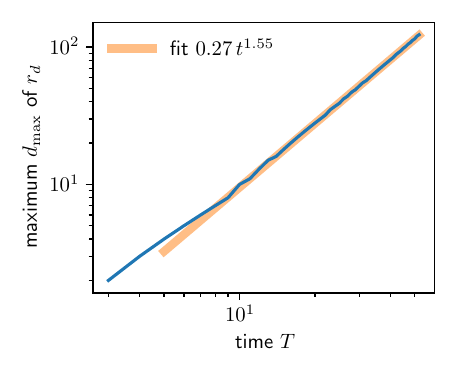}
    \caption{Left: Hamming distance $d$ resolved variance distribution $r_d$ for Haar-random circuits at different times with local Hilbert space dimension $q=2$. The Hamming distance $d$ resolved variance $\mathrm{Var}\langle O \rangle_{\rho_d}$ is divided by the full variance $\mathrm{Var}\langle O \rangle_{\rho}$ at every given timestep, the Hamming distance is rescaled with $t^{1.55}$, the scaling of the position of the coherence-maximum, see right panel. Right: Hamming distance $d_\mathrm{max}$ for which $r_d$ is maximal as a function of time (blue) and the fit (orange) indicating a scaling of $\sim t^{1.55}$. Results are obtained using state-vector simulations with system size $L=26$.}
    \label{fig:haar_random_trace_coherence_resolved}
\end{figure}

\begin{figure}[H]
    \centering
    \includegraphics{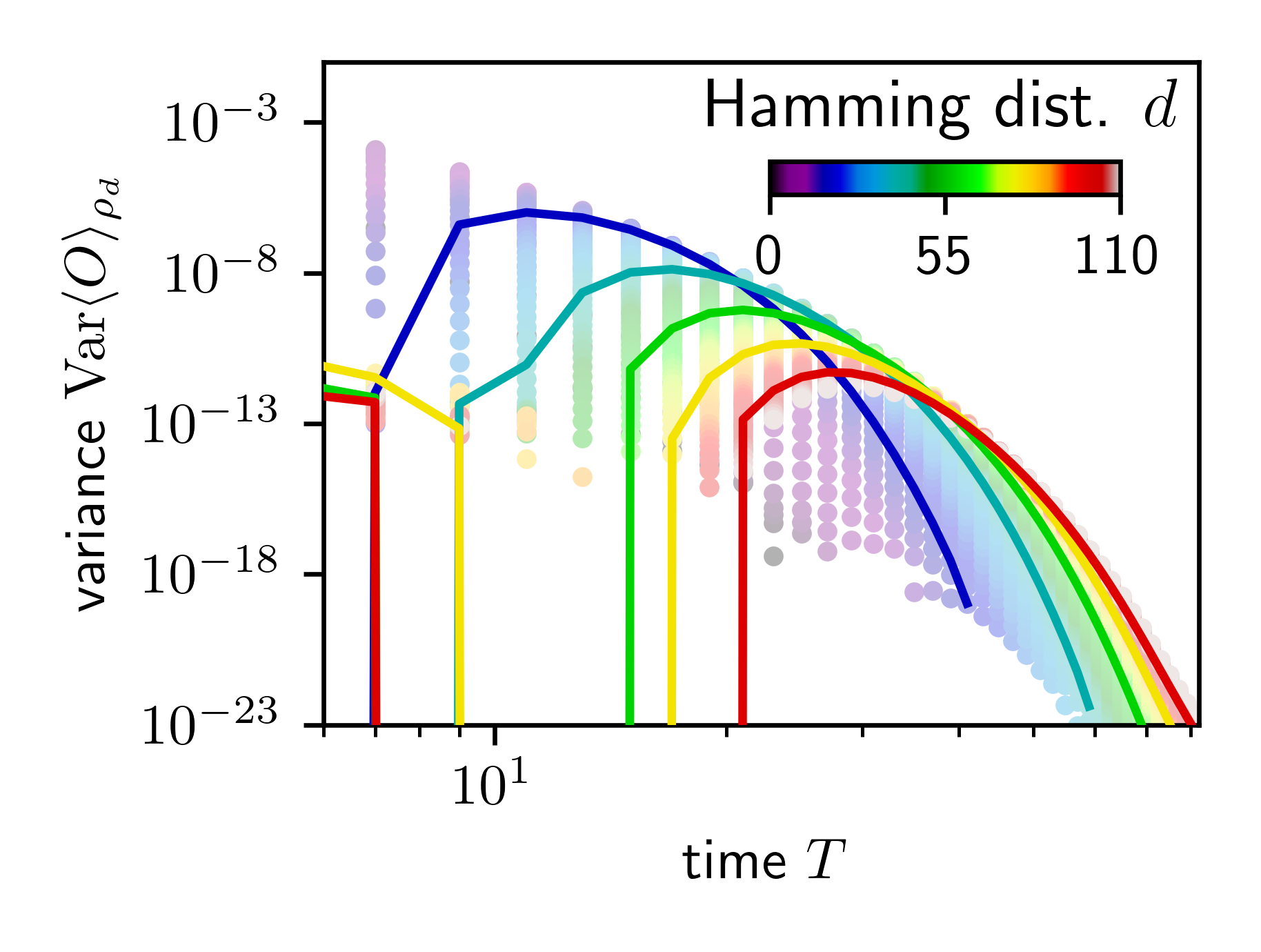}
    \includegraphics{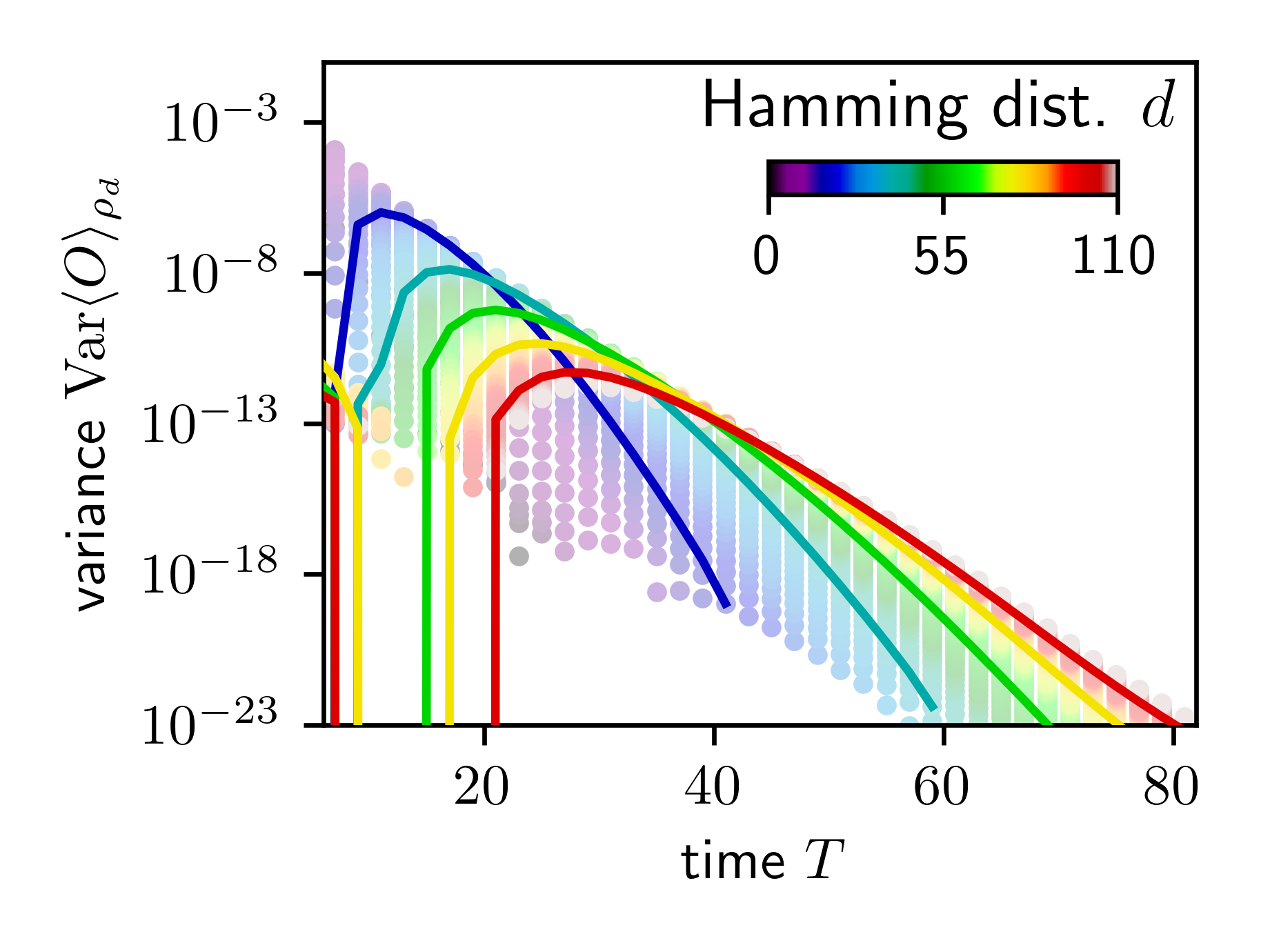}
    \caption{Variance of correlation values per Hamming distance $\mathrm{Var}\langle O \rangle_{\rho_d}$ for Haar-random circuits on log scale (left) and linear scale (right), the same data as in Fig.~\ref{fig:haar_random_trace_coherence_resolved}. In contrast to Haar-random circuits with a $U(1)$ conservation law, there is no long-time polynomial decay, $\mathrm{Var}\langle O \rangle_{\rho_d}$ appears to decay over exponentially fast with time.}
    \label{fig:variance_behaviour_haar}
\end{figure}

For Haar-random circuits, fluctuations of correlation values decay exponentially. In order to analyze the Hamming distance $d$ dependence, we again compute the variance contribution of all parts with Hamming distance $d$
\begin{align}
    \mathrm{Var} \langle O \rangle_{\rho_d(T)} = \sum_{X \text{ with }  d(X)=d} \overline{\langle O \rangle_{\rho_{X}(T)}^2}.
\end{align}
This corresponds to all possibilities of applying the projector $\mathcal{P}_0$ everywhere except for $d$-spacetime points where we apply the complementary projector $\mathcal{P}_1$. The results are shown in Fig.~\ref{fig:haar_random_trace_coherence_resolved} and Fig.~\ref{fig:variance_behaviour_haar}. For large times, the distribution appears to converge when rescaling the Hamming distance with $t^{1.55}$ being maximal around $d/t^{1.55}\sim 0.3$. Hence, for these circuits the Hamming distance $d$ of the parts with the largest contribution to the fluctuations is growing with time as $t^{1.55}$. %
This means that the number of largest contributions to the signal stems from those parts, which are off-diagonal at a macroscopic fraction of time points. In this sense, we can understand the signal as quantum: The contribution of the purely classical part $\rho_0$ vanishes; the amount of Hamming distance we have to keep for a faithful approximation to the expectation value scales roughly linear with the size of the butterfly cone.

The diffusive growth observed here may appear surprising given that information spreading measured by out-of-time ordered correlators is linear. This may be explained by us calculating $\mathrm{Tr}[\sigma^z_0(t) \sigma^z_0]$, the linear spreading information is projected out, leaving the part contributing to the signal to grow diffusively.

\tocless\section{Details for Haar-random Circuits with U(1) conservation\label{ssec:variance_behaviour_u1}}

\tocless\subsection{Maximum}
\begin{figure}[H]
    \centering
    \includegraphics{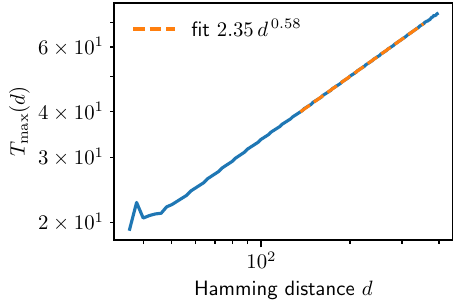}
    \caption{Position of the maximum $T_\mathrm{max}$ of $\mathrm{Var}\langle O\rangle_{\rho_d}$ for a given Hamming distance $d$ for the results shown in Fig.~\ref{fig:u1_coherence_analysis} (top). The maximum is found by fitting a parabola to the peak. Then, the center of the fitted parabola is used as $T_\mathrm{max}(d)$.}
    \label{fig:position_maximum_figure}
\end{figure}

In this section, we provide the numerical results for the position of the maximum $t_\mathrm{max}$ for a given Hamming distance part $\rho_d$ for the correlation values for U(1) Haar-random circuits. They are shown in Fig.~\ref{fig:position_maximum_figure}.

\tocless\subsection{Additional Figure}
For completeness, we show the required Hamming distance as a function of time given a relative error $\epsilon_\mathrm{r}$ in this section.

\begin{figure}[H]
    \centering
    
    \includegraphics[scale=1]{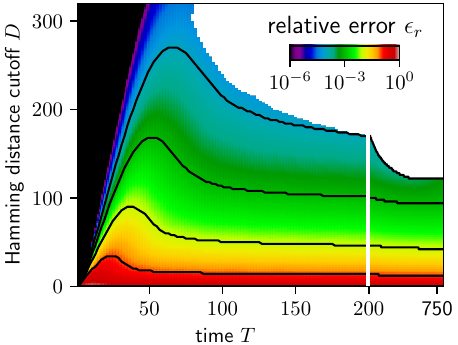}
    
    \caption{Needed Hamming distance cutoff $D$ as a function of time for different relative errors $\epsilon_\mathrm{r}$, Eq.~\eqref{eq:mse_coherence_resolved}, i.e., the  root-MSE over the standard deviation of $\langle O \rangle_\rho = \mathrm{Tr}[\sigma^z_0(t)\sigma^z_0]/2^L$ over the circuit ensemble. The black lines mark the results for the relative error in steps of $10^{-1}$, the right site of the figure is plotted on a different linear scale, separated by the white line from the left part.}
    \label{fig:variance_behaviour_u1}
\end{figure}

For short and intermediate times, the behavior of the relative error is quite different and shown in Fig.~\ref{fig:variance_behaviour_u1}. Here we find that to achieve a certain accuracy, the required Hamming distance cutoff $D$, Eq.~\eqref{eq:mse_coherence_resolved}, increases at first, then reaches a peak, and finally decreasing and levels off, converging to the discussed long-time value. This comes from an enhancement of the variance for intermediate times before it approaches the long-time behavior, see also Sec.~\ref{sec:result_charge_cons_circuits}. It highlights the long-time-limit suppressing the intricate intermediate time behavior and being, that way, simpler.

\tocless\subsection{Classical contribution to the Variance\label{ssec:convergence_classical_part}}
\begin{figure}[H]
    \centering
    \includegraphics{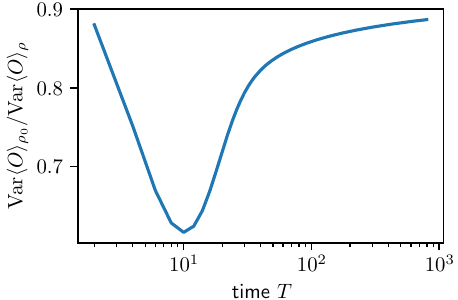}
    \includegraphics{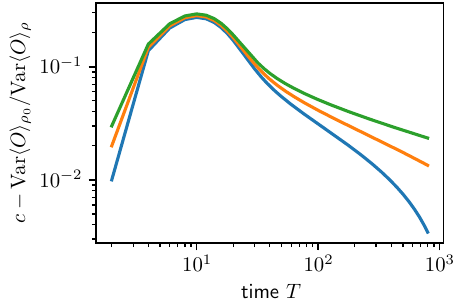}
    \caption{Ratio of the classical variance $\mathrm{Var}\langle O \rangle_{\rho_0}$, main text Eq.~\eqref{eq:coherence_decomp}, and the full variance $\mathrm{Var}\langle O \rangle_{\rho}$ (left). Assuming a polynomial convergence towards the long-time-limit, the long-time-limit has to be larger than about $0.89$, as can be deduced by the curves in the right panel not showing a polynomial decay when subtracting $0.89$.}
    \label{fig:classical_variance_contribution}
\end{figure}

In this section we show the evidence for the classical variance contributing at least about $90\%$ of the total variance. The numerical results are shown in Fig.~\ref{fig:classical_variance_contribution}

\tocless\subsection{Expectation Values\label{app:expectation_values_u1_circuits}}
In this section we present the results when calculating expectation values rather analogous to the analysis of correlators provided in the main text.

In particular, we consider the expectation value
\begin{align} \label{eq:expectation_values}
    \langle \Psi |U^\dagger(T)\sigma^z_0 U(T)|\Psi\rangle,
\end{align}
where $|\Psi\rangle$ is a domain wall state
\begin{align}
    |\Psi\rangle = |\!\!\uparrow\rangle_{-L/2+1}\otimes\cdots \otimes|\!\!\uparrow\rangle_{0}\otimes |\downarrow\rangle_{1}\otimes\cdots |\!\!\downarrow\rangle_{L/2}
\end{align}
and $U$ is a charge-conserving brickwall circuit as introduced in the main text.

\begin{figure}[H]
    \centering
    \includegraphics[scale=1]{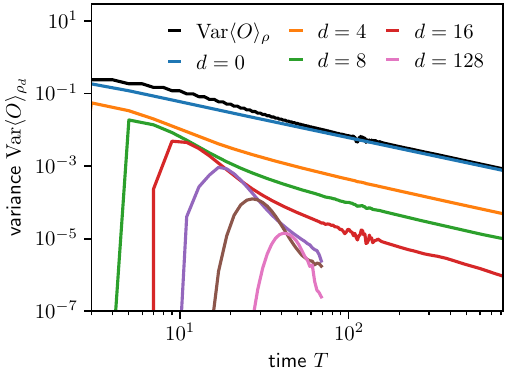}
    \includegraphics[scale=1]{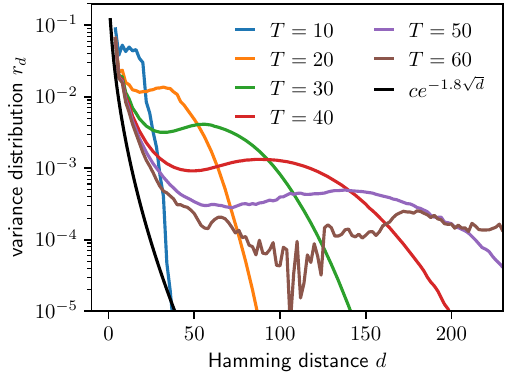}
    \caption{Results for the Hamming distance $d$ resolved circuit-to-circuit variance for expectation values, see \ref{eq:expectation_values}. These figures are analogous to the figure in the main text for correlators, see Fig.~\ref{fig:u1_coherence_analysis}. Again, we resolve the variance according to the Hamming distance $d$ by calculating the variance contribution of the density matrix $\rho_d(T)$, Eq.~\eqref{eq:coherence_decomp}. Left: variance as a function of time $T$ for different $d$. Black shows the full variance, i.e., the sum over $d$. Right: variance distribution $r_d$ as a function of Hamming distance $d$ for different times. Note that the normalization yields the sum over every curve to be $1$. Black shows for comparison the long-time-limit approached by the variance distribution for the correlators, taken from Fig.~\ref{fig:u1_coherence_analysis}. The results were obtained the same way like in Fig.~\ref{fig:u1_coherence_analysis}, with a system of size $200$ with the precision of itensor set to $10^{-20}$ and by simulating the average gate dynamics. Numerical approximation errors are clearly visible (fast oscillation around time $100$ in the left panel, large fluctuations for Hamming distance larger than $50$ in the right panel. Hence, we limit the shown results.}
    \label{fig:results_coherence_expectation_values}
\end{figure}

Again, we decompose the circuit-to-circuit variance of the expectation values according to the phase flip noise sensitivity in terms of the Hamming distance $d$, for which we show the results in Fig.~\ref{fig:results_coherence_expectation_values}.

First, we note that the numerical approximation through TEBD affect the results far more severely, preventing a large time analysis as done in the main text. Restricting ourselves to the faithful results, we find similar results like in the main text for correlators. The ``classical'' part of the variance dominates for large times, again yielding around $90\%$ of the full variance, see Fig.~\ref{fig:results_coherence_expectation_values} (left). Hence, one can reasonably approximate the real expectation value for an individual circuit using this classical approximation. Similarly to the results in the main text, the variance contributions decay quickly with an increasing amount of coherence, see Fig.~\ref{fig:results_coherence_expectation_values} (right). However, due to the numerical limitations, we cannot make any statements about the long-time scaling. Different to correlators, for finite times, we also observe a finite fraction of the variance to come from parts $\rho_X$ with a large amount of Hamming distance $d(X)$, scaling faster than linear with time. This is again reminiscent of the Haar-random case and cannot be captured efficiently through an expansion in terms of the coherence. However, the fraction of these variance contributions decay so that they become irrelevant for large times.

\begin{figure}[H]
    \centering
    \includegraphics[scale=1]{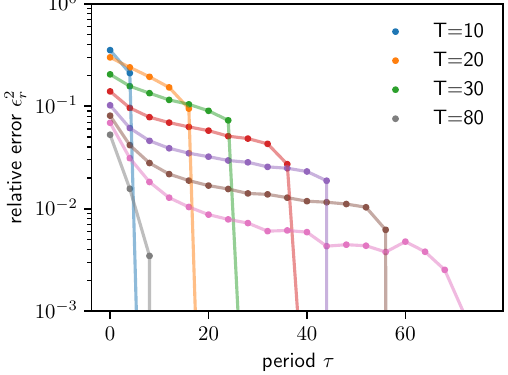}
    \caption{Performance of the Hamming distance based approximation algorithm for expecation values $\langle \Psi|\sigma^z_0(T)|\Psi\rangle$ taking four different diamond shaped partitions, analogous to Fig.~\ref{fig:performance_alg}. We show the relative error squared, i.e., the MSE divided by the full variance for different times.}
    \label{fig:expectation_values_algorithm}
\end{figure}

The performance of the Monte Carlo sampling algorithm is shown in Fig.~\ref{fig:expectation_values_algorithm}. Again, numerical inaccuracies prevent us from making statements about the scaling. Similar to correlators, the relative error decays quickly with increasing $\tau$. At finite times, it saturates, as can be expected from the analyzes of Fig.~\ref{fig:results_coherence_expectation_values}: there are large coherence contributions which require a large $\tau$ to be captured, but their overall contribution vanishes in time. As a result, using a fixed $\tau$, the relative error decreases in time. Hence, we expect a similar scaling like for correlators.

\tocless\subsection{Analytical Considerations\label{ssec:analytcal_u1_appendix}}

In this section we discuss the analytical calculation in detail. We start with the ensemble mean and then calculate the variance approximately.

\tocless\subsubsection{Ensemble averaged Correlator}

We aim to calculate

\begin{align}
    \overline{C}=\overline{\mathrm{Tr}[U \sigma^z_0 U^\dagger \sigma^z_0]}/2^L,
\end{align}
where $U$ is a brickwall circuit of depth $T$, $\sigma^z_0$ is a Pauli matrix and the overline denotes integrating every two-qubit gate $V_{j,j+1}$ over the Haar-random ensemble with $U(1)$ conservation, i.e., the ensemble average, App.~\ref{sec:charge_conserv}.

Taking the average commutes with the trace so that we can write
\begin{align}
    \overline{C}=\mathrm{Tr}[\overline{U \sigma^z_0 U^\dagger} \sigma^z_0]/2^L.
\end{align}

This can be solved iteratively. For the first time evolution, it holds
\begin{align}
    \overline{ V_{0,1} \sigma^z_0 V_{0,1}^\dagger} = \int\! \mathrm{d} V_{0,1} \, (V_{0,1} \sigma^z_0 V_{0,1}^\dagger) = \sigma^z_0/2 + \sigma^z_1/2,
\end{align}
where the integral denotes integrating $V_{0,1}$ over the Haar-random ensemble with $U(1)$ conservation, App.~\ref{sec:charge_conserv}.

Hence under the evolution, the $\sigma^z$ is just mapped to a superposition of $\sigma^z$s, the $\sigma^z$ just performs a random walk and it holds for any time
\begin{align}
    \overline{U\sigma^z_0 U^\dagger} = \sum_j p(j,T) \sigma^z_j.
\end{align}
The time evolution of $p(j,T)$ can then be written as
\begin{align}
    p(j,T+1) = \begin{cases} \frac{p(j,T)+p(j+(-1)^j,T)}{2} \qquad T \, \mathrm{even} \\ \frac{p(j,T)+p(j-(-1)^j,T)}{2} \qquad T \, \mathrm{odd} \end{cases}.
\end{align}
In order to arrive at a closed set of iteration equations, we coarse grain in space and time. We introduce $q(j,T)=p(j,T)/2+p(j+1,T)/2$ and obtain for $j$ and $T$ even
\begin{align}
   q(j,T+2) = \frac{q(j-2,T)+2q(j,T)+q(j+2,T)}{4},
\end{align}
which is a closed set of equations for the whole (coarse grained) time evolution. Hence, we apply the continuum limit and use 
\begin{align}
    q(j,T+2) &\approx q(j,T)+2\partial_T q(j,T) \\
    q(j\pm2,T) &\approx q(j,T)\pm 2 \partial_j q(j,T) +2\partial_j^2 q(j,T).
\end{align}
This yields the diffusion equation
\begin{align}
    \partial_T q(j,T) = \frac{1}{2} \partial_j^2 q(j,T).
\end{align}
It has the well known solution
\begin{align}
    q(j,T) = \frac{1}{\sqrt{2\pi T}}e^{\frac{-j^2}{2T}},
\end{align}
and required $\sum_j q(j,T) = 1$.

Hence, for large times we expect the continuum limit to be a valid approximation and obtain
\begin{align}
    \overline{C}=\mathrm{Tr}[\overline{U \sigma^z_0 U^\dagger} \sigma^z_0]/2^L = p(0,T) \approx q(0,T) = \frac{1}{\sqrt{2\pi T}}
\end{align}
and the propagator reported in the main text
\begin{align}
G(T,i-j)=\mathrm{Tr}[\overline{U \sigma^z_i U^\dagger} \sigma^z_j]/2^L = p(i-j,T) \approx q(i-j,T) = \frac{e^{\frac{-(i-j)^2}{2T}}}{\sqrt{2\pi T}}. \label{eq:derivation_propagator}
\end{align}

\tocless\subsubsection{Variance of the Correlator}
We proceed with calculating
\begin{align}
    \overline{C^2} = \overline{\mathrm{Tr}[U \sigma^z_0 U^\dagger \sigma^z_0]^2}/4^L =  \mathrm{Tr}[(\overline{U \sigma^z_0 U^\dagger \otimes U \sigma^z_0 U^\dagger} )(\sigma^z_0 \otimes \sigma^z_0)]/4^L,
\end{align}
where we doubled the size of the Hilbert space so that we have two copies, denoted with $\otimes$.

Similar to the mean, we start with calculating the update rules. We start with the initial condition

\begin{align}\label{eq:first_average}
    \int \!\mathrm{d} V_{0,1} \, (V_{0,1} \sigma^z_0 V_{0,1}^\dagger) \otimes (V_{0,1} \sigma^z_0 V_{0,1}^\dagger) = &\frac{\sigma^z_0+\sigma^z_1}{2} \otimes \frac{\sigma^z_0+\sigma^z_1}{2}\\ 
    &+ \frac{1}{12}\left(\sigma^z_0 \otimes \sigma^z_0+\sigma^z_1 \otimes \sigma^z_1\right)- \frac{1}{12}(\sigma^z_0 \otimes \sigma^z_1+\sigma^z_1 \otimes \sigma^z_0)\label{eq:first_average_part2}\\
    &+\frac{2}{3} ( \sigma^+_0 \sigma^-_1 \otimes \sigma^-_0 \sigma^+_1+\sigma^-_0 \sigma^+_1 \otimes \sigma^+_0 \sigma^-_1).
\end{align}
The first line is the ``non-interaction'' diffusion and corresponds to the dynamics of $\overline{C}^2$. The second line is a classical ``interaction'' correction---it is phase flip insensitive. The third line is a quantum ``interaction'' and creates the $\sigma^q_j = \sigma^+_j\otimes \sigma^-_j$ walkers. These are phase flip sensitive and projected out by $\mathcal{P}_0[\sigma^{\pm}] = 0$. In order to fully calculate the second moment, we need the full set of evolution rules, which we provide in a shortened notation. We write everything in blocks of four: the first row is the first copy of the Hilbert space, the second row is the second copy of the Hilbert stapce, the left column corresponds to site $j$ and the right column to site $j+1$.
\begin{align}
    \mymatrix{A}{B}{C}{D}  \equiv A_j B_{j+1} \otimes C_j D_{j+1}.
\end{align}
Furthermore, we use $Z\equiv \sigma^z$, for the identity $I$, and the arrow $\mapsto$ represents the evolution under the averaged two-qubit gate.

In this formulation, Eq.~\ref{eq:first_average} reads
\begingroup %
\setlength\arraycolsep{0.1pt}
\begin{align}
    \mymatrix{Z}{I}{Z}{I} \mapsto \frac{1}{4}\left(\mymatrix{Z}{I}{Z}{I} + \mymatrix{I}{Z}{Z}{I} + \mymatrix{Z}{I}{I}{Z} + \mymatrix{I}{Z}{I}{Z}\right)
+ \frac{1}{12}\left(\mymatrix{Z}{I}{Z}{I} + \mymatrix{I}{Z}{I}{Z}\right)- \frac{1}{12}\left(\mymatrix{Z}{I}{I}{Z} + \mymatrix{I}{Z}{Z}{I}\right)
+ \frac{2}{3}\left(\mymatrix{\sigma^+}{\sigma^-}{\sigma^-}{\sigma^+} + \mymatrix{\sigma^-}{\sigma^+}{\sigma^+}{\sigma^-}\right). \label{eq:first_fusion_rule}
\end{align}
\endgroup

All other needed evolution rules are
\begingroup %
\setlength\arraycolsep{0.1pt}
\begin{align}
\mymatrix{I}{Z}{I}{I} &\mapsto \frac{1}{2}\left(\mymatrix{Z}{I}{I}{I} + \mymatrix{I}{Z}{I}{I}\right),\\
    \mymatrix{I}{Z}{Z}{I} &\mapsto \frac{1}{4}\left(\mymatrix{Z}{I}{Z}{I} + \mymatrix{I}{Z}{Z}{I} + \mymatrix{Z}{I}{I}{Z} + \mymatrix{I}{Z}{I}{Z}\right)
- \frac{1}{12}\left(\mymatrix{Z}{I}{Z}{I} + \mymatrix{I}{Z}{I}{Z}\right)+ \frac{1}{12}\left(\mymatrix{Z}{I}{I}{Z} + \mymatrix{I}{Z}{Z}{I}\right)
- \frac{2}{3}\left(\mymatrix{\sigma^+}{\sigma^-}{\sigma^-}{\sigma^+} + \mymatrix{\sigma^-}{\sigma^+}{\sigma^+}{\sigma^-}\right),\label{eq:ana_class_inter}\\
\mymatrix{\sigma^+}{I}{\sigma^-}{I} &\mapsto \frac{1}{4}\left(\mymatrix{\sigma^+}{I}{\sigma^-}{I} + \mymatrix{\sigma^+}{Z}{\sigma^-}{Z} +\mymatrix{I}{\sigma^+}{I}{\sigma^-}+\mymatrix{Z}{\sigma^+}{Z}{\sigma^-}\right),\label{eq:quant_rd_walka}\\
\mymatrix{\sigma^+}{Z}{\sigma^-}{Z} &\mapsto \frac{1}{4}\left(\mymatrix{\sigma^+}{I}{\sigma^-}{I} + \mymatrix{\sigma^+}{Z}{\sigma^-}{Z} +\mymatrix{I}{\sigma^+}{I}{\sigma^-}+\mymatrix{Z}{\sigma^+}{Z}{\sigma^-}\right),\label{eq:quant_rd_walkb}\\
\mymatrix{\sigma^+}{\sigma^+}{\sigma^-}{\sigma^-} &\mapsto \mymatrix{\sigma^+}{\sigma^+}{\sigma^-}{\sigma^-},\\
\mymatrix{\sigma^+}{\sigma^-}{\sigma^-}{\sigma^+} &\mapsto + \frac{1}{24}\left(\mymatrix{Z}{I}{Z}{I} + \mymatrix{I}{Z}{I}{Z}\right)- \frac{1}{24}\left(\mymatrix{Z}{I}{I}{Z} + \mymatrix{I}{Z}{Z}{I}\right)
+ \frac{1}{3}\left(\mymatrix{\sigma^+}{\sigma^-}{\sigma^-}{\sigma^+} + \mymatrix{\sigma^-}{\sigma^+}{\sigma^+}{\sigma^-}\right). \label{eq:quant_int_corr}
\end{align}
All other needed evolution rules follow by the symmetry of swapping the left and right column, the top and bottom row or by multiplying the above rules with $\mymatrix{Z}{Z}{Z}{Z}$.
\endgroup

These rules are the ones used in the main text to obtain the second moment of the correlator and the expectation value via numerical simulation with TEBD presented in Fig.~\ref{fig:u1_coherence_analysis}.

These rules also yield the statements made in the main text: We have classical and quantum ``interactions'', the quantum interactions yield quantum particles $\sigma^q_j = \sigma^+_j\otimes\sigma^-_j$, sensitive to phase flips. Due to the brickwall structure, the quantum particles are created in pairs and live for at least two time steps, hence, the minimal amount of Hamming distance is $4$. These quantum particles then perform a random walk while emitting or absorbing $Z$ random walkers, Eq.~\eqref{eq:quant_rd_walka} and~\eqref{eq:quant_rd_walkb}. These random walkers can then fuse again, yielding a similar correction to the non-interacting $Z$-walkers like the classical one, compare Eq.~\eqref{eq:ana_class_inter} the terms with prefactor $1/12$ and Eq.~\eqref{eq:quant_int_corr} with prefactor $1/24$. Classical corrections are on the order of $1/12$, quantum interactions are further supressed by the prefactors $2/3$ and $1/24$ as well as the $1/4$ in the random walk. Furthermore, the quantum processes generate many-body dynamics, since the $\sigma^q_j$ emit and absorb $Z$-random walkers. Since the final probe in the correlator only permits one $Z$ random walker per Hilbert space, all of these emitted random walkers during a quantum process have to be absorbed again.

Now, we can perform an analytical calculation of the second moment using the above set of rules and the propagator $q$ from the previous section. As an approximation, we just consider processes with a single classical interaction at site $a$ at time $t$ so that we split up the whole circuit time evolution $T$ into three parts: the circuit up to $t$, the layer $\mathcal{U}_t$ generating the interaction and finally all the layers after $t$ till $T$. We approximate the time evolution of the first and last part as random walks for which we derived the propagator $G$, Eq.~\eqref{eq:derivation_propagator}. 

\paragraph{Single Interaction.} It yields for the process $\sigma^z_a \otimes \sigma^z_a \rightarrow \sigma^z_a \otimes \sigma^z_a/12$ of a single classical interaction
\begin{align}
    \left.\overline{C^2}\right|_{{\sigma^z_a \otimes \sigma^z_a \rightarrow \sigma^z_a \otimes \sigma^z_a/12 \text{ at } t}} &=  \mathrm{Tr}[ \sigma^z_0(t)\sigma^z_a]\mathrm{Tr}[ \sigma^z_0(t)\sigma^z_{a}]\frac{1}{12}\mathrm{Tr}[ \sigma^z_a(T-t-1)\sigma^z_0]\mathrm{Tr}[ \sigma^z_{a}(T-t-1)\sigma^z_{0}]/2^{4L} \nonumber \\
    &\approx G(t,a)^2 G(T-t-1,a)^2/12.
\end{align}
All possible classical processes from Eq.~\eqref{eq:ana_class_inter} for a two-qubit gate acting on sites $a$ and $a+1$ at time $t$ are
\begin{align}
    Z_a \otimes Z_a \rightarrow \frac{1}{12}\left( Z_a \otimes Z_a - Z_a \otimes Z_{a+1}-Z_{a+1} \otimes Z_a + Z_{a+1} \otimes Z_{a+1}\right) \\
    Z_{a+1} \otimes Z_a \rightarrow \frac{-1}{12}\left( Z_a \otimes Z_a - Z_a \otimes Z_{a+1}-Z_{a+1} \otimes Z_a + Z_{a+1} \otimes Z_{a+1}\right)\\
    Z_a \otimes Z_{a+1} \rightarrow \frac{-1}{12}\left( Z_a \otimes Z_a - Z_a \otimes Z_{a+1}-Z_{a+1} \otimes Z_a + Z_{a+1} \otimes Z_{a+1}\right) \\
    Z_{a+1} \otimes Z_{a+1} \rightarrow \frac{1}{12}\left( Z_a \otimes Z_a - Z_a \otimes Z_{a+1}-Z_{a+1} \otimes Z_a + Z_{a+1} \otimes Z_{a+1}\right),
\end{align}
which sum up to
\begin{align}
    (Z_a-Z_{a+1}) \otimes (Z_a-Z_{a+1}) \rightarrow \frac{1}{12}(Z_a-Z_{a+1}) \otimes (Z_a-Z_{a+1}).
\end{align}
This yields in terms of the propagators
 and approximate $T-t-1\approx T-t$
\begin{align}
    \left.\overline{C^2}\right|_{\text{classical interaction at $t,a$}} &= \frac{1}{12}\left[G(t,a)-G(t,a+1)\right]^2\left[G(T-t-1,a)-G(T-t-1,a+1)\right]^2 \\
    &\approx \frac{1}{12}\left[2 \partial_a q(a,t)\right]^2\left[2\partial_a q(a,T-t-1)\right]^2, \label{eq:app_exact_ana_u1}
\end{align}
where we use that the gradients of the discrete random walk fulfill $G(t,a)-G(t,a+1)=2 \partial_a q(a,t)$. The reason is that after every two qubit gate, the difference of $G(t,a)-G(t,a+1)$ is zero for the $(a,a+1)$ after the two qubit gate, but since the discrete and continuous distributions agree on average, the gradient of the discrete process, when not vanishing, is the sum of the two gradients of the continuous process, which is, roughly, twice the gradient.

Now we sum over all $t$ and all even or odd $a$ (depending on $t$ being odd or even to attribute for the brickwall structure) and approximate $T-t-1\approx T-t$
\begin{align}
    \left.\overline{C^2}\right|_{\text{single classical interaction}} &\approx  \frac{4}{3} \int\!\mathrm{d}t \int\!\frac{\mathrm{d}a}{2} \,\left[\partial_a G(t,a)\right]^2\left[\partial_a G(T-t,a)\right]^2. \label{eq:classical_inter_ana_prop}
\end{align}
Defining the variables $\tilde{t}=t/T$ and $\tilde{a}=a/\sqrt{T}$ yields the desired scaling $T^{-5/2}$ and a $T$ independent integral. We may proceed with the analytical calculation as
\begin{align}
    \left.\overline{C^2}\right|_{\text{single classical interaction}} &= \frac{2}{3}\frac{1}{4\pi^2}\int\!\mathrm{d}t \int\!\mathrm{d}a \frac{a^4}{t^3(T-t)^3} e^{-\frac{a^2}{t}-\frac{a^2}{(T-t)}}\\
    &= \frac{2}{3}\frac{1}{4\pi^2}\int\!\mathrm{d}t \int\!\mathrm{d}a \frac{a^4}{t^3(T-t)^3} e^{-\frac{a^2 T}{t(T-t)}}.
\end{align}
Defining $\sigma^2 = \frac{t(T-t)}{2T}$ we use
\begin{align}
    \int\!\mathrm{d}a\, a^4 e^{-\frac{a^2}{2 \sigma^2}} = \sqrt{2\pi\sigma^2} 3 \sigma^4
\end{align}
and get
\begin{align}
    \left.\overline{C^2}\right|_{\text{single classical interaction}} &= \frac{1}{8}\frac{1}{\pi^{3/2}}\frac{1}{T^{5/2}}\int_0^T\!\mathrm{d}t \frac{1}{\sqrt{t(T-t)}}\stackrel{r=t/T}{=} \frac{1}{8}\frac{1}{\pi^{3/2}}\frac{1}{T^{5/2}}\int_0^1\!\mathrm{d}r \frac{1}{\sqrt{r(1-r)}} \\
    &= \frac{1}{8\sqrt{\pi}}\frac{1}{T^{5/2}}, \label{eq:ana_calc}
\end{align}
where we used the known solution $\pi$ for the integral over $r$. The prefactor agrees very well with the numerical result for the purely classical contribution, which we find to be roughly $3/(16\sqrt{\pi})\pm0.001$. %
The missing $3/2$ can be traced back to a renormalization effect---we only allowed for a single interaction process. 

\begin{figure}[H]
    \centering
    \includegraphics[width=0.5\linewidth]{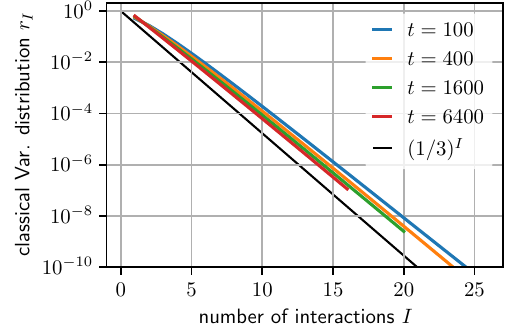}
    \quad
    \includegraphics[width=0.4\linewidth]{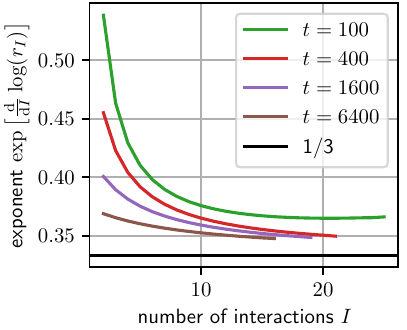}
    \caption{Left: Distribution of the classical variance $r_I = \mathrm{Var} \langle \sigma^z_0(T) \rangle_{\rho_I}/\mathrm{Var} \langle \sigma^z_0(T) \rangle_{\rho}$ as calculated from the first two parts Eqs.~\eqref{eq:first_average} and~\eqref{eq:first_average_part2} over the number of interactions, i.e., number of corrections Eq.~\eqref{eq:first_average_part2} considered. The results were obtained through direct simulation of Eq.~\eqref{eq:first_average} and multiplying the term in Eq.~\eqref{eq:first_average_part2} with $e^{i\omega}$ and, then, performing a Fourier transformation in $\omega$ to obtain the variance resolved by the number of interactions $I$. We can see a clear exponential decay, justifying the expansion in the number of interactions or corrections. Right: Log derivative of the results shown left, indicating $r_I$ converges to about $\exp(-I/3)$ for large times and large $I$.}
    \label{fig:classical_expansion}
\end{figure}

\paragraph{Multiple Interactions.} We can now consider the case of two interactions. For that, we have to consider all processes starting at site $0$ going to sites $a,a+1$ and then to sites $b,b+1$ and finally back to $0$

\begin{align}
    &Z_0 \otimes Z_0 \\
    &\stackrel{\text{evolution}\,t_1}{\rightarrow}(Z_a-Z_{a+1}) \otimes (Z_a-Z_{a+1}) \\
    &\stackrel{\text{interaction}}{\rightarrow}(Z_a-Z_{a+1}) \otimes (Z_a-Z_{a+1})\stackrel{\text{evolution}\,t_2}{\rightarrow} (Z_b-Z_{b+1}) \otimes (Z_b-Z_{b+1}) \\ &\stackrel{\text{interaction}}{\rightarrow} (Z_b-Z_{b+1}) \otimes (Z_b-Z_{b+1})\\
    &\stackrel{\text{evolution}\,T-t_1-t_2}{\rightarrow}Z_0 \otimes Z_0.
\end{align}
\begin{align}
    \left.\overline{C^2}\right|_{\text{two separate classical interactions}} &\approx  \frac{2^6}{12} \int\!\mathrm{d}t_1 \int\!\frac{\mathrm{d}a}{2}\int\!\mathrm{d}t_2 \int\!\frac{\mathrm{d}b}{2} \,\left[\partial_a G(t_1,a)\right]^2\left[\partial_a^2 G(t_2,a-b)\right]^2\left[\partial_b G(T-t_1-t_2,b)\right]^2\nonumber \\
    &\propto \frac{1}{T^4},
\end{align}
where the scaling in the last step follows from pulling $T$ out of the integral by defining the dimensionless integration variables $\tilde{t}_1=t_1/T$, $\tilde{t}_2= t_2/T$ and $\tilde{a}=a/\sqrt{T}$, $\tilde{b}=b/\sqrt{T}$. Iterating this argument shows that higher order interaction processes with $k$ interactions, where all interactions a separated in time by $\mathcal{O}(T)$, are suppressed as $T^{-1-3k/2}$. Hence, in the long-time-limit those processes are dominant where the interactions occur in vicinity to each other. We show the classical part of the variance resolved according to the number of interactions in Fig.~\ref{fig:classical_expansion}. We observe an exponential decay, which justifies the perturbative treatment in the number of interactions and focusing on the single interaction contribution as the dominant contribution. The exponential decay exponent is close to $(1/3)^I$. The value $(1/3)$ is consistent with the additional factor of $1.5$ between the numerical result and the calculation Eq.~\eqref{eq:ana_calc}: Assume that the interaction strength is renormalized by multiple interactions in a row; summing those would yield an enhancement by a factor $\sum_I (1/3)^I = 3/2$, consistent with the numerical result.

\paragraph{Single interaction at the boundary.} When the interaction occurs at the boundary, i.e., within the first or last time steps, the summation over processes does not scale with $T$, and the first propagator is $T$ independent. We hence get via dimensional analysis
\begin{align}
    \left.\overline{C^2}\right|_{\text{single classical boundary interaction}} &\approx  \frac{2^4}{12} \sum_{(t,a) \ll T} \,\left[\partial_a G(t,a)\right]^2\left[\partial_a G(T-t,a)\right]^2\propto \frac{1}{T^3},
\end{align}
where we used $\partial_a G(T-t,a) \propto T^{-3/2}$. Hence, the correction $T^{-3}$ to the $T^{-5/2}$ decay originates from boundary effects.

\begin{figure*}
    \centering
    \includegraphics[scale=0.9]{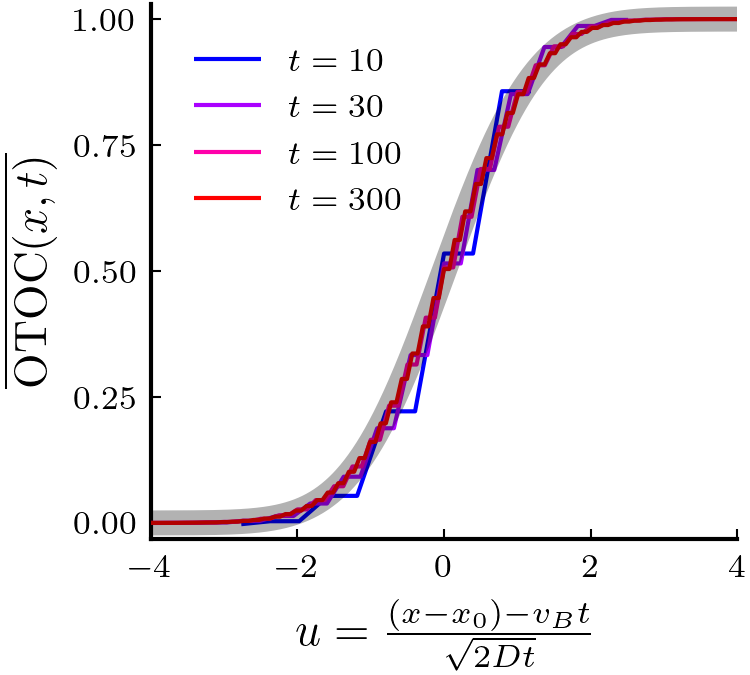}
    \includegraphics[scale=0.9]{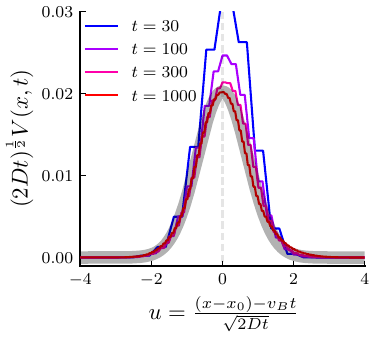}
    \includegraphics[scale=0.9]{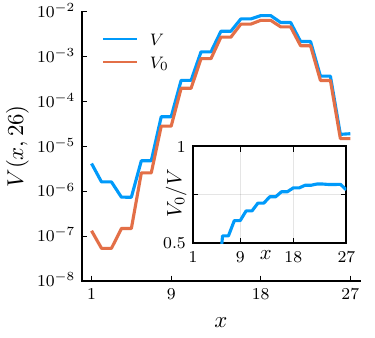}
    \caption{
    Comparison of the classical and exact quantum signal for random circuit OTOCs. (a):
    The mean value of the OTOC signal drops from $1$ to $0$ across the operator front, which signifies the rightward edge of the growing operator $\sigma^z(x_0, t)$. The operator front grows outwards at the butterfly velocity $v_B$ and takes the form of an error function, with a width that spreads diffusively as the front propagates. The mean value of the OTOC signal in the diagonal approximation is exactly the same by construction.
    (b): The circuit-to-circuit variance of the OTOC signal, denoted $V(x, t)$. The size of the variance asymptotically takes the form of a Gaussian centered around the operator front with a height that decays like $t^{-\frac12}$.
    (c): The variance $V(x, t)$ as a function of position in the front for a depth $t=26$ circuit, compared to the variance $V_0(x, t)$ of the diagonal approximation for the same depth. (Inset) The ratio of the fluctuations, which determines the correlation between the diagonal approximation and the exact signal. The diagonal approximation captures a large fraction of the OTOC fluctuations in the operator front --- roughly 80\% at the peak --- but does not accurately capture fluctuations far behind the front where the signal is very small.
    }
    \label{fig:otoc_overview}
\end{figure*}

\tocless\section{Free fermionic circuits\label{ssec:free_ferm}}

\begin{figure}[H]
    \centering
    \includegraphics[scale=1]{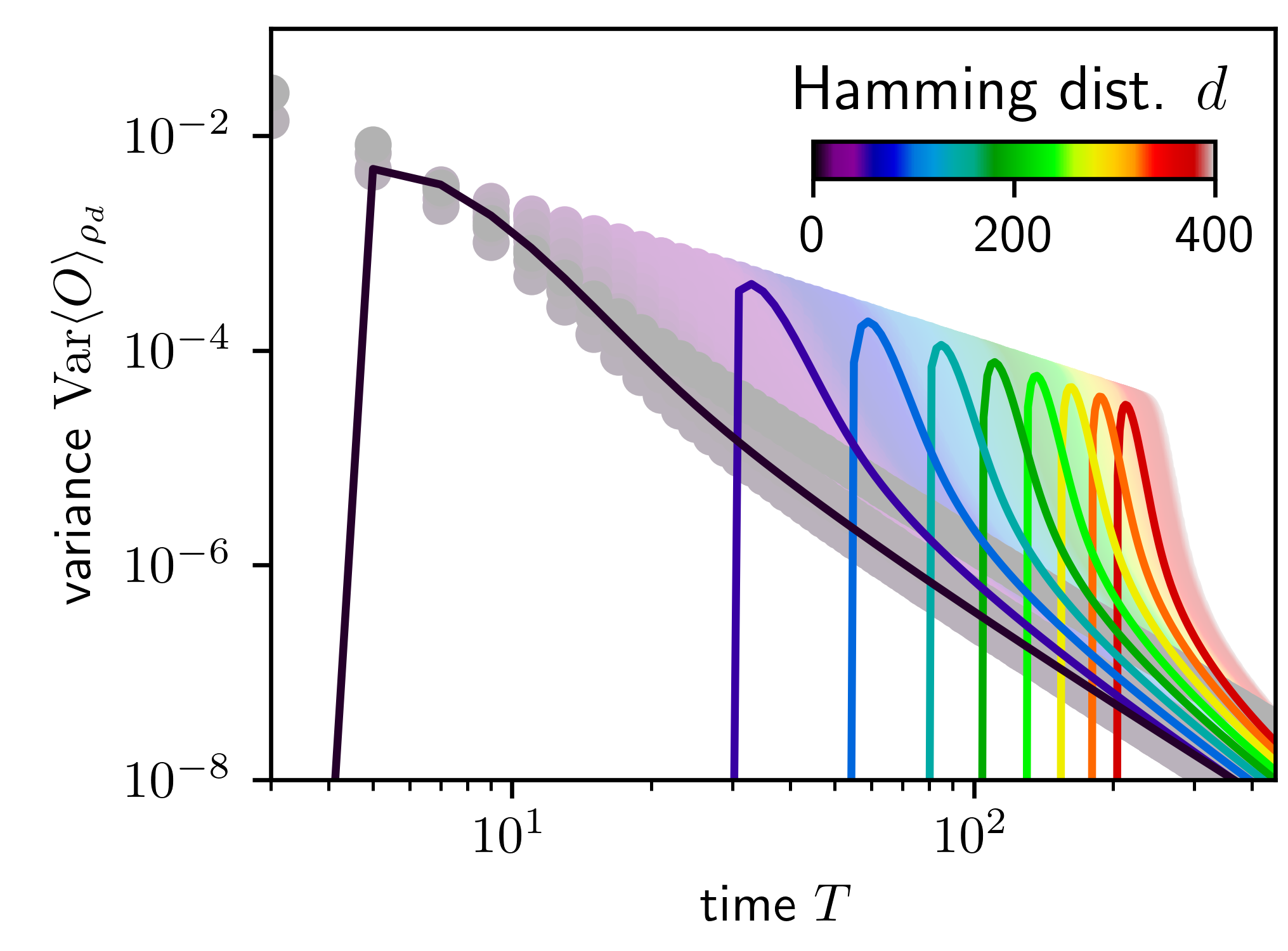}
    \includegraphics[scale=1]{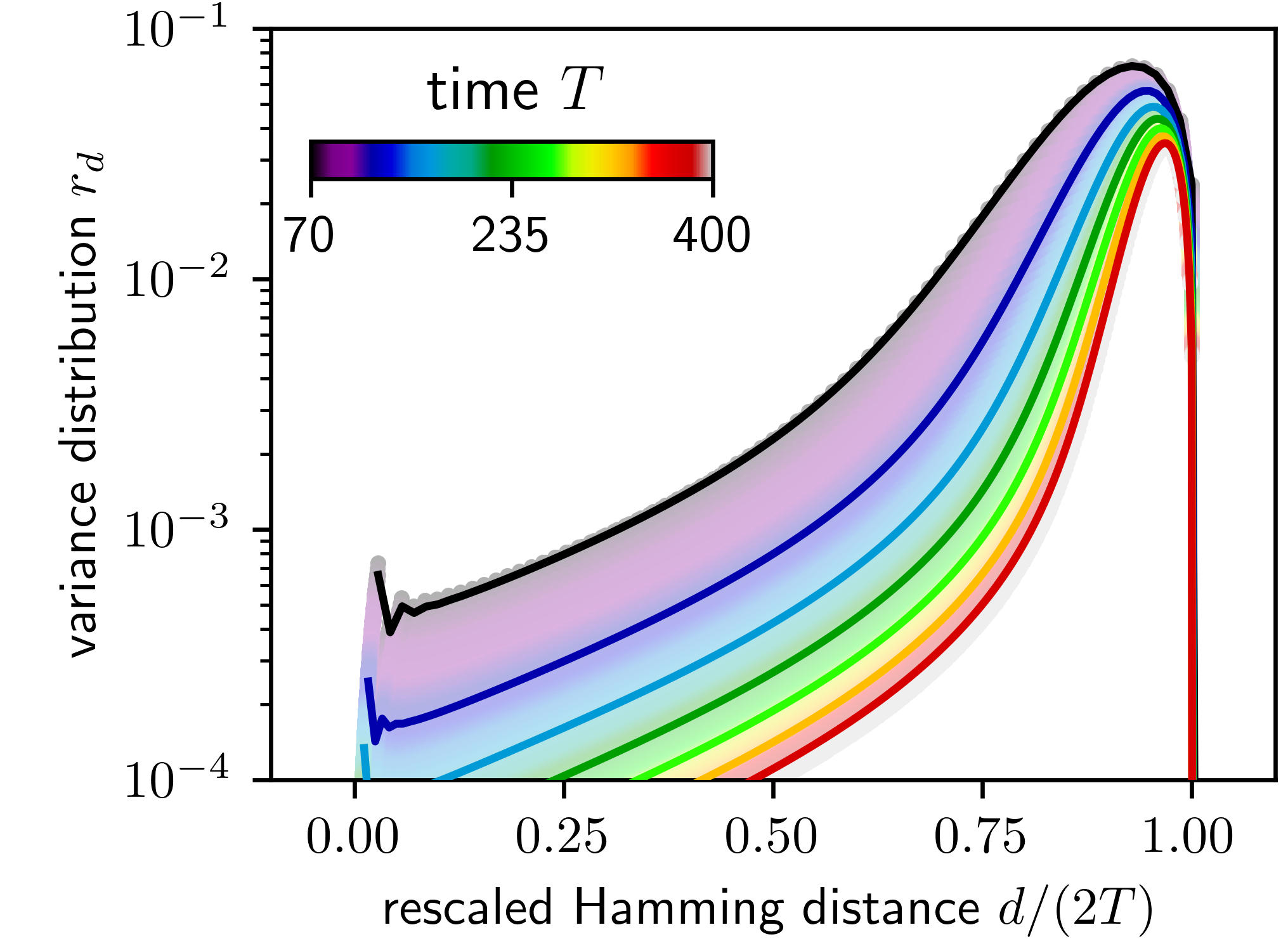}
    \caption{Similar to Fig.~\ref{fig:u1_coherence_analysis} and~\ref{fig:variance_behaviour_haar} we show left the Hamming distance resolved variance $\mathrm{Var}\langle O \rangle_{\rho_d}$ for the correlator with $\rho_\mathrm{init}=O=\sigma^z_0$ and right the distribution of variance for different times. The lines are added as guide to the eyes for every $50$th Hamming distance starting at $d=8$. We find all $\mathrm{Var}\langle O \rangle_{\rho_d}$ to decay as $T^{-2.5}$ for large times, while their sum decays as $\mathrm{Var}\langle O \rangle_{\rho}\propto T^{-1.5}$}
    \label{fig:variance_free_fermions}
\end{figure}

Another instructive set of circuits to study are free fermionic U(1) Haar-random circuits, obtained by removing the interaction term $e^{i \alpha \sigma^z_j \sigma^z_{j+1}}$ from U(1) Haar-random circuits. Here, we find the most significant contribution to the variance to come from those terms with a coherence scaling linear in time, see Fig.~\ref{fig:variance_free_fermions}.

A pictorial explanation based on the analytical perturbative calculation, App.~\ref{ssec:analytcal_u1_appendix}, can be devised as follows, analyzing
\begin{align}
    \mathrm{Var}\langle O \rangle_\rho = \overline{ \mathrm{Tr}\left[ \sigma^z_0(t) \sigma^z_0 \otimes \sigma^z_0(t) \sigma^z_0 \right]}/4^L - \overline{\mathrm{Tr} \left[ \sigma^z_0(t) \sigma^z_0\right]}^2/4^L.
\end{align}
The rules are almost the same as stated for Haar-random circuits with $U(1)$ conservation, the only rule which changes is Eq.~\eqref{eq:quant_rd_walka} and Eq.~\eqref{eq:quant_rd_walkb}, which are replaced by
\begingroup %
\setlength\arraycolsep{0.1pt}
\begin{align}
\mymatrix{\sigma^+}{I}{\sigma^-}{I} &\mapsto \frac{1}{2}\left(\mymatrix{\sigma^+}{I}{\sigma^-}{I}+\mymatrix{Z}{\sigma^+}{Z}{\sigma^-}\right),\label{eq:quant_rd_walka_free_ferm} \\
\mymatrix{\sigma^+}{Z}{\sigma^-}{Z} &\mapsto \frac{1}{2}\left(\mymatrix{\sigma^+}{Z}{\sigma^-}{Z}+\mymatrix{I}{\sigma^+}{I}{\sigma^-}\right).\label{eq:quant_rd_walkb_free_ferm}
\end{align}
\endgroup
This is easy to understand: in the free fermionic case, the particles are connected via a Jordan-Wigner string rather than creating additional $\sigma^z$ random walkers. This enhances the quantum correction to the variance since only the two quantum random walkers have to annihilate again compared to the interacting case, where many particles are created and have all to be annihilated again to yield a non-vanishing contribution to the variance.

In detail, the dynamics can again be understood in terms of the probability distribution describing the position of the two $\sigma^z$'s in $\sigma^z_0(t)\otimes \sigma^z_0(t)$. Initially, we start with two Z-random walkers on the same site $\sigma^z_0\otimes \sigma^z_0$, whose position is hence correlated. To lowest order, these walkers spread independently, yielding a decay of $\overline{ \mathrm{Tr}\left[ \sigma^z_0(t) \sigma^z_0 \otimes \sigma^z_0(t) \sigma^z_0 \right]}$. This initial correlation and its decay is the same as that of the term $\overline{\mathrm{Tr} \left[ \sigma^z_0(t) \sigma^z_0\right]}^2$ and this way subtracted away.

Now we can have two different interaction processes yielding a non-vanishing variance: one which we call ``classical'' as it is invariant under phase flip noise and one which we call ``quantum'' which is destroyed by phase flip noise, as discussed in App.~\ref{ssec:analytcal_u1_appendix} Eq.~\eqref{eq:first_average} to~\eqref{eq:quant_int_corr}. 

The ``classical'' interaction process occurs when the two random walkers meet yielding correction terms to their independent spreading. This interaction is suppressed by a factor $1/12$, allowing for a perturbative treatment, as discussed around Eq.~\eqref{eq:ana_class_inter}, yielding these classical interaction variance contributions scale as $T^{-2.5}$.

We can now compare this ``classical'' variance to the ``quantum'' variance. This is formed by the two $\sigma^z$-random walkers fusing to two $\sigma^\pm ,\sigma^\mp$-random walkers, see Eq.~\eqref{eq:ana_class_inter} and~\eqref{eq:first_fusion_rule}. These are susceptible to phase flip noise and mapped to $0$ under action of $\mathcal{P}_0$. The ``quantum'' random walkers preserve the initial fusion information as a sign, compare Eq.~\eqref{eq:ana_class_inter} and~\eqref{eq:first_fusion_rule}, over time, so that when interacting at a later time, they can decouple again into two random Z-random walkers biasing the configuration which yield to their creation. This then contributes to the variance. This allows the ``quantum'' random walker to propagate the fusion information over long periods of time and that way, stabilizing the variance over long periods of time. This can also be seen in Fig.~\ref{fig:variance_free_fermions} (right): at any time, most of the variance is created by the parts with Hamming distance $d$ close to $2T$, where $2T$ is the maximum Hamming distance created by the quantum walkers being created at time $t=0$ for the whole duration of the process. Hence, most of the variances comes from parts where the quantum walkers live for most of the time. 

A possible explanation for these processes to be dominant follows again from the evolution equations, Eq.~\eqref{eq:first_average} to~\eqref{eq:quant_int_corr}. While the classical contributions to the variance scale as $T^{-2.5}$ due to the propagators being effectively $\partial_a G$, Eq.~\eqref{eq:classical_inter_ana_prop}, the free fermionic quantum walkers perform a random walk which may yield a slower decay of the variance as $T^{-1.5}$.

\tocless\section{Details of the statistical model for random circuit OTOCs \label{sec:appendix_otoc}}

In this section, we develop an analytical model of circuit-to-circuit fluctuations of OTOCs that is solvable in the limit of large local Hibert space dimension, $q \gg 1$. 
The model provides an interpretation to OTOC fluctuations being the strongest close to the operator wavefront, $x = v_B t$, and explains why the fluctuations decay as a slow power law of circuit depth $T$, ${\rm Var\,OTOC}(x = x_0 + v_B T, T) \propto 1/\sqrt{T}$, where $x$ is the qubit position and $v_B$ is the front velocity.
We use the model to assess how well the diagonal approximation captures the circuit specific part of the OTOC.
Our results indicate that the diagonal approximation becomes exact in the limit $q \rightarrow \infty$:
\begin{equation}\label{eq:main_result_da_vs_no_da}
\frac{|{\rm OTOC}(x, T) - {\rm OTOC_{\rm DA}}(x, T)|}{\sqrt{{\rm Var}[{\rm OTOC}(x, T)]}} = {\cal O}(1 / q),
\end{equation}
where ${\rm OTOC_{\rm DA}}(x, T)$ is the OTOC value computed with the help of the diagonal approximation.

\tocless\subsection{Mean OTOC\label{sec:mean_OTOC}}
First, we evaluate the value of OTOC averaged over the realizations of a random circuit $U$. 
As shown in Ref.~\cite{nahum2018}, for a brickwork circuit composed of two-qubit Haar-random gates, the problem can be mapped onto drift-diffusive motion of a Brownian particle.
The Brownian particle represents a wavefront of a spreading operator $\sigma^z_{x_0}(T)$.
We review the mapping in detail to make the discussion self-contained.

To start with, we use Eq.~\eqref{eq:OTOC_definition} to represent $\overline{{\rm OTOC}(x, T)}$ as 
\begin{equation}\label{eq:otoc_from_pop_dyn}
    \overline{{\rm OTOC}(x, T)} = [{\Sigma_2}(T)]^{mn}_{qp} (\sigma^z_x)_{nq} (\sigma^z_x)_{pm},
\end{equation}
where indices $m, n, p, q$ label states in the system's Hilbert space, and a summation over the repeated indices is assumed. 
Operator ${\Sigma}_2(T)$ acts in an extended space equivalent to the two copies of the system's Hilbert space, with the upper and lower pairs of indices corresponding to the two respective copies. It contains information about circuit realization-averaged dynamics of $\sigma_{x_0}^z(T)$:
\begin{equation}\label{eq:sigma_T_definition}
    {\Sigma}_2(T) = \overline{U(T)\sigma_{x_0}^z U^\dagger (T) \otimes U(T) \sigma_{x_0}^z U^\dagger (T)}.
\end{equation}
Here the tensor product is between the two copies of the Hilbert space. 

A convenient way to evaluate $\Sigma_2(T)$ is to track the evolution of this operator as the circuit is constructed gate by gate.
Let us first consider an individual pair of qubits. 
A two-copy operator ${\cal O}$ acting on the pair transforms under the application of a two-qubit gate $V$ and subsequent averaging as
\begin{equation}\label{eq:evolution_step}
{\cal O}_{qp}^{mn} \rightarrow \overline{{V_{mm^\prime} V_{qq^\prime} {\cal O}^{m^\prime n^\prime}_{q^\prime p^\prime} V^\dagger_{n^\prime n} V^\dagger_{p^\prime p}}}.
\end{equation}
In the case of Haar-random gates, the averaging implements an orthogonal projection of ${\cal O}$ onto the subspace of operators that are invariant under the action of $V \otimes V$ for {\it any} $V$ \footnote{To introduce a notion of an orthogonal projection, one has to define a dot product operation. The dot product of two-copy operators ${\cal O}_1$ and ${\cal O}_2$ is $[{\cal O}_1^\dagger]^{mn}_{pq}[{\cal O}_2]^{nm}_{qp}$.}. 
There are only two such invariant operators.
These are the permutation operators ${\cal I} {\cal I}$ and ${\cal S} {\cal S}$.
Here ${\cal I}$ and ${\cal S}$ act on individual qubits and are defined via
\begin{equation}\label{eq:I_and_S}
    {\cal I}_{qp}^{mn} = \frac{\delta_{qp}\delta_{mn}}{q},\quad\quad {\cal S}_{qp}^{mn} = \frac{\delta_{qn}\delta_{mp}}{q^2}.
\end{equation}
Operator ${\cal S}$ implements the exchange the two Hilbert space copies, while ${\cal I}$ leaves the two copies intact.
We introduced normalization factors $1 / q$ and $1 / q^2$ in Eq.~\eqref{eq:I_and_S} for convenience; they guarantee that the basis operators satisfy the condition ${\cal I}^{pq}_{qp} = {\cal S}^{pq}_{qp} = 1$.

Because the averaging acts as a projector on the subspace spanned by ${\cal I}$ and ${\cal S}$, $\Sigma_2(T)$ can be represented as a superposition of strings of these operators at all times $T$:
\begin{equation}\label{eq:sigma_decomposition}
    \Sigma_2(T) = \sum_\nu p^{(2)}_\nu(T)\,\nu,
\end{equation}
where $\nu = \prod_{j = 1}^N \Pi_{j}$ with $\Pi_j \in \{{\cal I}, {\cal S}\}$. Together with the normalization condition for ${\cal I}$ and ${\cal S}$, definition \eqref{eq:sigma_T_definition} can be used to show that 
\begin{equation}\label{eq:prob_normalization}
    \sum_{\nu} p^{(2)}_\nu(T) = 1.
\end{equation}
This allows one to interpret $p^{(2)}_\nu$ as ``probabilities'' of different operator configurations $\nu$.

To obtain $p^{(2)}_\nu(T)$, we need to specify the rules by which operator configurations evolve with $T$. 
Operators ${\cal II}$ and ${\cal SS}$ are invariant under any unitary $V \otimes V$; therefore, 
\begin{equation}
    {\cal II} \rightarrow {\cal II},\quad\quad {\cal SS} \rightarrow {\cal SS}
\end{equation}
under the update in Eq.~\eqref{eq:evolution_step}.
The two configurations that evolve non-trivially are ${\cal SI}$ and ${\cal IS}$. 
Rules for these configurations follow from the fact that the right hand side of Eq.~\eqref{eq:evolution_step} amounts to an orthogonal projection onto an invariant subspace.
A straightforward calculation shows
\begin{equation}\label{eq:update_si}
    {\cal SI} \rightarrow \frac{q^2}{q^2 + 1}{\cal SS} + \frac{1}{q^2 + 1}{\cal II}.
\end{equation}
The update rule for ${\cal IS}$ is the same. 

Equations~\eqref{eq:sigma_decomposition}--\eqref{eq:update_si} define a Markov process in the space of operator configurations. 
The initial condition for the Markov process follow from applying a single layer of two-qubit gates to $\sigma_{x_0}^z \otimes \sigma_{x_0}^z$ and performing the averaging:
\begin{align}\label{eq:initial_conditions}
    \Sigma_2(1) =\, &(\dots {\cal I}{\cal I}{\cal S}{\cal S}{\cal I}{\cal I}\dots) \times \frac{q^4}{q^4 - 1} - (\dots {\cal I}{\cal I}{\cal I}{\cal I}{\cal I}{\cal I}\dots)\times \frac{1}{q^4 - 1}.
\end{align}
In the first term, the two ${\cal S}$ operators are on sites $x_0$ and~$x_0 + 1$.

Let us consider the evolution of the first term in Eq.~\eqref{eq:initial_conditions}. 
According to Eq.~\eqref{eq:update_si}, an initially localized domain of ${\cal S}$ operators starts growing in time in a stochastic way.
Consider, for example, the right boundary of the ${\cal S}$-domain.
This boundary can be viewed as a random walker.
A walker steps to the right with probability $p = q^2 / (q^2 + 1)$ and steps to the left with probability $1 - p$.
Because the two probabilities are unequal, the motion of the walker is a biased diffusion.
At sufficiently large times, $T \gg q^2$, a continuum description of the motion becomes appropriate. 
In this case, the probability distribution of the random walker position $x$ is a standard diffusion kernel:
\begin{equation}\label{eq:diffusion_kernel}
    G(x, T) = \frac{1}{\sqrt{4\pi D T}}\exp\Bigl[-\frac{(x - x_0 - v_B T)^2}{4D T}\Bigr].
\end{equation}
The diffusion coefficient $D$ and the drift velocity $v_B$ are given by
\begin{equation}\label{eq:D_vB}
    D = \frac{2q^2}{(q^2+1)^2},\quad\quad v_B = \frac{q^2 - 1}{q^2 + 1}.
\end{equation}
The left boundary of the ${\cal S}$-domain follows an equation similar to Eq.~\eqref{eq:diffusion_kernel}, with $v_B \rightarrow -v_B$.

The second term in Eq.~\eqref{eq:initial_conditions} deserves a note. This term ensures that at late times, the weight of a configuration without random walkers is {\it zero}. 
Indeed, despite the left and the right boundaries of the ${\cal S}$ domain, on average, move in opposite directions, there is a small chance that they meet and collapse (i.e., the term $\dots {\cal I}{\cal I}{\cal S}{\cal S}{\cal I}{\cal I}\dots$ evolves into $\dots {\cal I}{\cal I}{\cal I}{\cal I}{\cal I}{\cal I}\dots$). It is straightforward to show that the collapse probability is $1 / q^4$. Therefore, the weight of the configuration $\dots {\cal I}{\cal I}{\cal I}{\cal I}{\cal I}{\cal I}\dots$ at late times is $\tfrac{1}{q^4}\frac{q^4}{q^4 - 1} - \tfrac{1}{q^4 - 1} = 0$, where the first and the second terms originate from the first and the second term in Eq.~\eqref{eq:initial_conditions}, respectively. 

The above results contain the full information about the time evolution of $\Sigma_2(T)$. 
They can be used to access the realization-averaged dynamics of the operator spreading.
Indeed, the use of Eqs.~\eqref{eq:diffusion_kernel} and~\eqref{eq:otoc_from_pop_dyn} together with relations
${\cal S}_{qp}^{mn} (\sigma^z)_{nq} (\sigma^z)_{pm} = 0$ and ${\cal I}_{qp}^{mn} (\sigma^z)_{nq} (\sigma^z)_{pm} = 1$ yields (we assume $x > x_0$):
\begin{align}\label{eq:average_otoc}
    \overline{{\rm OTOC}(x, T)} &= \int_{-\infty}^x dx'\,G(x', T) = \frac{1}{2}\left(1 + {\rm erf}\Bigl[\frac{x - x_0 - v_B T}{\sqrt{4D T}}\Bigr]\right).
\end{align}
This equation shows that the initially localized operator spreads as a ballistically propagating wavefront. The wavefront velocity is $v_B$. 
The width of the front grows in time as $\sim \sqrt{DT}$.

\tocless\subsection{Circuit-to-circuit fluctuations of OTOC\label{sec:analytics_fluctuations}}
Equation~\eqref{eq:average_otoc} gives the OTOC value averaged over the realizations of the random circuit.
To quantify deviations from the average, we compute the OTOC variance, ${\rm Var\,OTOC}(x, T)$. 
We focus on the limit of large local Hilbert space dimension, $q \gg 1$, in which the problem is analytically tractable.

Evaluation of circuit-to-circuit fluctuations naturally requires one to average the OTOC product, $\overline{{\rm OTOC}(x, T)\,{\rm OTOC}(x, T)}$. 
The leading in $1 / q$ approximation corresponds to an independent averaging,
\begin{equation}\label{eq:independent}
\overline{{\rm OTOC}(x, T)\,{\rm OTOC}(x, T)}\approx \overline{{\rm OTOC}(x, T)}\cdot \overline{{\rm OTOC}(x, T)}.
\end{equation}
Each of the two terms in the latter expression can be visualized using the random walker picture of Sec.~\ref{sec:mean_OTOC}. 
With the considered accuracy, though, the random walkers are independent of each other and OTOC variance is zero.
Thus, to find the variance, we need to account for the interaction between the random walkers.

The interaction processes can be captured by extending the approach of Sec.~\ref{sec:mean_OTOC} to the case of four-copy operators. 
Instead of a two-copy operator $\overline{U \sigma_{x_0}^z U^\dagger \otimes U \sigma_{x_0}^z U^\dagger}$, to find ${\rm Var\,OTOC}(x,T)$, we need to evaluate the following operator:
\begin{equation}
    \label{eq:projection_4}
    \Sigma_4(T) = \overline{U \sigma_{x_0}^z U^\dagger \otimes U \sigma_{x_0}^z U^\dagger \otimes U \sigma_{x_0}^z U^\dagger \otimes U \sigma_{x_0}^z U^\dagger}.
\end{equation}
Similarly to $\Sigma_2(T)$ [Eq.~\eqref{eq:sigma_T_definition}], $\Sigma_4(T)$ can be constructed by starting with $\sigma_{x_0}^z\otimes\sigma_{x_0}^z\otimes\sigma_{x_0}^z\otimes\sigma_{x_0}^z$, and then evolving this operator in time as the gates are included in the circuit.

Let us first consider the effect of one gate.
Application of a two-qubit gate $V$ to a four-copy operator ${\cal O}$ and a subsequent averaging over the realizations of the gate projects ${\cal O}$ onto the subspace that is invariant under the action of $V\otimes V \otimes V \otimes V$ for any $V$. 
The invariant subspace is spanned by $24$ operators of the form $\Pi\Pi$, where $\Pi$ are the operators permuting the Hilbert space copies [in other words, $\Pi$ are the elements of the permutation group $S_4$].
The orthogonal projection onto the permutation subspace can be carried out using the rules of Weingarten calculus (see, e.g., Ref.~\cite{zhou_2019}). 

The above shows that $\Sigma_4(T)$ can be represented as
\begin{equation}\label{eq:configurations_4}
    \Sigma_4(T) = \sum_\nu p^{(4)}_\nu(T)\,\nu
\end{equation}
at any time $T$, where $\nu = \prod_{j=1}^N \Pi_j$ with $\Pi_j \in S_4$. 
We assume that the basis elements $\Pi_j$ are normalized according to $[\Pi_j]_{mn, nm,kp,pk} = 1$ [the four pairs of indices correspond to the four copies]. 
With this normalization, their weights $p^{(4)}_\nu$ satisfy, $\sum_\nu p^{(4)}_\nu = 1$, in full similarity to Eq.~\eqref{eq:prob_normalization}.

The $S_4$ group is significantly richer than the two-element permutation group encountered in Sec.~\ref{sec:mean_OTOC}.
However, it contains a $S_2 \times S_2$ subgroup which plays a special role.
The elements of this subgroup are permutations that operate {\it independently} on the first and the second pair of Hilbert space copies:
\begin{equation}\label{eq:independent_manifold}
    \begin{matrix}
    {\cal I}\\
    {\cal I}
    \end{matrix}\,\,, \quad\quad
    \begin{matrix}
    {\cal S}\\
    {\cal I}
    \end{matrix}\,\,, \quad\quad
    \begin{matrix}
    {\cal I}\\
    {\cal S}
    \end{matrix}\,\,, \quad\text{and}\quad
    \begin{matrix}
    {\cal S}\\
    {\cal S}
    \end{matrix}\,\,,
\end{equation}
where upper and lower elements corresponds to the first and the second pair, respectively.
In the limit $q \gg 1$, the operator configurations contributing to Eq.~\eqref{eq:configurations_4} are predominantly formed out of these four permutations.
Indeed, were the averaging independent in the two pairs of copies, all of the configurations entering Eq.~\eqref{eq:configurations_4} would have a form
\begin{equation}\label{eq:configuration_typical}
    \dots\,\, 
    \begin{matrix}
    {\cal S}\\
    {\cal S}
    \end{matrix}
    \begin{matrix}
    {\cal S}\\
    {\cal S}
    \end{matrix}
    \begin{matrix}
    {\cal S}\\
    {\cal S}
    \end{matrix}
    \begin{matrix}
    {\cal S}\\
    {\cal S}
    \end{matrix}
    \begin{matrix}
    {\cal I}\\
    {\cal S}
    \end{matrix}
    \begin{matrix}
    {\cal I}\\
    {\cal S}
    \end{matrix}
    \begin{matrix}
    {\cal I}\\
    {\cal S}
    \end{matrix}
    \begin{matrix}
    {\cal I}\\
    {\cal S}
    \end{matrix}
    \begin{matrix}
    {\cal I}\\
    {\cal S}
    \end{matrix}
    \begin{matrix}
    {\cal I}\\
    {\cal I}
    \end{matrix}
    \begin{matrix}
    {\cal I}\\
    {\cal I}
    \end{matrix}
    \begin{matrix}
    {\cal I}\\
    {\cal I}
    \end{matrix}
    \begin{matrix}
    {\cal I}\\
    {\cal I}
    \end{matrix}\,\dots,
\end{equation}
with different positions of random walkers (i.e., the ``domain walls'' between ${\cal S}$ and ${\cal I}$ operators) in the upper and lower pairs of copies. 
The random walkers would move according to Eq.~\eqref{eq:update_si}.
What corrections does a finite $q$ bring to this picture?

Remarkably, we find that even when $q$ is finite, the two random walkers move independently---as long as they are situated on different sites. Specifically, application of the Weingarten calculus rules for averaging over the $U(2)$ group yields the following update rules:
\begin{subequations}
\begin{align}
    \begin{matrix}
    {\cal S}\\
    {\cal S}
    \end{matrix}\begin{matrix}
    {\cal I}\\
    {\cal S}
    \end{matrix}\,&\rightarrow\,\,
    \frac{q^2}{q^2 + 1}\,\begin{matrix}
    {\cal S}\\
    {\cal S}
    \end{matrix}\begin{matrix}
    {\cal S}\\
    {\cal S}
    \end{matrix}\,\,+\,\,\frac{1}{q^2+1}\,\begin{matrix}
    {\cal I}\\
    {\cal S}
    \end{matrix}\begin{matrix}
    {\cal I}\\
    {\cal S}
    \end{matrix}\,,\\
    \begin{matrix}
    {\cal I}\\
    {\cal S}
    \end{matrix}\begin{matrix}
    {\cal S}\\
    {\cal S}
    \end{matrix}\,&\rightarrow\,\,
    \frac{q^2}{q^2 + 1}\,\begin{matrix}
    {\cal S}\\
    {\cal S}
    \end{matrix}\begin{matrix}
    {\cal S}\\
    {\cal S}
    \end{matrix}\,\,+\,\,\frac{1}{q^2+1}\,\begin{matrix}
    {\cal I}\\
    {\cal S}
    \end{matrix}\begin{matrix}
    {\cal I}\\
    {\cal S}
    \end{matrix}\,.
\end{align}
\end{subequations}
On the other hand, when the two random walker are on the same site, there are corrections to update rule leading to the deviations from independent random walker picture:
\begin{align}\label{eq:collision}
    \begin{matrix}
    {\cal S}\\
    {\cal S}
    \end{matrix}\begin{matrix}
    {\cal I}\\
    {\cal I}
    \end{matrix}\,\rightarrow\,\,\frac{q^4}{(q^2+1)^2}\,\begin{matrix}
    {\cal S}\\
    {\cal S}
    \end{matrix}\begin{matrix}
    {\cal S}\\
    {\cal S}
    \end{matrix}\, + \frac{q^2}{(q^2+1)^2}\Bigl[\,\begin{matrix}
    {\cal S}\\
    {\cal S}
    \end{matrix}\begin{matrix}
    {\cal I}\\
    {\cal S}
    \end{matrix} + \begin{matrix}
    {\cal S}\\
    {\cal S}
    \end{matrix}\begin{matrix}
    {\cal S}\\
    {\cal I}
    \end{matrix}\,\Bigr] +
    \frac{1}{(q^2+1)^2}\,\begin{matrix}
    {\cal I}\\
    {\cal I}
    \end{matrix}\begin{matrix}
    {\cal I}\\
    {\cal I}
    \end{matrix} + \sum_{\Pi\in S_4} w(\Pi)\,\Pi \Pi.
\end{align}
The first three terms on the right hand side correspond to the independent motion of the two random walkers; however, there is an additional term that involves a summation over $24$ permutations operators. The total weight in this term is zero, $\sum_\Pi w(\Pi) = 0$.
As follows from the Weingarten calculus rules, the weights satisfy $w(\Pi) = {\cal O}(1 / q^6)$. 

The last term in Eq.~\eqref{eq:collision} brings the system out of the manifold corresponding to two pairs of independent copies [Eq.~\eqref{eq:independent_manifold}].
The detour from the independent pair manifold is, however, transient: to the leading order in $1 / q$, the system returns back to manifold~\eqref{eq:independent_manifold} in $\sim 1$ time steps.
After the return, the weights of configurations $p_\nu^{(4)}(T)$ start to differ from the independent random walkers prediction.
Specifically, the weight of configurations in which the random walkers stay at the same site increases by $4 / q^8$, whereas the weight of configurations in which the walkers separate decreases by $4 / q^8$.
The redistribution of weights can be interpreted as an attractive interaction between the random walkers.
The interaction ``amplitude'' is parametrically small, $\propto 1 / q^8$; this makes the problem of finding the OTOC variance amenable to perturbative treatment.

The OTOC variance can be represented in terms of the joint distribution function $P(x_1, x_2)$ of the two random walkers as
\begin{align}\label{eq:var_otoc_1}
    {\rm Var}&\,{\rm OTOC}(x,T) = 
    \int^{+\infty}_xdx_1dx_2\,[P(x_1,x_2, T)-P_0(x_1,x_2, T)].
\end{align}
Here $P_0(x_1,x_2, T) = G(x_1,T)G(x_2,T)$ is the distribution function is the absence of interaction [see Eq.~\eqref{eq:diffusion_kernel} for the definition of $G(x, T)$]. 
Let us use the perturbation theory to find $P(x_1, x_2)$; first, we account for a single interaction process.
In this approximation, the difference $\delta P = P - P_0$ can be represented as
\begin{align}\label{eq:distr_function_correction}
 \delta P(x_1,x_2,T) = \frac{8}{q^8} \int_{-\infty}^{+\infty} \frac{dx^\prime}{2} \int_0^{T}dT^\prime\,G^2(x^\prime, T^\prime) 2\partial_{x^\prime} G(x_1 - x^\prime, T -T^\prime) 2\partial_{x^\prime} G(x_2 - x^\prime, T - T^\prime),
\end{align}
where the integral over $x'$ corresponds to summing over every second site reflecting the brickwork structure and the derivatives are enhanced by a factor $2$ due to the discretization, see the discussion below Eq.~\eqref{eq:app_exact_ana_u1}. The term $G^2(x^\prime, T^\prime)$ describes propagation of two random walkers to a point $x^\prime$ in time $T^\prime$, where an interaction process occurs. In turn, $G(x_i - x^\prime, T - T^\prime)$ with $i\in \{1, 2\}$ describes propagation of the $i$th random walker from the interaction point to the walker's final point $x_i$ at time $T$.
Derivatives $\partial_{x^\prime}$ reflect the character of the interaction between the random walkers.
They stem from coarse-graining the rule that locally increases the weight of configurations in which the two walkers stay together, at the expense of the configurations in which the walkers separate.
Using Eq.~\eqref{eq:distr_function_correction} in Eq.~\eqref{eq:var_otoc_1} and integrating by parts, we represent the OTOC variance as
\begin{align}
    {\rm Var}\,{\rm OTOC}(x,T) =  \frac{16}{q^8} \int_{-\infty}^{+\infty} dx^\prime \int_0^{T}dT^\prime G^2(x^\prime,T^\prime)G^2(x-x^\prime
,T-T^\prime).
\end{align}
By computing this integral with the help of Eq.~\eqref{eq:diffusion_kernel}, we find (assuming $x > x_0$):
\begin{equation}\label{eq:var_result}
    {\rm Var\,OTOC}(x,T) = \frac{1}{q^6} \frac{\exp\Bigl[-\frac{(x - x_0-v_BT)^2}{2DT}\Bigr]}{\sqrt{2\pi DT}}.
\end{equation}
The wavefront velocity $v_B$ and the diffusion coefficient $D$ are defined in Eq.~\eqref{eq:D_vB}.

Equation~\eqref{eq:var_result} shows that the circuit-to-circuit fluctuations of OTOC are only appreciable at the wavefront, and decrease exponentially away from it. 
At the front, though, the fluctuations decay in a slow, algebraic way with the circuit depth, ${\rm Var\,OTOC}(x_0 + v_B T,T) \propto 1 / \sqrt{T}$.
This conclusion agrees with the phenomenological discussion in Ref.~\cite{nahum2018}.

Expression \eqref{eq:distr_function_correction} for $\delta P(x_1, x_2, T)$ can be straightforwardly extended to account for several interaction processes. However, we find that such processes produce a subleading contribution to the OTOC variance (i.e., with a higher power of $1 / T$).

\tocless\subsection{Comparison of OTOC and OTOC$_{\rm DA}$}
It is natural to ask: do circuit-to-circuit fluctuations of OTOC described by Eq.~\eqref{eq:var_result} have quantum origin (same as, e.g., the universal conductance fluctuations in the physics of mesoscopic transport), or they can be captured by purely classical laws?
Here, we show that the diagonal approximation to the calculation of OTOCs becomes exact in the limit $q \gg 1$. This allows one to view OTOC as being a classical quantity in this limit.

To quantify the closeness of the diagonal approximation to the exact OTOC value, we focus on
\begin{align}\label{eq:difference_MC}
    &\overline{\bigl[{\rm OTOC}(x,T) - {\rm OTOC}_{\rm DA}(x,T)\bigr]^2}\equiv {\rm Var\,OTOC}(x,T)-{\rm Var\,OTOC}_{\rm DA}(x,T),
\end{align}
where ${\rm OTOC}_{\rm DA}(x,T)$ stands for the value of OTOC in the diagonal approximation. The first term on the right hand side of Eq.~\eqref{eq:difference_MC} is given by Eq.~\eqref{eq:var_result}; thus, what remains to do is to find ${\rm Var\,OTOC}_{\rm DA}(x,T)$.

A calculation of ${\rm Var\,OTOC}_{\rm DA}(x,T)$ is close to the one in Sec.~\ref{sec:analytics_fluctuations}; the only difference is that we need to apply a diagonal projection after each layer of two-qubit gates.
To specify the action of the diagonal projection, it is useful to start with a general expansion of an arbitrary single-qubit four-copy operator ${\cal O}\equiv {\cal O}_{mn,pq,kl,rt}$ into the Pauli operator basis. The expansion reads
\begin{equation}\label{eq:Pauli_expansion}
    {\cal O} = \sum_{\alpha \beta \gamma \delta} c_{\alpha\beta\gamma\delta}\,\sigma^\alpha\otimes \sigma^\beta \otimes \sigma^\gamma \otimes\sigma^\delta,
\end{equation}
where $c_{\alpha\beta\gamma\delta} = {\cal O}_{mn,pq,kl,rt} \sigma^\alpha_{nm} \sigma^\beta_{qp}\sigma^\gamma_{lk} \sigma^\delta_{tr} / q^4$. 
Pauli operators satisfy ${\rm Tr}\,\sigma^\alpha\sigma^\beta =q\,\delta_{\alpha\beta}$ and $\sigma_\alpha^2 = 1$. 
The diagonal projection is a transformation of an operator ${\cal O}$ that pairs Pauli matrices in expansion \eqref{eq:Pauli_expansion} within the first and the second pairs of copies:
\begin{equation}\label{eq:diagonal_approximation}
    {\cal O}\,\overset{\rm DA}{\rightarrow}\,\sum_{\alpha \gamma} c_{\alpha\alpha\gamma\gamma}\sigma^\alpha \otimes\sigma^\alpha \otimes \sigma^\gamma\otimes \sigma^\gamma.
\end{equation}
We are interested in how the continuous application of the diagonal projection affects the dynamics of $\Sigma_4(T)$ [Eq.~\eqref{eq:projection_4}] compared to its dynamics without projections.

To understand this, we evaluate transformation rules for $\Pi \in S_4$.
First, by using Eq.~\eqref{eq:diagonal_approximation} together with Eq.~\eqref{eq:I_and_S}, we find that elements of $S_2 \times S_2$ subgroup are {\it invariant} under the diagonal approximation:
\begin{equation}\label{eq:DA_invariance}
    \begin{matrix}
    {\cal I}\\
    {\cal I}
    \end{matrix} \,\overset{\rm DA}{\rightarrow}\,\, 
    \begin{matrix}
    {\cal I}\\
    {\cal I}
    \end{matrix}\,\,, \quad
    \begin{matrix}
    {\cal S}\\
    {\cal I}
    \end{matrix} \,\overset{\rm DA}{\rightarrow}\,\, 
    \begin{matrix}
    {\cal S}\\
    {\cal I}
    \end{matrix}\,\,, \quad
    \begin{matrix}
    {\cal I}\\
    {\cal S}
    \end{matrix}\,\overset{\rm DA}{\rightarrow}\,\, 
    \begin{matrix}
    {\cal I}\\
    {\cal S}
    \end{matrix}\,\,, \quad
    \begin{matrix}
    {\cal S}\\
    {\cal S}
    \end{matrix}\,\overset{\rm DA}{\rightarrow}\,\,
    \begin{matrix}
    {\cal S}\\
    {\cal S}
    \end{matrix}\,\,.
\end{equation}
By contrast, the remaining $20$ operators in $S_4$ transform non-trivially. We summarize transformation rules for them in App.~\ref{sec:digonal_projection_permutation_rules}.

Rules \eqref{eq:DA_invariance} imply that the random walker model is applicable not only for evaluation of ${\rm Var\,OTOC}(x,T)$, but also for evaluation of ${\rm Var\,OTOC}_{\rm DA}(x,T)$:
the free motion of random walkers is not affected by the diagonal approximation.
The only object in the random walker model that is, in principle, affected by the diagonal approximation is the amplitude of interaction between the walkers.
This follows from the fact that only the last term in Eq.~\eqref{eq:collision}---i.e., the term resulting from the random walker interaction---transforms non-trivially under the diagonal projection.

We obtain the amplitude by starting with a configuration $\sum_\Pi w(\Pi)\Pi\Pi$ that emerges after the interaction process [cf.~Eq.~\eqref{eq:collision}], and then explicitly evaluating the evolution of this configurations over several subsequent applications of the gate layers.%
We find that the amplitudes with and without the diagonal projections {\it coincide} within the leading approximation [i.e., with the accuracy $\propto 1 / q^8$]. The difference only appears in the subleading corrections.
This means that 
\begin{equation}\label{eq:var_OTOC_DA}
    {\rm Var\,OTOC}_{\rm DA}(x,T) = \Bigl(1 - \frac{c}{q^2}\Bigr)\cdot {\rm Var\,OTOC}(x,T),
\end{equation}
where $c$ is a numeric coefficient $\sim 1$.
The value of $c$ can be obtained by comparing results of matrix product state simulations for the OTOC variance with and without the diagonal projections.
The use of Eq.~\eqref{eq:var_OTOC_DA} in Eq.~\eqref{eq:difference_MC} results in 
\begin{align}\label{eq:difference_MC_final}
    \frac{\overline{\bigl[{\rm OTOC}(x,T) - {\rm OTOC}_{\rm DA}(x,T)\bigr]^2}}{{\rm Var\,OTOC}(x, T)} = \frac{c}{q^2}.
\end{align}
This conclusion is the main result of this section.
It shows that the diagonal approximation becomes exact in the limit $q \rightarrow \infty$. 
As discussed in Sec.~\ref{sec:numerical_results_for_OTOCs}, though, the diagonal approximation works well even for $q = 2$ capturing at least $80\%$ of the OTOC variance.

\subsubsection{Diagonal projection rules for the elements of $S_4$}\label{sec:digonal_projection_permutation_rules}
Here we specify how the diagonal projection affects the four-copy permutation operators. The rules quoted below follow from a direct application of Eqs.~\eqref{eq:Pauli_expansion} and \eqref{eq:diagonal_approximation}.

We first set up the notations. We use standard cycle notation to denote the elements of $S_4$. For example, the identity permutation is $(\,)$; the transpositions are $(1,2)$, $(2,3)$, $(3,4)$, $(1,4)$, $(1,3)$, $(2, 4)$; the three cycles are $(1, 2,3)$, $(2,3,4)$, $\dots$, and so on. We normalize the respective operators $\Pi$ via $\Pi_{mn, nm,kp,pk} = 1$, where the four pairs of indices correspond to the four Hilbert space copies, see the discussion after Eq.~\eqref{eq:configurations_4}. We note that in these general notation, the identity and swap operators of Sec.~\ref{sec:analytics_fluctuations} are represented by 
\begin{equation}\label{eq:more_general_notation_for_permutations}
    \begin{matrix}
    {\cal I}\\
    {\cal I}
    \end{matrix} \equiv (\,)\,\,, \quad\quad
    \begin{matrix}
    {\cal S}\\
    {\cal I}
    \end{matrix} \equiv (1,2)\,\,, \quad\quad
    \begin{matrix}
    {\cal I}\\
    {\cal S}
    \end{matrix}\equiv (3,4)\,\,, \quad\text{and}\quad
    \begin{matrix}
    {\cal S}\\
    {\cal S}
    \end{matrix}\equiv (1,2)(3,4)\,.
\end{equation}

We are now in position to summarize the diagonal projection rules. 
The identity and swap permutations are invariant:
\begin{equation}
    (\,) \,\overset{\rm DA}{\rightarrow}\, (\,), \quad\quad (1,2)\,\overset{\rm DA}{\rightarrow}\,(1,2), \quad \quad(3,4)\,\overset{\rm DA}{\rightarrow}\,(3,4),\quad\quad(1,2)(3,4)\,\overset{\rm DA}{\rightarrow}\,(1,2)(3,4).
\end{equation}
The four remaining transpositions turn into an identity:
\begin{equation}
    (1,3),\,(1,4),\,(2,3),\,(2,4)\,\overset{\rm DA}{\rightarrow}\,(\,).
\end{equation}
Four out of six four-cycles transform into a double-transposition:
\begin{equation}
    (1,2,3,4),\,(1,2,4,3),\,(2,1,3,4),\,(2,1,4,3)\,\overset{\rm DA}{\rightarrow}\,(1,2)(3,4).
\end{equation}
All eight three-cycles turn into transpositions:
\begin{equation}
    (1,2,3),\,(1,3,2),\,(1,2,4),\,(1,4,2)\,\overset{\rm DA}{\rightarrow}\,(1,2)
\end{equation}
and 
\begin{equation}
    (1,3,4),\,(1,4,3),\,(2,3,4),\,(2,4,3)\,\overset{\rm DA}{\rightarrow}\,(3,4).
\end{equation}
What remains is two four-cycles and two double-transpositions. The rules for these elements are more cumbersome. For double-transpositions we find
\begin{align}
(1,3)(2,4),&\,(1,4)(2,3)\,\overset{\rm DA}{\rightarrow}\,\frac{1}{q^2-9}\Bigl[(1-6/q^2)(\,)-(1,2)-(3,4)+(q^2-6)(1,2)(3,4)\Bigr]+\frac{3}{q^2(q^2-9)}[\text{all three-cycles}]\notag\\
&-\frac{1}{q^2(q^2-9)}[\text{four remaining transpositions}] +\frac{1-6/q^2}{q^2 - 9}[\text{two remaining double-transpositions}]\\
&+\frac{1}{q^2-9}\Bigl[-(1,2,3,4) - (1,2,4,3) - (2,1,3,4) - (2,1,4,3)-\frac{1}{q^2}(1,3,2,4) -\frac{1}{q^2} (1,4,2,3)\Bigr].
\end{align}
For the two remaining four-cycles:
\begin{align}
(1,3,2,4),&\,(1,4,2,3)\,\overset{\rm DA}{\rightarrow}\,\frac{1}{q^2-9}\Bigl[-(\,)+(q^2-6)(1,2)+(q^2-6)(3,4)-q^2(1,2)(3,4)\Bigr]-\frac{1}{q^2-9}[\text{all three-cycles}]\notag\\
&+\frac{3}{q^2(q^2-9)}[\text{four remaining transpositions}] -\frac{1}{q^2 - 9}[\text{two remaining double-transpositions}]\\
&+\frac{3}{q^2-9}\Bigl[(1,2,3,4) + (1,2,4,3) + (2,1,3,4) + (2,1,4,3)+\frac{1-6/q^2}{3}(1,3,2,4) + \frac{1-6/q^2}{3}(1,4,2,3)\Bigr].\notag
\end{align}

\tocless\section{Sampling Algorithm}

\tocless\subsection{Stochastic Independence of Partitions}

In this section we discuss the linear dependence of partitions. We omit the time dependence on $T$ in the following. A partition $Y$ of a circuit refers to inserting at every space time location a projector, either $\mathcal{P}_0$ or $\mathcal{P}_2$. Note that for a part $\rho_X$ we inserted either $\mathcal{P}_{0}$ or $\mathcal{P}_1$. A partition $Y$ can be written as a sum over all parts it contains as
\begin{align}
    \rho_Y &= \sum_{ \{X_{t,l} \text{ so that } Y_{t,l}=2\}} \rho_X,
\end{align}
which essentially encodes the constraint that a partition imposes on all parts it contains: if a partition has the projector $\mathcal{P}_0$ at some point $(t,l)$, i.e., $Y_{t,l}=0$, all parts contained in that collection also have to have $\mathcal{P}_0$ at $(t,l)$, i.e., $X_{t,l}=0$ for all $(t,l)$ for which $Y_{t,l}=0$, and otherwise we sum over $X_{t,l}\in\{0,1\}$.

We have shown before that different parts are stochastically independent. Using the decomposition of partitions into parts yields the stochastic dependence of partitions
\begin{align}\label{eq:stoch_dep_coll}
    \overline{\rho_{Y_1}\rho_{Y_2}} &= \sum_{ \{X_{1,t,l} \text{ so that } Y_{1,t,l}=2\}}\sum_{ \{X_{2,t,l} \text{ so that } Y_{2,t,l}=2\}} \overline{\rho_{X_1}\rho_{X_2}}\\
    &= \sum_{ \{X_{1,t,l} \text{ so that } Y_{1,t,l}=2\}}\sum_{ \{X_{2,t,l} \text{ so that } Y_{2,t,l}=2\}} \delta_{X_1,X_2}\overline{\rho_{X_1}\rho_{X_1}}\\
    &= \sum_{ \{X_{1,t,l} \text{ so that } Y_{1,t,l}=2 \text{ and } Y_{2,t,l}=2\}}\overline{\rho_{X_1}\rho_{X_1}}
\end{align}
We can rewrite the condition $Y_{1,t,l}=2 \text{ and } Y_{2,t,l}=2$ by defining
\begin{align}
    (Y_1 \odot Y_2)_{t,l} = Y_{1,t,l}\,Y_{2,t,l}/2.
\end{align}
Note that $Y_1 \odot Y_2$ is again a partition.
So that we can write
\begin{align}
    \sum_{ \{X_{1,t,l} \text{ so that } Y_{1,t,l}=2 \text{ and } Y_{2,t,l}=2\}}{\rho_{X_1}} = \rho_{Y_1 \odot Y_2}.
\end{align}
Together with Eq.~\eqref{eq:stoch_dep_coll} it follows that $\rho_{Y_1 \odot Y_2}$ are all parts which are both, in $Y_1$ and $Y_2$ so that $\rho_{Y_2}- \rho_{Y_1 \odot Y_2}$ are all paths which are solely in $Y_2$. Hence, $\rho_{Y_1}$ and \begin{align}
     \rho_{Y_2}' = \rho_{Y_2}- \rho_{Y_1 \odot Y_2}
\end{align} 
are stochastically independent
\begin{align}\label{eq:stoch_dep_coll_result}
    \overline{\rho_{Y_1}\otimes\rho_{Y_2'}} =0.
\end{align}
Hence, they can be added to reduce the MSE reliably
\begin{align}
    \overline{(\langle O \rangle_\rho-\langle O \rangle_{\rho_{Y_1}+\rho_{Y_2}'})^2} &=
    \overline{(\langle O \rangle_\rho-\langle O \rangle_{\rho_{Y_1}}-\langle O \rangle_{\rho_{Y_2}'})^2} \\
    &=\overline{\langle O \rangle_\rho^2} -  \overline{\langle O \rangle_{\rho_{Y_1}}^2} -  \overline{\langle O \rangle_{\rho_{Y_2'}}^2} \leq \overline{(\langle O \rangle_\rho -  \langle O \rangle_{\rho_{Y_1}})^2} 
\end{align}
where we used
\begin{align}
    \overline{\rho \otimes \rho_Y} = \overline{\rho_Y \otimes \rho_Y}.
\end{align}

\tocless\subsection{Detailed Explanation Sampling for Partitions\label{ssec:sampling_grouped_gates}}

\tocless\subsubsection{Partitioning by Grouping Gates}

\begin{figure}[H]
    \centering
    \includegraphics[scale=0.85]{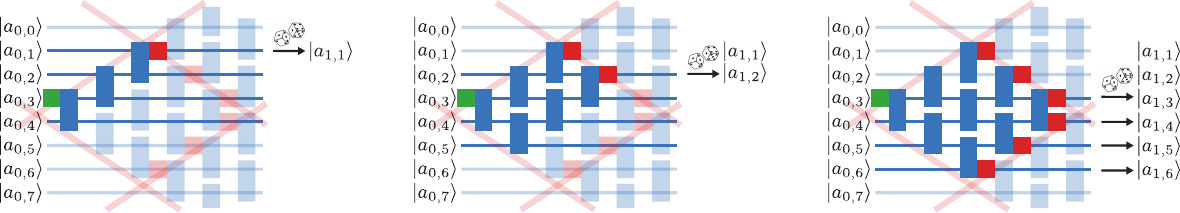}
    \caption{Initial iterative sampling procedure for getting Monte Carlo sampling estimates for partitioned circuits yielding $\rho_{Y_1}$, see Fig.~\ref{fig:diamond_slicing_eff}. Left: We start with a random product state $|a_{0,j}\rangle$ and apply the first three gates (blue, solid). Then, the first projector $\mathcal{P}_0$ (red, solid) is equivalent to a marginal sampling using the probabilities $p_{a_{1,1}} = \langle a_0| a_{1,1}\rangle \langle a_{1,1,} |a_0\rangle$. Note that this operation only involves $4$ sites and in general $\tau/2+1$ sites. Center: We iterate the procedure, again applying three more gates to the tensor $\langle a_{a,a}|a_0\rangle$ obtained from the previous iteration. Again, we use the equivalence of the $\mathcal{P}_0$ projection to sample, calculating the probabilities via a tensor network contraction, again involving only $4$ sites. Right: Final step, where the last set of gates is applied and all remaining sites are sampled. We end again with a product state. Note that previously, the operations depicted were meant to be superoperators. Here, these superoperators act on states of the form $|a\rangle \langle a|$, so that we can treat them as ordinary operators acting on $|a\rangle$.}
    \label{fig:detailed_exp_partition_sampling}
\end{figure}

In this section we discuss the details of the Monte Carlo sampling algorithm for partitions. For that, we consider the diamond-shaped partitions, Fig.~\ref{fig:diamond_slicing_eff}. The procedure is shown in Fig.~\ref{fig:detailed_exp_partition_sampling}.
For the correlators, for every sample, we start by inserting a random product state $|a_0\rangle$ next to $\rho_{\mathrm{init}}$. This yields a $s_\mathrm{init} = \pm 1$ sign. In the first iteration, we apply the first set of gates, as shown in Fig.~\ref{fig:detailed_exp_partition_sampling} (left). This involves $\tau/2+1$ sites and yields a statevector for $\tau/2+1$ sites, for now, we ignore all sites no gates are applied to. Using that the action $\mathcal{P}_{0}$ on site $1$ can be written as 
\begin{align}
    \mathcal{P}_{0,1}[\rho] = \sum_{a_{1,1}}|a_{1,1}\rangle\langle a_{1,1}|\mathrm{Tr}[\rho |a_{1,1}\rangle\langle a_{1,1}|], \label{eq:sampling_partition_det}
\end{align}
we use $\mathrm{Tr}[\rho |a_{1,1}\rangle\langle a_{1,1}|]=p_{a_{1,1}}$ as probabilities to select one element of the sum in Eq.~\eqref{eq:sampling_partition_det} and sample. We apply the selected state $|a_{1,1}\rangle$ to the statevector. In this example with qubits, this yields a tensor with $3$ "open" legs, i.e., a vector of size $2^3$ corresponding to the sites $2,3$ and $4$. Then, we can apply the next set of gates, Fig.~\ref{fig:detailed_exp_partition_sampling} (center). This can be done via tensor network contraction. Again, we are left with a vector $v$ of size $2^4$ from $4$ open legs corresponding to the sites $2,3,4$ and $5$. We use the equivalence of the projector $\mathcal{P}_{0}$ on site $2$ as in Eq.~\eqref{eq:sampling_partition_det} and use the absolute values squared $|v_i|^2$ as probabilities to calculate the marginal probabilities $p_{a_{1,2}}$. Again, we select one $a_{1,2}$ via sampling and apply the state to the tensor network, yielding a vector of size $2^3$ for the three remaining open sites $3,4$ and $5$. In the last step, we apply the remaining gates and select the states for the remaining sites by sampling. Hence, we again have a product state on all sites and can repeat the procedure with the next group of gates. After having applied the final group of gates, we obtain the product state $|a_T\rangle$. Then, we calculate the expectation value of $\sigma^z_0$ for this product state yielding a final sign $s_\mathrm{final}=\pm 1$. Repeating the sampling $M$ times and getting $M$ values $s_{\mathrm{init},m},s_{\mathrm{final},m})$, we calculate
\begin{align}
    \mathrm{Tr}[\rho_{Y_1}\sigma^z_0]\approx\frac{1}{M}\sum_m s_{\mathrm{init},m}s_{\mathrm{final},m}.
\end{align}
When combining different partitions, we use this scheme to get estimates for all required values $\mathrm{Tr}[\rho_{Y_1}\sigma^z_0]$ and combine these estimates via Eq.~\eqref{eq:combining_collections}.

\tocless\subsubsection{Partitioning by Dividing the Circuit}

\begin{figure}[H]
    \centering
    \includegraphics[scale=0.85]{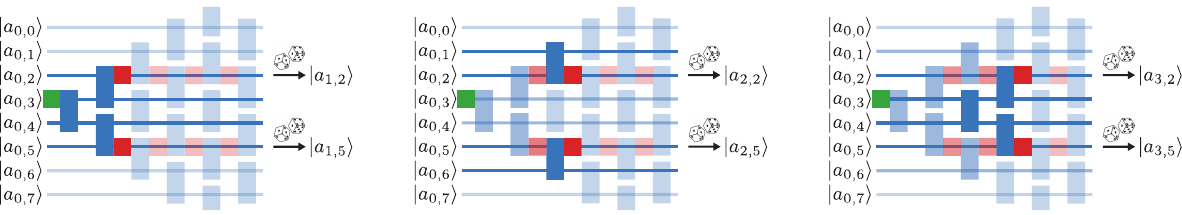}
    \caption{Iterative sampling procedure for getting Monte Carlo sampling estimates for partitioned circuits such as the shown horizontal decomposition yielding $\rho_{Y_1}$ with $\ell=3$, see Fig.~\ref{fig:horiz_slicing_eff}. As opposed to the previous scheme, this requires to keep the state of all the patches in parallel in memory. Left: Like before, we start with a random product state $|a_{0,j}\rangle$ and apply the first three gates (blue, solid). Then, the projectors $\mathcal{P}_0$ (red, solid) on the projected qubits $2$ and $5$ is again equivalent to a marginal sampling using the Born rule with probabilities $p_{a_{1,2},a_{1,5}} = \langle a_0| a_{1,2}a_{1,5}\rangle \langle a_{1,2}a_{1,5} |a_0\rangle$. Note that this operation only involves $4$ sites and in general $\ell+1$ sites. After sampling, the qubits $2$ and $5$ are not entangled and in computational basis states. Center: Next, we apply one two qubit gate to the computational basis state $|a_{0,1}\rangle|a_{2,2}\rangle$ and similar the other two qubit gate. After their application, we sample again the qubits $2$ and $5$ using the Born rule. Now we have the state of the top patch, the central patch and the bottom patch in memory. Right: We iterate the procedure again applying three gates to the central patch. The edge qubits are in computational basis states, while the two center qubits are in the state we were left with after the previous sampling procedure. Hence, when applying those gates, the qubits in the top and bottom patch are not needed, yielding a computational complexity of this operator of in general $2^\ell$. After the application of the gates, we sample again qubits $2$ and $5$. This yields the state to factorize and allowing to apply the next set of gates to the top and bottom patch without considering the qubits in the central patch.}
    \label{fig:detailed_exp_partition_sampling_hor}
\end{figure}

Next, we consider partition schemes which are based onto dividing the circuit into disconnected pieces. This means, that if we would remove all the connections on which we sample, the resulting circuit would consists of disconnected patches of the circuit with various shapes. This yields a slightly more complicated algorithm. In particular, we have to simulate multiple patches in parallel, previously a sequential simulation was possible. Here, we consider the horizontal partitioning algorithm discussed in the main text as an example. The procedure is shown in Fig.~\ref{fig:detailed_exp_partition_sampling_hor}.
In general, this scheme works exactly the same way like the sampling scheme outlined in Sec.~\ref{ssec:sampling_grouped_gates}. The only difference is that for such decomposition of the circuit (or in general decomposition which partition the circuit into disjoint pieces), we need to keep the state of all patches in memory in parallel - we have to simulate all patches in parallel. Then, we select one patch to simulate further until we would need the sampling input from the next patch. This way, we simulate the patches in an alternating fashion, using the sampling output from one patch as input to the next simulated gates of the next patch. This allows to simulate any partitioning of the circuit with a complexity scaling as the maximal tensor-network contraction cost of an individual patch.

\tocless
\subsection{Diamond partitioning\label{ssec:numerical_results_algorithm_diamond}}

In this section we discuss the results for an alternative partitioning scheme compared to the one presented in the main text - namely partitioning the circuit into diamond shaped patches. This has the advantage of being algorithmically a little bit simpler since every patch can be simulated on its own while all the remaining qubits are in a product state of computational basis states, i.e., in the $Z$-eigenbasis.

Interestingly, we find this scheme to be also efficient in terms of the scaling of the computational cost with accuracy, although less efficient than the horizontal partitioning presented in the main text. Being able to capture quantum corrections efficiently, this shows, that the diamond shape matches the shape of the quantum correction.

\begin{figure}
    \centering
    \includegraphics[scale=0.49]{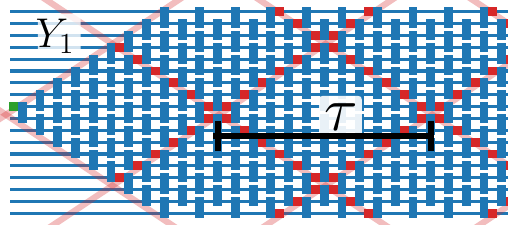}
    \includegraphics[scale=0.49]{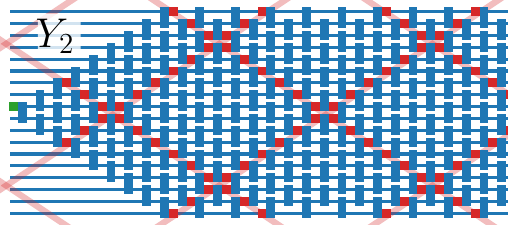}\\
    \includegraphics[scale=0.49]{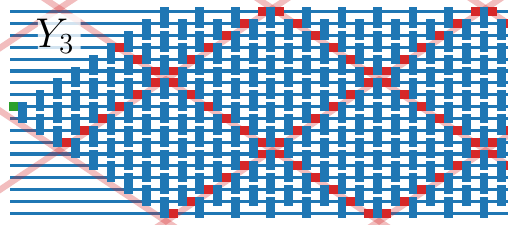}
    \includegraphics[scale=0.49]{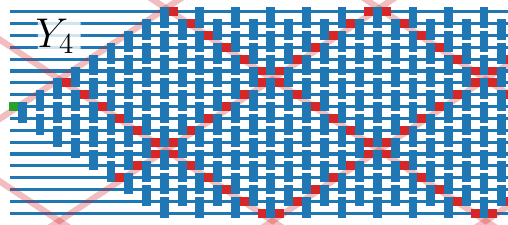}\\
    \includegraphics[scale=0.49]{fig/Figure_slicing_algorithm_heuristic_arrow.pdf}
    \includegraphics[scale=0.49]{fig/Figure_slicing_algorithm_heuristic_legend.pdf}
    \caption{One possibility for circuit partitions in 1+1D depicted similar to Fig.~\ref{fig:overview_results} (b). The initial density matrix $\rho_\mathrm{init}=\sigma^z_0$ (green) is evolved with the superoperators $\mathcal{V}_{j,j+1}$ (blue) consisting of the two-qubit gates $V_{j,j+1}$. We insert projections $\mathcal{P}_0$ (red) along a diamond pattern with periodicity $\tau$ and shift the pattern in time and space. %
    Partition $Y_2$ is created by shifting $Y_1$ in time by $\tau/2$, and $Y_3$ and $Y_4$ in time by $3\tau/4$ and space by $\pm \tau/4$. These four partitions are then combined via iterating Eq.~\eqref{eq:combining_collections}.}
    \label{fig:diamond_slicing_eff}
\end{figure}

\begin{figure}
    \centering
    \includegraphics[scale=0.75]{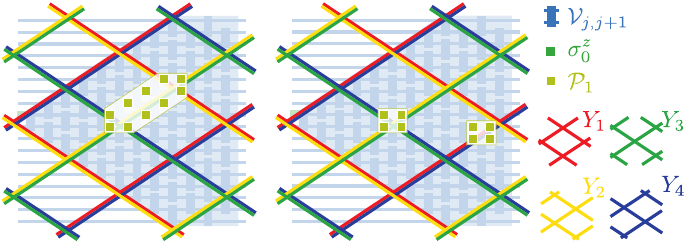}
    \caption{Two examples for parts $\rho_X$, Eq.~\eqref{eq:path_resolved_density_matrix}, which are not captured when combining the partitions shown in Fig.~\ref{fig:diamond_slicing_eff} via Eq.~\eqref{eq:combining_collections}. Left: parts with one large region of off-diagonal projections $\mathcal{P}_1$ having overlap with $\mathcal{P}_0$s in all four $Y_1,\ldots,Y_4$, so that the part is not captured. Right: parts which have two separate regions of $\mathcal{P}_1$s with overlap with $\mathcal{P}_0$s in all four $Y_1,\ldots,Y_4$. We find the large regions to be the dominant contributions up to $\tau \sim 50$, Fig.~\ref{fig:performance_alg} and App.~\ref{app:detailed_perform_heuristic_slicing}. Hence, the four partitions $Y_1,\ldots,Y_4$ are sufficient up to $\tau \sim 50$.}
    \label{fig:failure_parts}
\end{figure}

We focus on just combining four different partitions, shown in Fig.~\ref{fig:diamond_slicing_eff}, so that sampling has to be applied to a total of $15$ partitions. We find combining only four partitions to be sufficient for a large range of $\tau$. By taking only four partitions, we do not capture low Hamming distance parts where at least two quantum interactions in the form of a $\mathcal{P}_1$ occur on least two lines with the same orientation, see Fig.~\ref{fig:failure_parts}. However, based on the analytical insights, we expect their MSE contribution to be small, the MSE is dominated by contributions with a single region of $\mathcal{P}_1$s as shown in the left panel of Fig.~\ref{fig:failure_parts}. This expectation is verified numerically in the following.

\begin{figure}
    \centering
    \includegraphics%
    {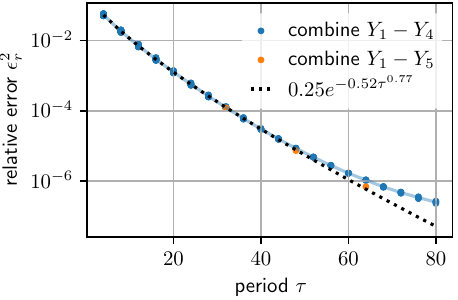}
    \caption{Relative error squared $\epsilon_\mathrm{r}^2$, Eq.~\eqref{eq:truncation_error}, for the addition algorithm with the diamond shaped partitions for different $\tau$ in the long-time-limit $T\rightarrow \infty$, restricted to $T$ being a multiple of $\tau$. Results for four partitions $Y_1,\ldots,Y_4$, Fig.~\ref{fig:diamond_slicing_eff}, (blue) and an additional fifth partition $Y_5$ (orange), which has the diamond pattern with period $\tau/2$ and is shifted in space and time by $\tau/4$ with respect to $Y_1$. Fit to 4 partition data including only $\tau\leq 40$ (dotted line).}
    \label{fig:performance_alg}
\end{figure}

The performance of the algorithm in terms of the relative error $\epsilon_\mathrm{r}$ is shown in Fig.~\ref{fig:performance_alg}.
For large times, we find the relative error of the algorithm to scale roughly as $\epsilon_\mathrm{r}^2 \propto e^{-0.5\tau^{0.77}}$. This is consistent with the expectation from a diffusive processes combined and the observed stretched exponential decay found in the main text: We found a scaling with the Hamming distance of a quantum interaction as $e^{-\mathrm{const} \sqrt{d}}$, assuming that $d$ is the area of a diffusive process of length $\tau$ and width $\sqrt{\tau}$, yields an exponent of $\sqrt{d}\sim \sqrt{\tau \sqrt{\tau}}\sim \tau^{0.75}$.

As a result, for the diamond shaped partitioning, we find the computational cost to scale in terms of the accuracy as
\begin{align}
    N_\mathrm{samples}  L T e^{\lambda' \log(\epsilon_\mathrm{r})^{1.3}}, \label{eq:scaling_simplified_alg}
\end{align}
which is significantly better than Eq.~\eqref{eq:computation_cost_scaling} but worse than Eq.~\eqref{eq:scaling_simplified_alg_hor}. For large $\tau \gtrsim 50$, four partitions appear insufficient; adding a fifth partition improves the results, Fig.~\ref{fig:performance_alg} (orange dots). It appears that for $\tau \gtrsim 50$, parts with small total Hamming distance generated in multiple locations, Fig.~\ref{fig:failure_parts} (right) dominate the error again rather than parts with a large Hamming distance generated by a single interaction location, Fig.~\ref{fig:failure_parts} (left). These parts with distributed $\mathcal{P}_1$ can be more efficiently captured by increasing the number of partitions $R$ combined, so that in practice, the number of partitions used may be adapted heuristically to $\tau$. Further details are provided in App.~\ref{app:detailed_perform_heuristic_slicing}.

The performance of this variation of the algorithm as a function of $\tau$ confirms the picture suggested by the analytical calculation. Given a part with Hamming distance $d(X)$, the most likely parts or dominant parts in terms of the variance are indeed those where the decoherence sensitivity or $\mathcal{P}_1$s accumulate in space and time rather than dispersing, Fig.~\ref{fig:failure_parts}.

Note that all results presented here are calculated using an ensemble average. In practice, for individual circuits, the algorithm proposed here may only be feasible to run on current classical computers up to $\tau \approx 60$ without any further approximations. 

\tocless\subsection{Convergence of the Scaling Predictions\label{app:detailed_perform_heuristic_vert_slicing}}
\begin{figure}[H]
    \centering
    \includegraphics[]{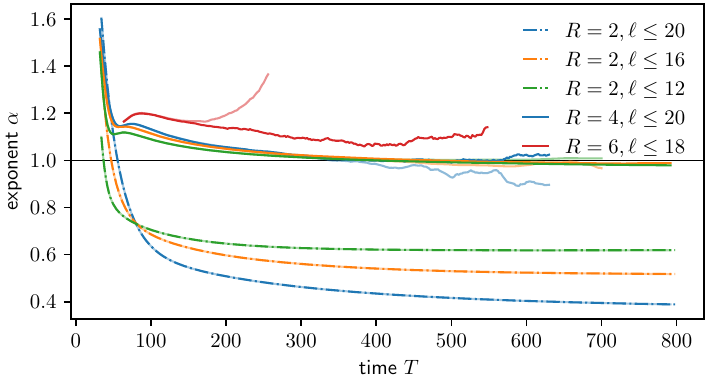}
    \caption{Results for $\alpha$ fitting $a e^{b \ell^\alpha}$ to $\epsilon_r$ for the data also presented in Fig.~\ref{fig:performance_alg_vert}. Results are shown for combining two partitions $R=2$ (dash-dot, blue, orange, green), four partitons $R=4$ (solid, blue, orange, green) and six partitions (red). We vary the fitting range and fit to the data with $\ell \leq 12$ (green),  $\ell \leq 16$ (orange), $\ell \leq 20$ (blue) and $\ell \leq 18$ (red). Different numerical accuracies are shown, namely an itensor cutoff parameter of $10^{-20}$ (semi-opaque) and $10^{-22}$ (opaque). This shows that the scaling exponent $\alpha$ drifts with time $T$ to smaller values. For $R=4$ it converges to a value close to $1$, while for $R=6$ for the simulated times a value above one seems likely, but the results are clearly impacted by numerical noise.}
    \label{fig:hor_part_exp}
\end{figure}

In this section we provided more details on the performance of the horizontal partitioning algorithm presented in the main text. In particular, we show in Fig.~\ref{fig:hor_part_exp} the results for fitting decays $a e^{b \ell^\alpha}$ to $\epsilon_r$ for the data also presented in Fig.~\ref{fig:performance_alg_vert}. This supports the expectation of the sampling algorithm scaling polynomial, but does not provide definite evidence as the results are limited by numerical noise. When combining four partitions $R=4$, the exponent $\alpha$ seems to converge with time $T$ to a value close to $1$. Since the results are not converged with $R$ at $R=6$, this is not the best scaling we can expect to observe. Increasing to $R=6$ indeed yields an exponent $\alpha>1$ for the times simulated, up to numerical noise. This suggests that it is indeed possible to increase $R$ with increasing $\ell$ such that $\epsilon_r$ decays exponentially, yielding a polynomial scaling algorithm.

\tocless\subsection{Detailed performance of the Diamond partitioning\label{app:detailed_perform_heuristic_slicing}}
In this section we discuss the details of the performance of the addition algorithm when using four partitions instead of $\tau/2$. The main result is, that even with four partitions, we get most of the variance contributions up to Hamming distance $d\sim \tau$.

\begin{figure}[H]
    \centering
    \includegraphics[scale=1]{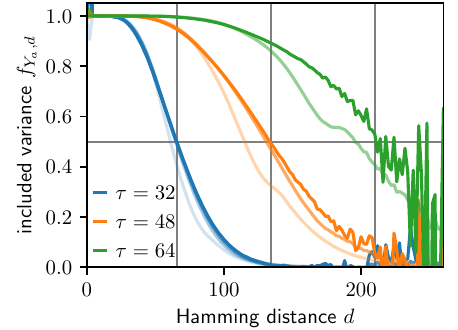}
    \includegraphics[scale=1]{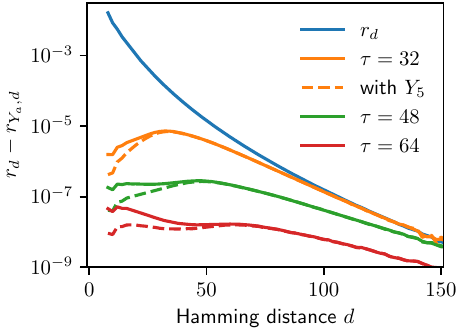}
    \caption{Left: Fraction of the variance $f_{Y_a,d}$, Eq.~\eqref{fraction_captured}, captured by the addition algorithm with four partitions, see Fig.~\ref{fig:diamond_slicing_eff} for different times up to about $T=150$, $T$ being a multiple of $\tau$, and different periods $\tau$ of the partition pattern. Right: Residual variance distribution $r_d - r_{Y_a,d}=(\mathrm{Var}\langle O \rangle_{\rho_{d}}-\mathrm{Var}\langle O \rangle_{\rho_{Y_a,d}})/\mathrm{Var}\langle O \rangle_{\rho}$ not captured by the algorithm for different $\tau$ (orange, green and red) for large times $T\approx 200$. For reference, the full variance distribution $r_d$ (blue) are also shown. For the dotted lines, we added one more partition $Y_5$ as explained in the main text, Fig.~\ref{fig:performance_alg}, in order to capture low-coherence paths more efficiently.}
    \label{fig:distribution_algorithm_coherence}
\end{figure}

For that, consider Fig.~\ref{fig:distribution_algorithm_coherence}. Here, we show as a function of the Hamming distance $d(X)$, the fraction of the variance 
\begin{align}\label{fraction_captured}
    f_{Y_a,d} = \frac{\mathrm{Var}\langle O \rangle_{\rho_{Y_a,d}}}{\mathrm{Var}\langle O \rangle_{\rho_{d}}}
\end{align} 
captured by the algorithm, where 
\begin{align}
    \rho_{Y_a} = \rho_{Y_1 \cup Y_2 \cup Y_3 \cup Y_4}
\end{align}
and $\rho_{Y_a,d}$ is the Hamming distance decomposed $\rho_{Y_a}$, see Eq.~\eqref{eq:coherence_decomp}.

This confirms the expectation of the Hamming distance coming from small regions. If the Hamming distance would be dispersed across the whole part, with a low local Hamming distance density, we would expect a failure of the algorithm and an exponential decay of the captured relative variance of the form $f_{Y_a,d}\sim e^{-\mu d(X)/\tau}$ with some $\mu$. The reason being, that the algorithm misses parts where the Hamming distance terms are on two lines with distance of at least $\tau$ to each other, see Fig~\ref{fig:failure_parts}. Now, increasing the Hamming distance in a low-Hamming-distance density setting would yield the probability to not hit any of those lines with a $\mathcal{P}_1$ to decay exponentially, the probability for an individual hit scales with the fraction of space-time locations where $Y_a$ has a $\mathcal{P}_0$, which scale as $\sim\mu/\tau$ with $\tau$, so that the amount of space-time locations without $\mathcal{P}_0$ in $Y_a$ scale as $(1-\mu/\tau)\approx e^{-\mu/\tau}$. Now, inserting $d$ projectors $\mathcal{P}_1$, we get a probability of $e^{-\mu d/\tau}$ that none of the $\mathcal{P}_1$ and $\mathcal{P}_0$ coincide, yielding an expected exponential decay of $f_{Y_a,d}$. Instead, we find in Fig.~\ref{fig:distribution_algorithm_coherence} that the algorithm captures most of the variance for parts with a Hamming distance $d$ up to very roughly $d(X)\sim \tau^{1.75}$ (the three different $\tau$ analyzed are insufficient to make precise statements about the exponent).

This further indicates that the Hamming distance does not only accumulate in space and time, it also forms regions with a smaller surface than volume. If the volume would scale with the surface, as if the regions forms lines, we would expect to not being able to capture any parts with coherence larger than $d(X)\sim\tau$. However, it seems like we capture most of the coherence up to $d(X)\sim \tau^{1.75}$. Based on the underlying dynamics being diffusive, this may not be too surprising.

In any way, the Hamming distance forming regions, i.e., the $\mathcal{P}_1$ typically accumulating in a single region, is a behavior making the simulation easier, it condenses the Hamming distance in space and time, allowing it to be captured by the algorithm more efficiently in practice compared to the more rigorous scaling analysis yielding Eq.~\ref{eq:computation_cost_scaling}. 

Last, one may then ask what parts are the most important ones left out by the algorithm. For that, we show the residual variance distribution $r_d - r_{Y_a,d}=(\mathrm{Var}\langle O \rangle_{\rho_{d}}-\mathrm{Var}\langle O \rangle_{\rho_{Y_a,d}})/\mathrm{Var}\langle O \rangle_{\rho}$ in Fig.~\ref{fig:distribution_algorithm_coherence} (right).

For small $\tau=32$ we observe that the most residual variance comes from parts with coherence $d(X)\sim\tau$. For larger $\tau$, more variance is captured, the residual variance decreases. However, the lower Hamming distance parts now start to dominate the residual variance. This is the trade-off one may want to optimize: By increasing the number of partitions these small Hamming distance parts can be captured at the cost of increasing the computational complexity, making the scaling for large desired accuracies slightly worse than stated in Eq.~\eqref{eq:scaling_simplified_alg}. This is shown with the dashed lines in Fig.~\ref{fig:distribution_algorithm_coherence} (right), where we added one more partition $Y_5$ to $Y_a$, capturing a larger fraction of the low-Hamming distance parts.

However, adding $Y_5$ does not help capturing any more of the large-Hamming distance parts. In particular, the maximum of the residual coherence at $d(X)=\tau$ is unchanged.
This suggests that most of the residual variance originates from parts of the form that $d(X)=2$ for a duration of $\tau/2$, the minimal distance between the edges of the diamonds Fig.~\ref{fig:diamond_slicing_eff}, and $d(X)=0$ otherwise. These can only be captured by increasing $\tau$ and not by adding more partitions with the same or smaller $\tau$.

\end{document}